\documentclass[useAMS,usenatbib]{mnras}

\usepackage{subfigure,wrapfig,placeins}
\usepackage{graphicx,amsmath,amssymb,multirow,units}
\usepackage{euscript}
\usepackage{zref-savepos}
\usepackage{xspace}
\usepackage{mathptmx}
\usepackage{blindtext}
\usepackage[colorinlistoftodos,prependcaption,textsize=small]{todonotes}
\usepackage{breakcites}
\usepackage{microtype}
\usepackage{ctable}
\usepackage{tabularx}
\usepackage{aas_macros}

\newcommand{\ie}{{i.e.\/}\xspace}
\newcommand{\eg}{{e.g.\/}\xspace}
                 
\newcommand{\asec}{\ensuremath{^{\prime\prime}}}
\newcommand{\amin}{\ensuremath{^{\prime}}}
\newcommand{\iband}{\ensuremath{i}-band\xspace}
\newcommand{\halpha}{H\ensuremath{\alpha}\xspace}
\newcommand{\hbeta}{H\ensuremath{\beta}\xspace}
\newcommand{\degree}{\ensuremath{^{\circ}}\xspace}
\newcommand{\angstrom}{\,\AA\xspace} 
\newcommand{\sextractor}{\texttt{Source Extractor}\xspace}
\newcommand{\javelin}{\texttt{Javelin}\xspace}
\newcommand{\kali}{\texttt{KALI}\xspace}
\newcommand{\fastkde}{\texttt{FastKDE}\xspace}
\newcommand{\lag}{\ensuremath{t_{lag}}\xspace}
\newcommand{\AGNMassCatalogue}{AGN Mass Catalogue\xspace}

\newcounter{whereq}
\newcommand{\location}[1]{\zsaveposy{#1}}
\newcommand{\where}[1]{  \stepcounter{whereq}  \location{whereq-\thewhereq}    \ifdim\zposy{whereq-\thewhereq}sp<\zposy{#1}sp
    above  \else
    below  \fi}

\graphicspath{{figures/}}
\DeclareGraphicsExtensions{.pdf}

\title[Photometric Reverberation Mapping]{The Performance of Photometric Reverberation Mapping at High Redshift and the Reliability of Damped Random Walk Models}
\author[S.\,C.\,Read et al.]
{S.\,C.\,Read$^{1,2}$\,\thanks{E-mail: shaun.c.read@gmail.com}, 
D.\,J.\,B. Smith$^1$, 
M.\,J. Jarvis$^{3,4}$
G. G\"urkan$^{1,5}$ \\
$^{1}$Centre for Astrophysics Research, School of Physics, Astronomy and Mathematics, University of Hertfordshire, Hatfield, Herts, AL10 9AB, UK\\
$^{2}$INAF -- Osservatorio Astronomico di Roma, via Frascati 33, I-00040 Monte Porzio Catone, Roma, Italy\\
$^{3}$Astrophysics, University of Oxford, Denys Wilkinson Building, Keble Road, Oxford, OX1 3RH, England\\
$^{4}$Physics and Astronomy Department, University of the Western Cape, Bellville 7535, South Africa\\
$^{5}$CSIRO Astronomy and Space Science, 26 Dick Perry Avenue, Kensington, Perth, 6151, WA, Australia}

\begin{document}
\date{\today}

\pagerange{\pageref{firstpage}--\pageref{lastpage}} \pubyear{2019}

\maketitle

\label{firstpage}

\newcommand{\TargetTenFullName}{\texttt{SDSSJ144645.44+625304.0}\xspace}
\newcommand{\TargetTen}{Target-10\xspace}
\newcommand{\TargetTenCadence}{14 days\xspace}
\newcommand{\TargetTenMinSNR}{\ensuremath{>20}\xspace}
\newcommand{\TargetTenRedshift}{0.351}
\newcommand{\SplineGradientTolerance}{$0.05$ mag\xspace}
\newcommand{\SDSSMagnitudeAggreement}{$0.05$ mag\xspace}
\newcommand{\AutocorrelationTimeTolerance}{1 per cent\xspace}
\newcommand{\AutocorrelationTimeMultiple}{50\xspace} \newcommand{\JavelinNWalkers}{200\xspace}
\newcommand{\NumberOfSimulatedLightCurves}{50\,000\xspace}
\newcommand{\TargetTenReferenceSpectrumRaDec}{\ensuremath{\alpha_{J2000} = 14^{h}46^{m}37^{s},  \delta_{J2000} = +62\degree 57\amin 36\asec}}
\newcommand{\TargetTenLagPeakLow}{63 days\xspace}
\newcommand{\TargetTenLagPeakHigh}{72 days\xspace}

\newcommand{\TargetTenUninformedLag}{\ensuremath{73^{+4}_{-13}} days\xspace}
\newcommand{\TargetTenInformedLag}{\ensuremath{65^{+6}_{-1}} days\xspace}
\newcommand{\TargetTenInformedSNR}{\ensuremath{18.6}\xspace}
\newcommand{\TargetTenLagUncertaintyWidth}{7 days\xspace}
\newcommand{\SNRIncrease}{\ensuremath{\sim 2.2}\xspace}
\newcommand{\TargetTenUninformedMass}{\ensuremath{10^{8.27_{-0.15}^{+0.13}} \mathrm{M_{\odot}}}\xspace}
\newcommand{\TargetTenInformedMass}{\ensuremath{10^{8.22^{+0.13}_{-0.15}}\mathrm{M_{\odot}}}\xspace}
\newcommand{\ffactor}{\ensuremath{\langle f \rangle = 4.3 \pm 1.1}\xspace}
\newcommand{\AverageLagRecoveryErrorLessThan}{6 per cent\xspace}
\newcommand{\ReliableLagsBelow}{170 days\xspace}
\newcommand{\IterativeReweightingTolerance}{0.001 mag\xspace}
\newcommand{\NumberOfBlackHoleMasses}{100\xspace}
\newcommand{\EfficiencyPercentIncrease}{218 per cent\xspace}
\newcommand{\RoughEfficiencyPercentIncrease}{$\sim$ 200 per cent\xspace}
\newcommand{\TargetTenBaseline}{330 days} 
\begin{abstract}
  Accurate methods for reverberation mapping using photometry are highly sought after since they are inherently less resource intensive than spectroscopic techniques.
  However, the effectiveness of photometric reverberation mapping for estimating black hole masses is sparsely investigated at redshifts higher than $z\approx0.04$.
  Furthermore, photometric methods frequently assume a Damped Random Walk (DRW) model, which may not be universally applicable.
  We perform photometric reverberation mapping using the \javelin photometric DRW model for the QSO \TargetTenFullName at $z=0.351$ and estimate the \hbeta lag of \TargetTenInformedLag and black-hole mass of \TargetTenInformedMass.
  An analysis of the reliability of photometric reverberation mapping, conducted using many thousands of simulated CARMA process light-curves, shows that we can recover the input lag to within \AverageLagRecoveryErrorLessThan on average given our target's observed signal-to-noise of \TargetTenMinSNR and average cadence of \TargetTenCadence (even when DRW is not applicable).
  Furthermore, we use our suite of simulated light curves to deconvolve aliases and artefacts from our QSO's posterior probability distribution, increasing the signal-to-noise on the lag by a factor of \SNRIncrease.
  We exceed the signal-to-noise of the Sloan Digital Sky Survey Reverberation Mapping Project (SDSS-RM) campaign with a quarter of the observing time per object, resulting in a \RoughEfficiencyPercentIncrease increase in SNR efficiency over SDSS-RM.
\end{abstract}


\section{Introduction}\label{sec:RM:intro}
All active galactic nuclei (AGN) are believed to be powered by an accretion disk around a central super-massive black-hole (SMBH) which is itself surrounded by a broad-line region \citep[BLR;][]{Antonucci1993Unified,Urry1995Unified,Ho2008Nuclear,Heckman2014Coevolution}. 
The mass of the SMBH has been observed to scale with the properties of its host galaxy (\eg \citealt{Magorrian1998Demography,Silk1998Quasars,Benson2003Shapes,Haering2004On,Croton2006Many,Guo2011From}; and \citealt{Kormendy2013Coevolution} for a full review) and so it is essential that accurate masses for the SMBH can be derived in order to investigate the effect AGN feedback has on their host galaxies.

In the absence of a direct black-hole mass measurement, there exist scaling relations based on emission line widths (\eg $\mathrm{H\beta}$: \citealt{Wandel1999Central} and $\mathrm{Mg_{II}}$: \citealt{Mclure2002Measuring}) and luminosity at 5100\,\AA\,(\eg \citealt{Bentz2013LowLuminosity}).
These relations are typically calibrated at low redshift and have not been extended to high redshift \citep{Hiner2015Probing,Barisic2017Stellar} despite wide-spread extrapolated use at high redshift \citep{Mclure2004Cosmological,Vestergaard2004Early,Vestergaard2006Determining,Netzer2007Black,Runnoe2013Rehabilitating,Feng2014SingleEpoch,Restrepo2016Active}.
Therefore, it is also for the purposes of validating these scaling relations that more black-hole mass measurements at higher redshifts are needed.

Reverberation mapping \citep{Blandford1982Reverberation,Gaskell1986Line,Gebhardt2000Relationship,Ferrarese2000Fundamental,Peterson2004Black} is a powerful technique for estimating black-hole masses. 
Assuming that the broad-line region is gravitationally dominated by the SMBH, it is possible to estimate the black-hole mass from the time delay between continuum emission from the accretion disk and the reprocessed emission from the broad-line region, also known as the ``lag'', from the Keplerian motion equation:
\begin{equation}\label{eq:RM:mass}
  M_{BH} = f\frac{R_{BLR} \sigma_{\mathrm{disp}}^{2}}{G},
\end{equation}
\noindent where the virial parameter $f$ describes the structure and orientation of a broad-line region with radius $R_{BLR} = ct_{\mathrm{lag}}$ and velocity dispersion, $\sigma_{\mathrm{disp}}$, of the broad-line region.
Assuming that the virial factor, $f$, is fully generated by the inclination, $\theta$, of the disc, $f = 1/4 \sin^2\theta$ and so at $\theta=30^{\circ}$, $f=1$ \citep{Mclure2001Black,Liu2017New}.
The $f$ can be determined on a case-by-case basis by modelling the BLR using spectroscopic measurements \citep{Pancoast2011Geometric,Pancoast2014Modelling,Williams2018Lick} or purely photometric means \cite{Nunez2014Modelling}, through gravitational redshift measurements \citep{Liu2017New}, or through combinations of independent black-hole mass estimators.
However, it is common to use an aggregated average for use in large data sets. 
\citet{Grier2013Structure}, \citet{Onken2004Supermassive}, \citet{Park2012Recalibration}, and \citet{Graham2011Expanded} have measured values of $\langle f \rangle = 4.3 \pm 1.1$, $5.5\pm1.8$, $5.1\pm1.3$, and $2.8\pm 0.6$ respectively from the independently measured stellar velocity dispersions.

So far, about \NumberOfBlackHoleMasses black-hole masses have been measured using spectroscopic reverberation mapping techniques \citep{Kaspi2000Reverberation,Bentz2009RadiusLuminosity,Bentz2009Lick,Denney2010Reverberation,Bentz2013LowLuminosity,Barth2015Lick,Grier2012Reverberation,Shen2015Sloana,Du2015Supermassive,Du2016Supermassive,Du2016Supermassivea,Grier2017Sloan}, which require long-term spectroscopic observations to recover their lags.
Since BLR radii can span up to several hundred light days \citep{Peterson2004Black,Bentz2014Mass,Fausnaugh2017Reverberation,Williams2018Lick} light curve observations need to take place over several months or years to match features in the continuum to the echoes from the BLR, with 3 times the observed-frame lag being the recommended baseline \citep{Shen2015Sloan}.
Cosmological time dilation increases the time-scale of observed variability and so high-redshift QSOs require much longer observational campaigns than low-redshift QSOs.
To compound this effect, higher-redshift QSOs are intrinsically more luminous than lower-redshift QSOs, which implies that they have longer lag time-scales than lower-redshift QSOs (given the lag-luminosity relation).

\citet{Fine2013Stacked} and then \citet{Brewer2014Hierarchical} have developed methods to recover lags from the stacked cross-correlations of photometric and spectroscopic observations to be used when individual lags are poorly constrained but there is a large sample of AGN.
This method allows for the detection of emission-line lags for a population of AGN at very high redshift (\citealt{Fine2013Stacked} use a sample of AGN with redshifts $z \lesssim 4.5$) and provides convincing evidence for the decreasing BLR radius for emission-lines with higher excitation energies.
However, stacked reverberation mapping is a statistical technique and cannot provide more signal-to-noise for individual objects.

An extra source of inefficiency for spectroscopic campaigns is the need to disperse the light and subsequent decreased signal-to-noise especially at high redshift.
Therefore, observing emission lines spectroscopically for reverberation mapping is expensive due to the required overhead, and restricted to bright or low redshift sources and so accurate photometric methods for reverberation mapping are highly sought after.

The variability of the BLR emission line can be captured within a redshifted narrow-band (or broad-band) photometric filter through the careful separation of the underlying, driving continuum \citep{Haas2011Photometric,Chelouche2012Photometric,Nunez2012Photometric,Zu2016Application}.
This can be done either by modelling the variability using a stochastic time-series model such as the Damped Random Walk \citep{Zu2011Alternative,Zu2013Quasar,Zu2016Application} or by more empirical measures such as cross-correlation analysis, which are model-independent \citep{White1994Comments,Rybicki1994Study,Peterson2004Black,Chelouche2012Photometric,Shen2015Sloan,Fausnaugh2017Reverberation}.

\javelin \citep{Zu2013Quasar,Zu2011Alternative,Zu2016Application} is a parametric Bayesian tool which models the variability of the QSO itself rather than extracting peaks from empirical cross-correlation functions. 
Modelling the continuum emission as a DRW has some advantages over cross-correlation in that it allows for natural inclusion of Bayesian inference techniques for noisy data from which parameter values and uncertainties can be estimated \citep{Zu2011Alternative,Zu2013Quasar}.
Stochastic DRW models of the accretion disk continuum emission are based on physical assumptions that can be tested by observations. 
The physical mechanism supporting the use of DRW models is the stochastic heating of the accretion disk by the central source and its subsequent variability due to thermal fluctuations \citep{Kelly2009Are}. 
However, there is growing evidence that DRW is not universally applicable and that more complex time-series models are necessary to explain the correlations at high frequency (e.g. \citealt{Kelly2014Flexible,Kasliwal2015Do,Kasliwal2017Extracting,Guo2017Far,Smith2018Kepler}).
If this is the case, then assuming a DRW when interpolating light-curves (in order to estimate the lag) may introduce artificial peaks in the posterior distribution. 
Therefore, it may be beneficial to estimate the lag without interpolation, as with a Von Neumann estimator \citep{Chelouche2017Methods} or ZCDF method \citep{Alexander2013Improved}.
However, these methods have their own problems, when binning with few data points, and biases due to the combined continuum and line light-curve in the narrow-band photometric filter.
Although the sample of reverberation mapped QSOs is becoming more representative (in terms of luminosity and redshift) with time, the current sample is biased to low redshift QSOs and a narrow range of emission line properties \citep{Shen2015Sloan,Grier2017Sloan}. 
If photometric reverberation mapping can recover precise lag estimates for SMBHs, then fewer resources would have to be spent on spectroscopic campaigns in order to fill in the parameter space of black-hole mass, luminosity and redshift.

Photometric reverberation mapping has been performed on both individual targets below $z=0.04$ \citep{Haas2011Photometric,Edri2012Broadband,Nunez2012Photometric,Ramolla2014Photometric,Nunez2014Modelling,Carroll2015Photometric,Hood2015Photometric,Nunez2015BroadLine} and for a sub-sample of the SDSS-RM \citep{Shen2015Sloan} catalogue \citep{Hernitschek2015Estimating,Zhang2017Broadband}.
However, the estimated uncertainties for these SDSS-RM sub-samples are typically larger than $100$ per cent.
Photometric reverberation mapping has also been applied to the continuum to measure the properties of the accretion disk \citep{Mudd2017Quasar,Cackett2018Accretion}, though not to estimate black-hole masses until recently \citep{Nunez2019Optical}.

This work sets out to demonstrate the efficacy and reliability of photometric reverberation mapping even for higher redshift targets and to test its accuracy when the DRW assumption is not applicable. 
We aim to produce the first robust photometric reverberation mapped black-hole mass with a redshift above $z=0.04$.

In Section~\ref{sec:RM:methods}, we carefully pre-select targets to give us the best possible chance of recovering precise lags. 
We specify that candidates must have redshifts that allow the use of a redshifted \halpha photometric filter and have expected observed lags (from the lag-luminosity relation \citealt{Bentz2013LowLuminosity}) such that they can be observed for $3t_{\mathrm{lag}}$ days over multiple semesters.
We then detail our observations and the methods used to produce photometric light-curves for use with \javelin.
Before fitting QSO variability models to our observations, we produce a suite of simulated light-curves in order to test how well \javelin can recover known lags for QSOs with the same cadence and signal-to-noise as our target observations.
In Section~\ref{sec:RM:results} we present the fitted BLR lag and black-hole mass distributions for our observations.
In order to test whether the slope is significantly affected by non-Gaussian errors, we also apply rigorous statistical analysis to the fitting of the \hbeta lag-luminosity relation by not assuming Gaussian uncertainties for either our targets or for the \citet{Grier2017Sloan} catalogue.
In Section~\ref{sec:RM:discussion} we compare the efficiencies of the SDSS-RM campaign \citep{Shen2015Sloan,Grier2017Sloan} and our own, in terms of signal-to-noise of the fitted lag.
We also discuss future potential applications of photometric reverberation mapping in upcoming surveys where such techniques can easily be applied.
Finally, we summarise our conclusions and outlook in Section~\ref{sec:RM:conclusions}.

\section{Methods}\label{sec:RM:methods}
Our intermittent requirements make RM observations of small samples of high redshift targets unsuited to continuous observing campaigns.
We observed our QSOs robotically with the Liverpool Telescope \citep{Steele2004Liverpool} since it can accommodate our discontinuous observation campaign.
We make use of the optical components of the infrared-optical (IO:O) suite of instruments available on the Liverpool Telescope since a range of $\halpha$ filters are available in addition to the SDSS $ugriz$ filters.
This allows us to observe the \hbeta emission lines of a wide range of high redshift QSOs, since their observed emission line will fall within the bandpass of one of the available \halpha filters.

\subsection{Target selection}\label{sec:RM:methods:targets}
We select our targets to have $i_{AB} < 18$, spectroscopically-confirmed in the SDSS DR12  \citep{York2000Sloan,Eisenstein2011SdssIii} or BOSS \citep{Dawson2013Baryon}, and have broad \hbeta emission lines with equivalent widths $>50$\angstrom.
We only select those QSOs whose redshifted $H\beta$ line will fall into one of the IO:O $H\alpha$ photometric filters. 
Additionally, using the 5100\angstrom luminosities from \citet{Shen2011Catalog} and the $R-L_{5100}$ relation from \citet{Bentz2013LowLuminosity}, we pre-select targets that are likely to have observed lags $\lag(1+z) < 95$ days. 

\citet{Shen2015Sloan} construct 2 metrics in order to determine which combination of properties of their simulated light-curves yield the most accurate lag detections.
They find that the ratio of the number of data points contributing to the calculation of the cross-correlation function to the number of data points that contribute to resolving the true lag is typically $\approx 2$ for detected lags.
In the limit of $N_{\textrm{epoch}} \gg 1$, this is equivalent to a requirement on the total observing run duration of 3 times the true observed lag, $t_{\textrm{span}} / t_{\textrm{lag}} = 3$.
We therefore imposed an additional criterion that the QSOs be observable for at least 3 times the length of their expected lag between the 14 months of the Liverpool Telescope extended 2015B and 2016A semesters. 
Applying these constraints yields 10 targets which we submitted for observation. 

\begin{table*}
\centering
\begin{tabular}{ccccccccccccc}
\toprule
 & RA & Dec & & $i_{AB}$ & $\lambda L_{5100}$ & $t_{H\beta}(1+z)$ & \multicolumn{2}{c}{visibility} & \multicolumn{2}{c}{epochs} & \multicolumn{2}{c}{observed baseline}\\
SDSS ID &          /\degree &        /\degree &       z &   /mag & /$\log[\mathrm{erg s^{-1}}]$ & /days & /days & /$t_{H\beta}(1+z)$ & \halpha & \iband & /days & /$t_{H\beta}(1+z)$\\
\midrule
J082905.01+571541.6 &   127.2708 &   57.2615 &   0.350 &   17.9 &     44.24 &       61.4 &             250 &          4.1 & 0 & 0 & 0 & 0\\
J153601.07+162838.4 &   234.0044 &   16.4773 &   0.382 &   17.9 &     44.35 &       72.2 &             301 &          4.2 & 16 & 14 & 208 & 2.9\\
J112600.00+304005.3 &   171.5000 &   30.6681 &   0.361 &   17.9 &     44.41 &       76.1 &             231 &          3.0 & 0 & 0 & 0 & 0\\
J122519.30+372053.6 &   186.3304 &   37.3482 &   0.388 &   17.9 &     44.45 &       81.9 &             249 &          3.0 & 11 & 11 & 180 & 2.2\\
J154246.51+194626.1 &   235.6938 &   19.7739 &   0.398 &   17.9 &     44.46 &       83.1 &             305 &          3.7 & 0 & 0 & 0 & 0\\
J150243.09+111557.3 &   225.6795 &   11.2659 &   0.390 &   17.5 &     44.56 &       93.4 &             278 &          3.0 & 0 & 0 & 0 & 0\\
J153057.45+304322.0 &   232.7393 &   30.7228 &   0.378 &   17.6 &     44.57 &       93.5 &             314 &          3.4 & 17 & 19 & 189 & 2.0\\
J164224.30+444509.8 &   250.6012 &   44.7527 &   0.368 &   17.8 &     44.58 &       94.5 &             334 &          3.5  & 16 & 15 & 179 & 1.9\\
J153729.23+272250.7  &  234.3717  &  27.3807  &  0.388  &  17.9  &     44.57 &       95.0 &             312 &          3.3 & 0 & 0 & 0 & 0\\
\textbf{J144645.44+625304.0} &   \textbf{221.6893} &   \textbf{62.8844} &   \textbf{0.351} &   \textbf{17.8} &     \textbf{44.45} &       \textbf{91.7} &             \textbf{337} &          \textbf{3.7} & \textbf{31} & \textbf{30} & \textbf{329} & \textbf{3.6}\\
\bottomrule
\end{tabular}
\caption{Our pre-observation targets selected based on their luminosity, redshift and visibility. 
$t_{H\beta}(1+z)$ is expected observed-frame lag derived from the \citet{Bentz2013LowLuminosity} $t_{\mathrm{rest}}-L_{5100}$ relation, where $L_{5100}$ is estimated from the SDSS spectrum. 
We quote the length of time each target is visible and the baseline of the observations in units of days and expected observed-frame lag.}
\label{tab:RM:targets}
\end{table*}
 
\begin{figure}
  \centering
  \includegraphics[width=\linewidth]{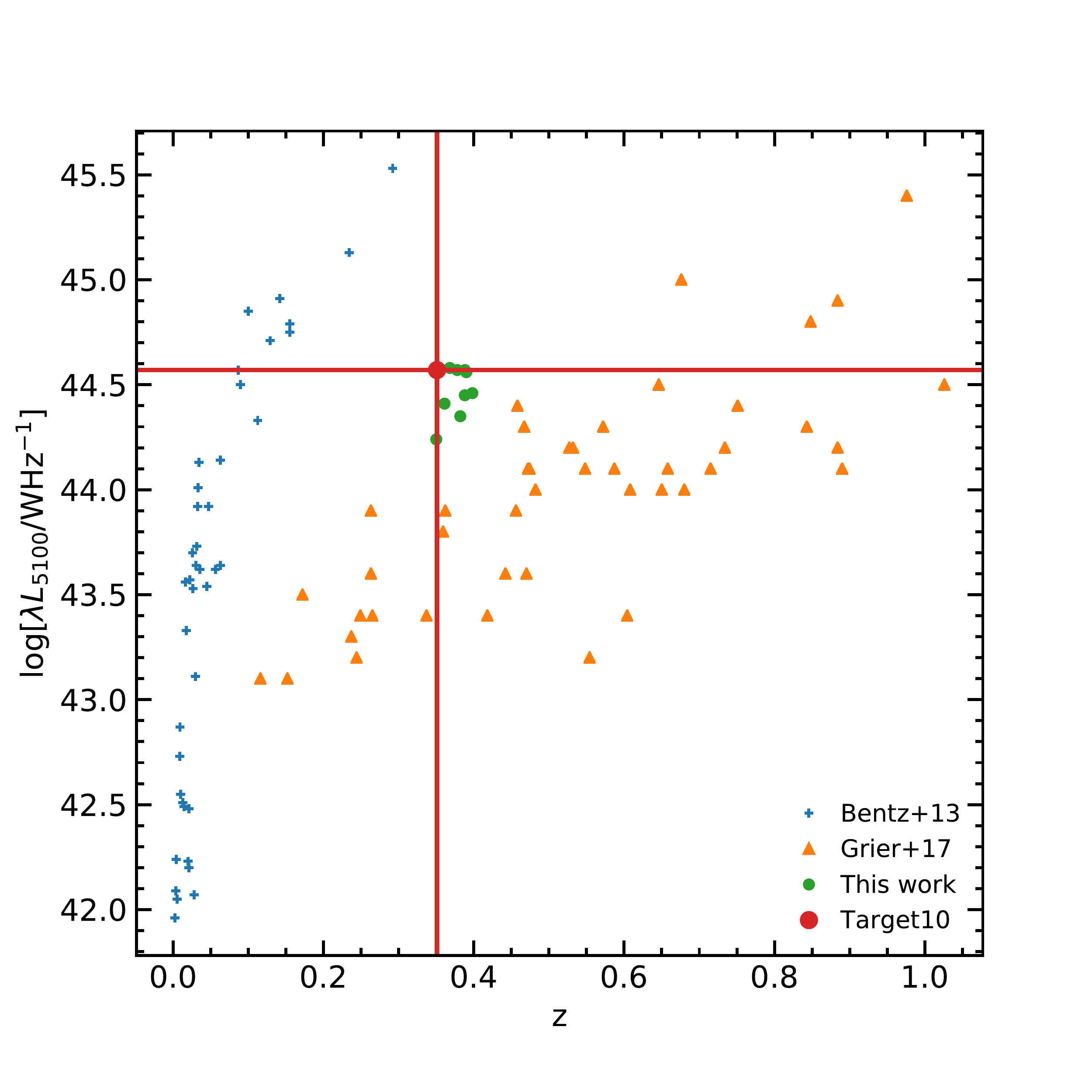}
  \caption{The distribution of luminosity versus redshift for the \citet{Bentz2013LowLuminosity} sample, shown in blue, and the \citet{Grier2017Sloan} sample, shown in orange.
  Our dataset is shown in green with \TargetTen, highlighted with red lines, between the two datasets.
  }
  \label{fig:RM:redshift_luminosity}
\end{figure}

Our targets, shown in Fig~\ref{fig:RM:redshift_luminosity} as green points, are positioned between the redshift-luminosity locations of the high-redshift spectroscopic sample from \citet{Grier2017Sloan} and the low-redshift sample from \citet{Bentz2013LowLuminosity}.

\subsection{Observations}
\begin{figure}
  \centering
  \includegraphics[width=\linewidth]{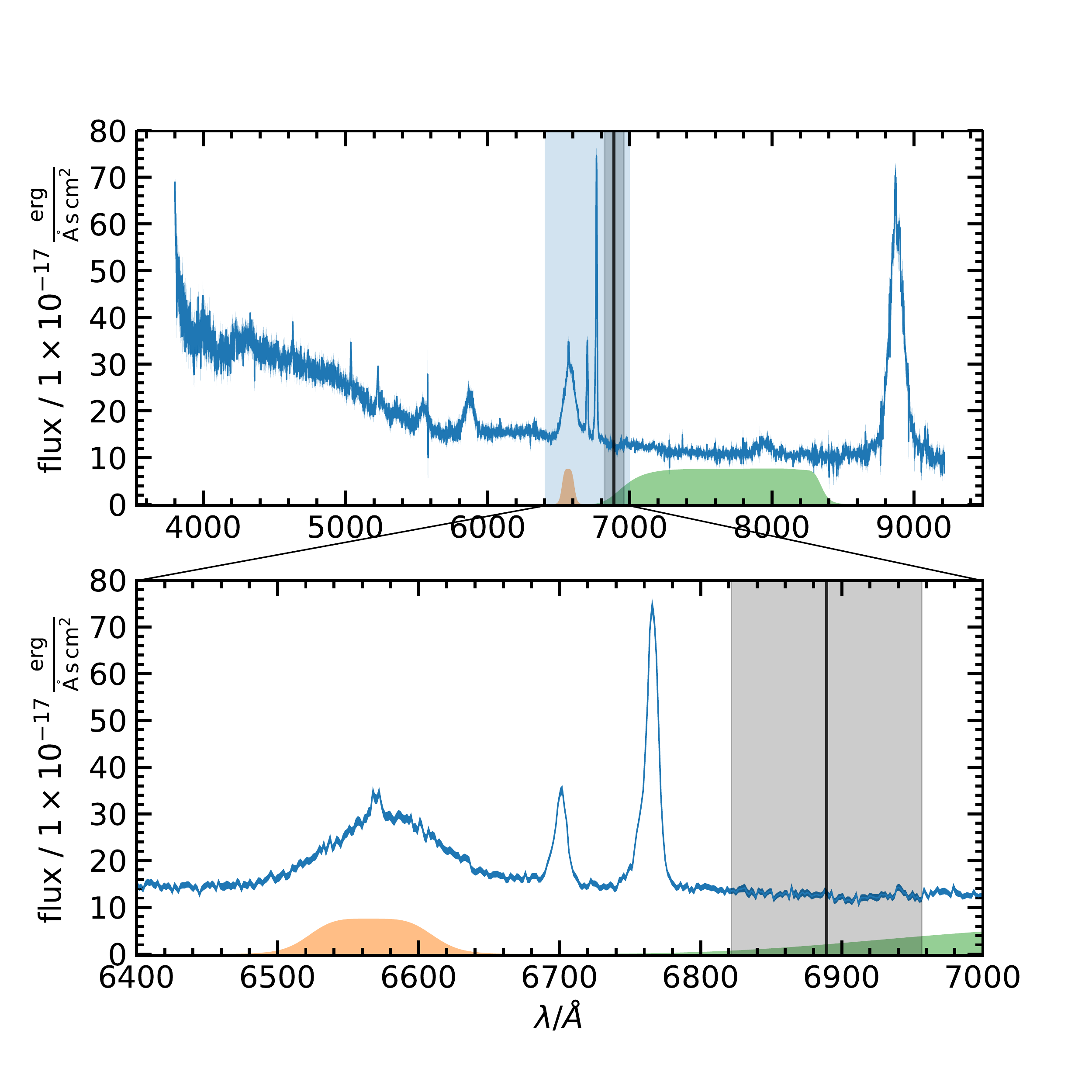}
  \caption{The SDSS-BOSS spectrum for \TargetTen. 
  The transmission curve for the \halpha filter used to measure the flux contained within the \hbeta line is shown in orange and the SDSS \iband filter is shown in green.
  The region between 6820\,\AA\xspace and 6960\,\AA\xspace used to determine the median rest-frame $L_{5100}$ for the SDSS spectrum is shown in grey.
  \textbf{Top:} The whole spectrum.
  \textbf{Bottom:} The region between 6400\,\xspace and 7000\,\AA\xspace which contains both the broad \hbeta line and the region used to measure the rest-frame $L_{5100}$.}
  \label{fig:RM:target10_spectrum}
\end{figure}

Since the expected variability of QSOs is of order 10-70 per cent \citep{Kaspi2007Reverberation}, we conservatively derive \iband exposure times, assuming an $\mathrm{SNR} > 20$ \citep[\eg][]{Bentz2013LowLuminosity,Shen2015Sloan} and seeing $< 2$ arcseconds, of 88s. 
This exposure time was calculated for our faintest target and so the SNR for the rest of our targets will be larger.
Using the SDSS BOSS observations of our targets (shown in Fig~\ref{fig:RM:target10_spectrum}) we detect no bright spectral features that would interfere with our ability to measure the continuum accurately. 
Accounting for the large equivalent widths of the \hbeta lines, we use a 600s integration time for broad-line (\ie narrow-band) observations.

Our targets span a range of redshifts between 0.350 and 0.398. 
Therefore, for each source, we use the appropriate \halpha photometric filter for which the redshifted \hbeta line dominates.
For \TargetTen, we use the \halpha-6566\AA\xspace narrow-band filter.

As seen in Table~\ref{tab:RM:targets}, we obtain the largest number of acceptable exposures with \TargetTenFullName (referred to as \TargetTen hereafter).
Indeed, \TargetTen is the only QSO for which we have obtained a baseline of observations longer than the recommended $3t_{H\beta} (1+z)$ needed to recover a lag. 
Thus, in what follows, we only discuss the analysis of \TargetTen and defer the rest to a future work.

\subsection{Ensemble Photometry and Flux Calibration}\label{sec:RM:methods:calibration}
In order to estimate lags between the broad-line region and the continuum-emitting region of the QSO, we must first calibrate the \iband and \halpha photometric magnitudes to a common magnitude system. 
We are then required to calibrate our \iband photometry using the known SDSS DR12 AB magnitudes of sources in the observed field.
We calibrate \halpha photometry by propagating available SDSS spectra through the transmission curve for the same narrow-band \halpha filter (6566\,\AA\xspace) used to observe the \hbeta line in \TargetTen, accounting for the fibre aperture.

We perform aperture photometry using \sextractor \citep{Bertin1996Sextractor} to estimate Petrosian magnitudes \citep{Petrosian1976Surface,Graham2005Total} for each detected source in the field for both \iband and \halpha exposures. 
We use Petrosian magnitudes in order to calibrate each exposure to the SDSS catalogue and to easily avoid the effects of differing seeing between our observations without modelling the PSF.
We consider only those sources which have SDSS \textsc{clean}$=$\textsc{True} and \sextractor \textsc{FLAGS}$=0$ for use as reference sources.
We can then apply a similar ensemble photometry method to that detailed by \citet{Honeycutt1992Ccd}, on the \iband exposures and calibrate those instrumental magnitudes to the SDSS absolute AB magnitude system. 
The details of our ensemble photometry method are described in Appendix~\ref{app:calibration}.

\subsection{Light-curve Calibration}

\begin{figure}
  \centering
  \includegraphics[width=\linewidth]{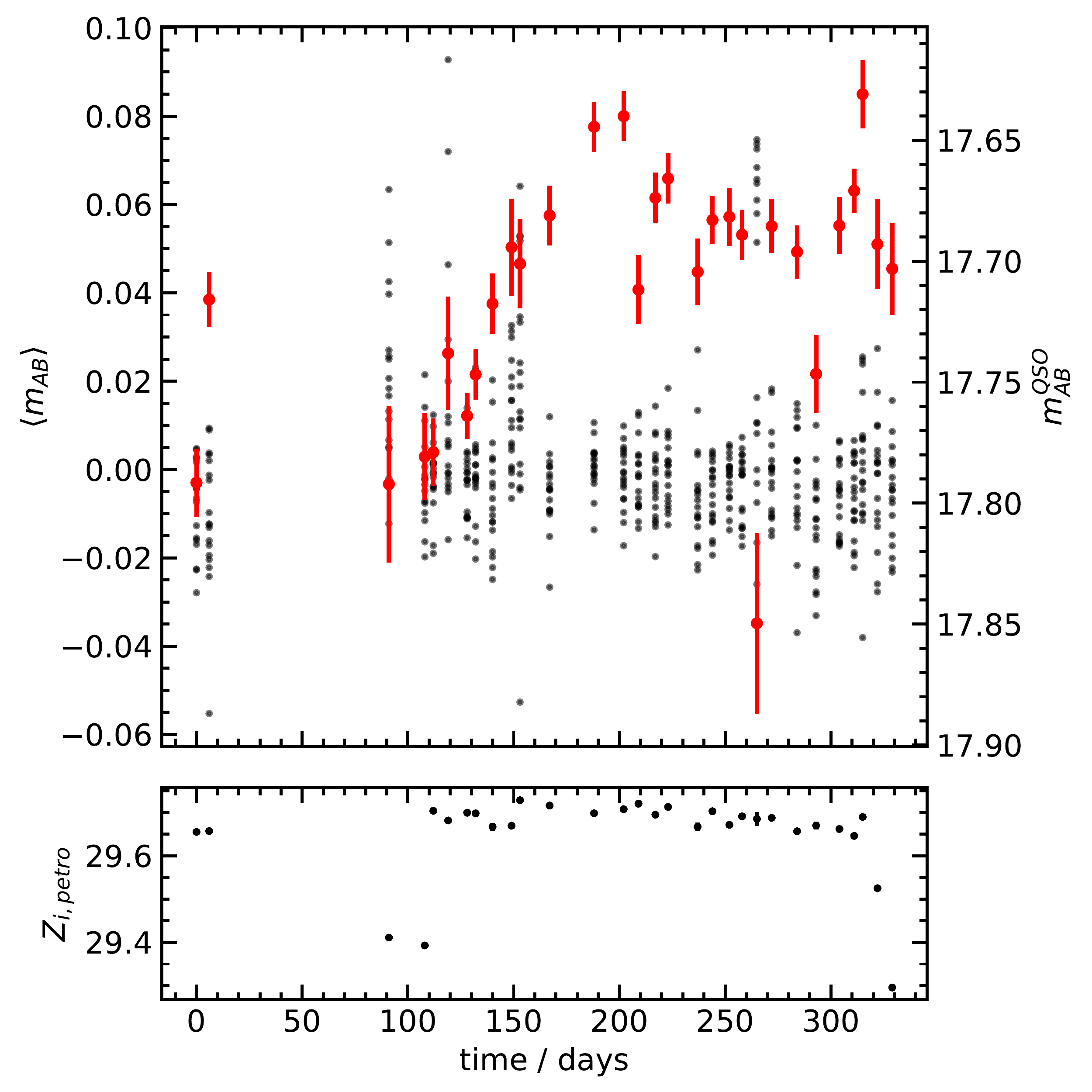}
  \caption{\textbf{Top:} The light curve for \TargetTen is shown in red with its calibrated \iband AB magnitudes labelled on the right axis. 
  The deviation from the mean magnitude for each of the reference sources for \TargetTen \iband are also shown on the left axis. 
  \textbf{Bottom:} The \iband AB zeropoint for each exposure calibrated to SDSS magnitudes using the Petrosian aperture.}
  \label{fig:RM:iband_zeropoints_and_qso_light_curve}
\end{figure}

Fig~\ref{fig:RM:iband_zeropoints_and_qso_light_curve} shows the calibrated light curve for \TargetTen in the \iband along with the deviation from the mean magnitude for its reference sources.
The average uncertainty for the AB magnitudes for \TargetTen is about $0.015$ mag with the largest being $0.040$ mag. 
The \iband magnitudes for \TargetTen therefore have signal-to-noise ratio of between 25 and 120, exceeding than the necessary SNR$> 20$ recommended by \citet{Bentz2013LowLuminosity} and \citet{Shen2015Sloan} to achieve reliable lags.

The SDSS DR12 catalogue lacks \halpha photometry and our observed fields contain few sources for which SDSS has spectra (only one of which is not a QSO).
Therefore, it is necessary to calibrate our \halpha exposures to the magnitudes obtained from propagating SDSS spectra through IO:O \halpha photometric filters.
We derive zeropoints, relative to the ``best'' exposure (\ie the exposure highest mean SNR for spectroscopic reference sources), for each of the \halpha exposures by using the same ensemble photometry method detailed above.
We make use of the SDSS spectroscopic catalogue to identify potential reference sources but find only one such source (\TargetTenReferenceSpectrumRaDec) observed by BOSS.

\begin{figure}
  \centering
  \includegraphics[width=\linewidth]{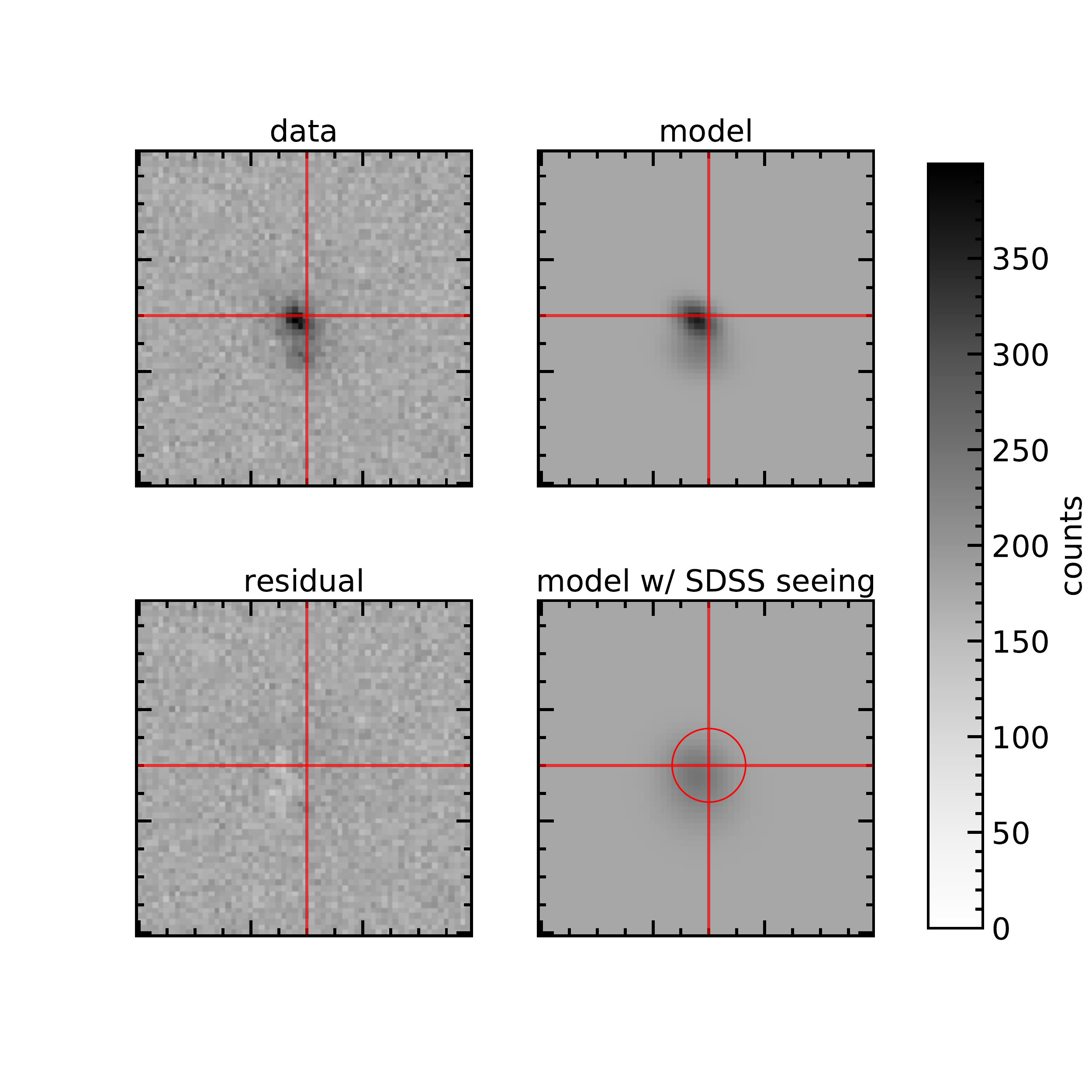}
  \caption{Modelling the spectral reference source at \TargetTenReferenceSpectrumRaDec in \halpha photometry. 
  \textbf{Top left:} Our original exposure of the spectral reference source in \halpha. 
  \textbf{Top right:} The model of the spectral reference source using two Gaussian components and a background.
  \textbf{Bottom left:} The residuals from our two component model.
  \textbf{Bottom right:} The model convolved to the SDSS seeing for the spectrum observation using a difference-of-two-Gaussians kernel.
  Overplotted in red crosshairs is the location of the centre of the 2 arcsecond BOSS aperture and the aperture is shown in the bottom right panel.
  \label{fig:RM:reference_modelling}
  }
\end{figure}

Our calibration depends upon the accurate measurement of the reference's flux within the \halpha filter.
Given that we find that the source is resolved into two components as shown in Fig~\ref{fig:RM:reference_modelling}, the effect of seeing and aperture corrections cannot be neglected.
We first fit a model consisting of two Gaussians to our best \halpha exposure, then transform the model to the same seeing as the BOSS observation, and finally extract the flux contained within the BOSS 2 arcsecond aperture.
The difference between the ensemble calibrated instrumental magnitude we obtain for our best exposure and the propagated BOSS spectrum is taken as our zeropoint, accounting for uncertainties in both magnitudes.

\begin{figure}
  \centering
  \includegraphics[width=\linewidth]{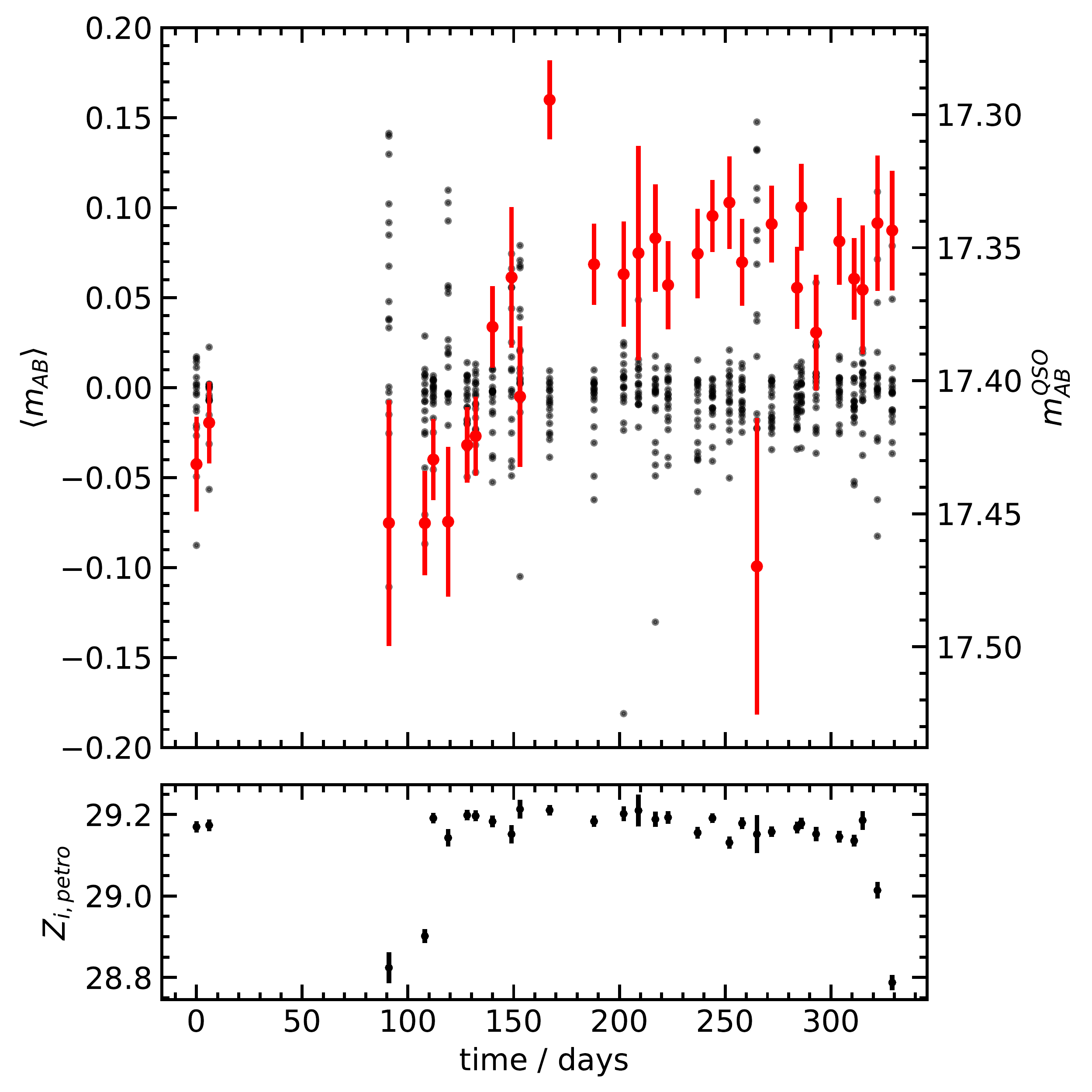}
  \caption{\textbf{Top:} The light curve for \TargetTen is shown in red with its calibrated \halpha AB magnitudes labelled on the right axis. 
  The deviation from the mean magnitude for each of the reference sources for \TargetTen \halpha are also shown on the left axis. 
  \textbf{Bottom:} The \halpha AB zeropoint for each exposure calibrated to SDSS magnitudes using the Petrosian aperture.}
  \label{fig:RM:halpha_zeropoints_and_qso_light_curve}
\end{figure}

Fig~\ref{fig:RM:halpha_zeropoints_and_qso_light_curve} shows the the resultant light curve for \TargetTen in the \halpha waveband along with the deviation from the mean magnitude for its references sources.
Due to the necessary intermediate step of calibrating differential magnitudes to the AB magnitude system via the spectral reference source, the signal-to-noise ratio of the \halpha magnitudes is smaller than those in the \iband.
We measure signal-to-noise ratios for the \halpha fluxes of \TargetTen range between 19.5 and 80.0. 

The zeropoint for both \iband and \halpha exposures can change by about 0.4 mag and the exposures where this occurs are the ones the highest uncertainty for the QSO magnitude.
Upon inspection, it is clear that these exposures have increased cloud cover or worse-than-normal seeing.
Our ensemble calibration method above takes into account the instantaneous deviation of reference sources from their inferred mean magnitudes and updates their weightings accordingly (see Appendix~\ref{app:calibration}). 
We therefore do not exclude these exposures from further analysis.

\subsection{Reliability simulations}\label{sec:RM:reliability}
\javelin \citep{Zu2013Quasar} can be used to model quasar variability with either spectroscopic \citep{Zu2011Alternative} or purely photometric measurements \citep{Zu2016Application}. 
\javelin supports a number of random walk covariance kernels which control the strength of the correlation between any two flux observations given the time between them.
\citet{Zu2013Quasar} finds that the exponential covariance kernel is appropriate on time-scales, $\tau$, between months and years, and we therefore adopt their recommendation for fitting with Javelin.
Below a time-scale of a few months, the correlation becomes stronger than can be accounted for by the exponential covariance kernel \citep{Mushotzky2011Kepler,Zu2013Quasar} and the characteristics of stochastic behaviour at time-scales longer than a few years are not well known due to lack of data.
There is further evidence that the DRW is not sufficient to explain high frequency light-curve variance as seen by Kepler and SDSS \citep{Kasliwal2015Are,Guo2017Far,Smith2018Kepler}.
The impact of fitting non-DRW light curves assuming the DRW model is not well understood.
Furthermore, current variability-modelling techniques are not physically motivated and attempt to interpolate gaps in the light curve by assuming some correlated time-series model. 
If this model is too inaccurate and the gaps in the light curve too long, we risk producing artificial peaks in the lag posterior distribution, which can be indistinguishable from peaks describing the physical lag.

To test whether reverberation mapping with interpolated models can be trusted in the presence of such model-dependent problems we employ 3 techniques:
\begin{enumerate}
  \item Use a non-parametric Von Neumann estimator of narrow band + continuum time-series as demonstrated with spectroscopic measurements in \citet{Chelouche2017Methods}. This allows model-independent verification.

  \item Generate a suite of simulated light-curves each with different generative parameters to test whether a given method can reliably retrieve a known lag under different models.

  \item Use the newly reprocessed Kepler light-curves \citep{Smith2018Kepler} as the basis for realising the simulated light-curves by fitting the Kepler data with the Continuous Auto-Regressive Moving Average process (see below) with order $p=2,q=1$, now indicated as CARMA(2,1), using \kali \citep{Kasliwal2015Are,Kasliwal2017Extracting}. 
  This allows a test of the performance of DRW fitting procedures with non-DRW light-curves.
\end{enumerate}

Ideally, we would generate these light-curves from a physically-motivated hydrodynamic self-consistent model of the BLR. 
However, this is beyond the scope of this work and so we settle on a suite of light-curves informed only by the reprocessed Kepler database \citep{Smith2018Kepler}; a prior distribution of BLR window parameters; and the observed SNR, cadence, and spectrum of our target QSO.
In this BLR model (which is the same model that \javelin uses), the continuum light-curve is first smoothed by a top hat window of width $w$, then scaled by line-scale $s$.
To generate the emission seen through the \halpha photometric filter, the contribution of the continuum over the \halpha filter is added to the simulated emission line flux.
The \halpha photometric light-curve, $n(t)$ is therefore described by
\begin{equation}
  n(t) = \alpha c(t) + s\int_{-w/2}^{w/2} c(t - \tau_{lag} - t') dt',
\end{equation}
\noindent where $\alpha$ is the ratio of the continuum measured in the \iband, $c(t)$, relative to that in the \halpha filter.

\begin{figure*}
    \centering
    \includegraphics[width=\linewidth]{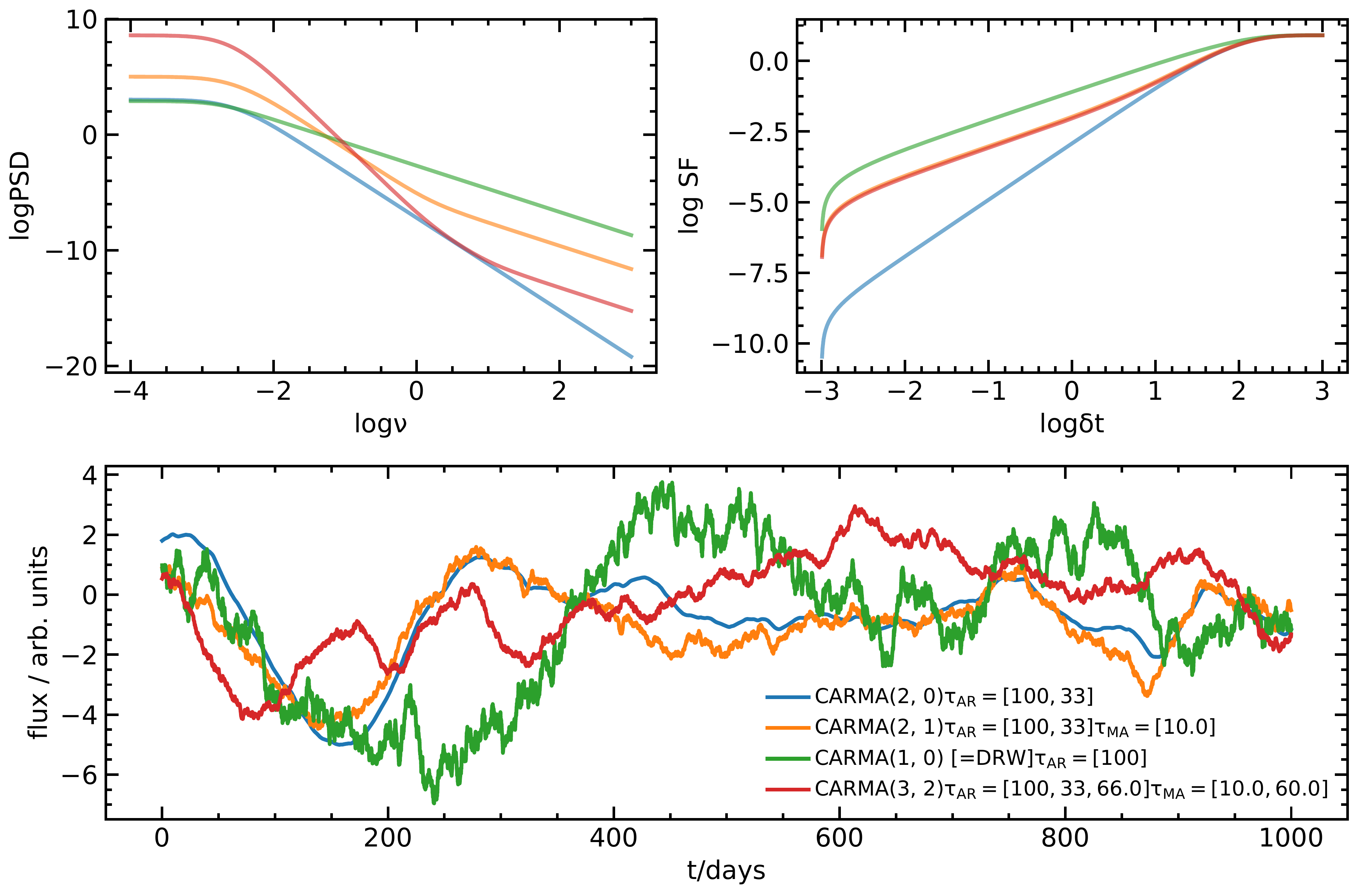}
    \caption{The difference between different CARMA process orders (whose matched parameters are described in the legend). 
    {\bf Top left:} The power spectra of the CARMA processes.
    {\bf Top right:} The structure function of the CARMA processes.
    {\bf Lower}: One realisation for each CARMA process generated from the same random seed.}
    \label{fig:RM:CARMA_DRW}
\end{figure*}

We fit the CARMA(2,1) process to the \citep{Smith2018Kepler} light-curves using \kali (\citealt{Kasliwal2017Extracting}). 
A CARMA process is a stationary time-series model consisting of auto-regressive components and moving-average components \citep{Kelly2014Flexible}. 
Following \citet{Kelly2014Flexible}, a CARMA(p,q) process, $y(t)$, is defined as solution to the stochastic differential equation 
\begin{align}
\begin{split}
  \frac{\mathrm{d}^p y(t)}{\mathrm{d}t^p} + \alpha_{p-1}\frac{\mathrm{d}^{p-1} y(t)}{\mathrm{d}t^{p-1}} + ... + \alpha_{0}y(t) &= \\
  \beta_q \frac{\mathrm{d}^q\epsilon(t)}{\mathrm{d}t^q} + \beta_{q-1}\frac{\mathrm{d}^{q-1}\epsilon(t)}{\mathrm{d}t^{q-1}} + ... + \epsilon(t)
\end{split}
\end{align}
\noindent where $p$ is the total number of auto-regressive time-scales, $q$ is the number of moving-average time-scales, $y(t)$ is a small flux perturbation from the mean at time $t$, $\epsilon(t)$ is a white noise process drawn as $\sim \mathcal{N}(\mu=0, \sigma^2)$ and $\alpha$ \& $\beta$ are constants. 
We define $\alpha_p = \beta_0 = 1$ and the CARMA process is only stationary around a mean if $p > q$.
A DRW, or CARMA(1,0) process is therefore defined as a solution to 
\begin{equation}
      \frac{\mathrm{d}y(t)}{\mathrm{d}t} + \tau y(t) = \epsilon(t),
\end{equation}
\noindent where $\tau$ is the time-scale of the variations, bringing the total number of parameters to 2 ($\tau$ and $\sigma$). 
Similarly, a CARMA(2,1) process is defined as a solution to 
\begin{equation}
  \frac{\mathrm{d}^2 y(t)}{\mathrm{d}t^2} + \alpha_{1}\frac{\mathrm{d}y(t)}{\mathrm{d}t} + \alpha_{0}y(t) = \beta_1 \frac{\mathrm{d}\epsilon(t)}{\mathrm{d}t} + \epsilon(t),
\end{equation}
\noindent which is equivalent to a damped harmonic oscillator
\begin{equation}
\mathrm{d}^{2}y + 2\zeta\omega\mathrm{d}y + \omega^{2} y = \beta_{1} \mathrm{d}\epsilon(t) + \epsilon(t),
\end{equation}
\noindent where $\zeta$ is the forcing ratio, $\omega$ is the angular frequency of oscillation and $\beta_1$ controls the frequency dependence, "colour", of the noise (i.e. if $\beta_1 \neq 0$, the noise power spectral density, PSD, is not flat). 
The CARMA(2,1) process therefore has 4 parameters inclusive of the amplitude of the variations, $\sigma$.
The differences between different CARMA processes are shown in Fig~\ref{fig:RM:CARMA_DRW}, which also highlights that the DRW is a CARMA(1,0) process, i.e. it is lacking a moving average component.

The second order differential equation underlying the CARMA(2,1) process is familiar to many branches of physics and is therefore more easily interpretable than higher order processes. 
Indeed, the thermal motion with a fluid produces sound waves described by a PSD $\sim \nu^2$ \citep{Mellen1952Thermal}, which suggests that CARMA(2,1) can be physically motivated by such distortions in the accretion disk.
For more information concerning the statistics and physical applicability of CARMA processes to astronomical light-curves see \citet{Kelly2014Flexible} and \citet{Kasliwal2017Extracting}.

The CARMA process has been shown to more accurately match PSDs of AGN which experience deviations from the DRW model \citep{Kelly2014Flexible,Kasliwal2017Extracting} since it has more degrees of freedom and is therefore more flexible than its lower order counterpart (DRW). 
Here we fit CARMA(2,1) to all 20 of the \citet{Smith2018Kepler} reprocessed Kepler light-curves that have spectroscopic redshifts with \kali.
Sampling from the time-scale probability distributions of each of the fits, we can produce light-curves whose structure functions and power-spectra resemble that of Kepler light-curves.
We also perform this analysis for simulated light-curves generated by a DRW process, as a comparison, still using the same template Kepler light-curves.

These simulations allow us to estimate the degree to which we can trust lag parameter estimations for a given QSO target and fitting method.
They will reveal the nature of any artefacts which can occur due to the cadence, generative model, or interpolation of the input light curve.
Furthermore, it allows us to test whether the DRW model predicts lags that too optimistic and therefore estimate a more robust uncertainty for the lag.
We perform such analysis with \NumberOfSimulatedLightCurves\footnote{We arrived at this number simply by tracking the stability of our results as the number of simulations increases. At around \NumberOfSimulatedLightCurves, the reduction in the lag uncertainty from deconvolution (see Section~\ref{sec:RM:results}) reaches a plateau. The optimal number of simulations may be different for different objects.} simulated light curves constructed by sampling from the time-scale distributions fit to the \citet{Smith2018Kepler} Kepler light-curve and priors for $s$, $\alpha$, and variability amplitude $\sigma$ informed by the target's spectrum, for a given target's cadence. 

\begin{figure}
  \centering
  \includegraphics[width=\linewidth]{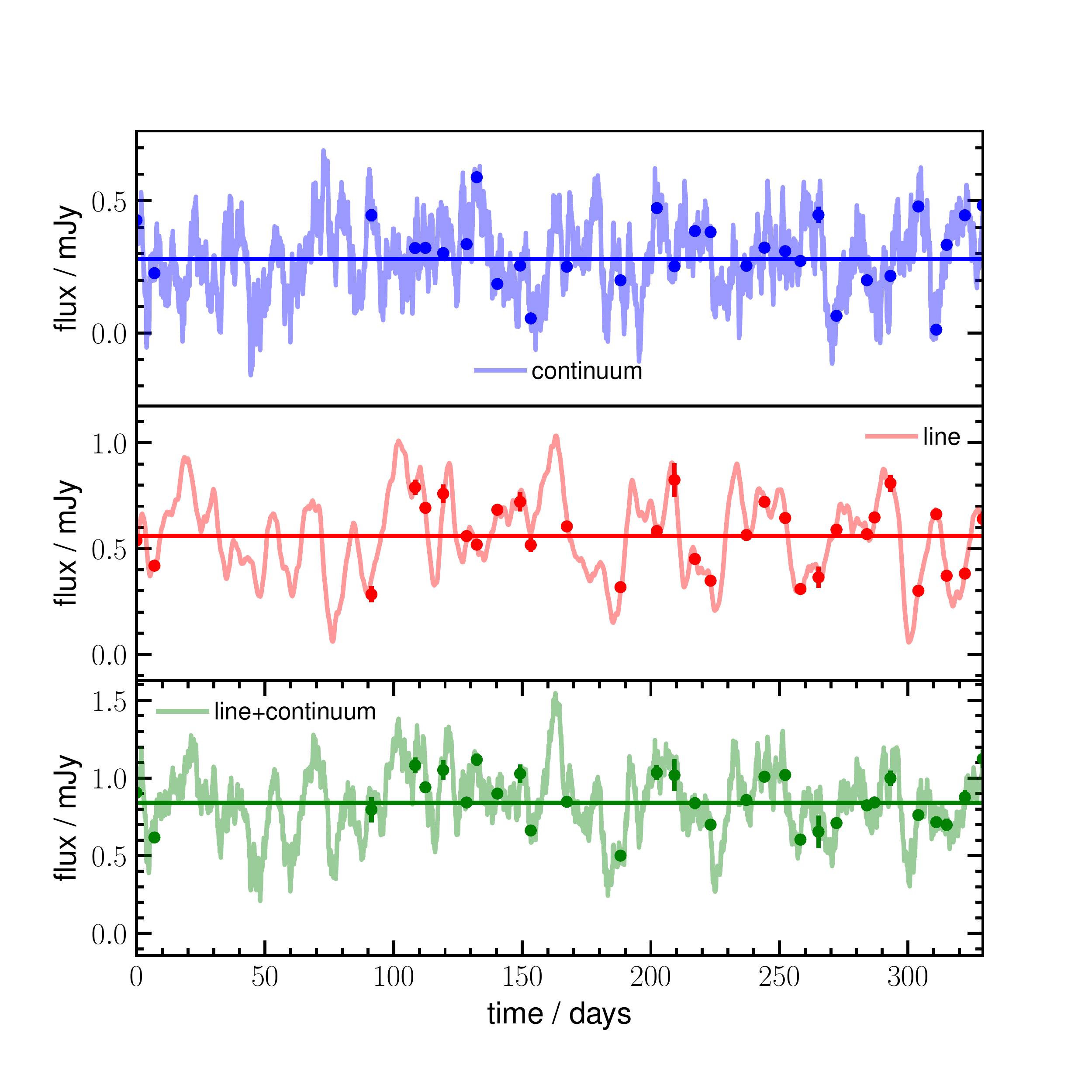}
  \caption{One of the \NumberOfSimulatedLightCurves light curves generated from a grid of parameters based on the Kepler light-curves. 
  The continuum, pure line, and line with continuum light curves are shown in blue, red, and green respectively.
  The lines depict the intrinsic light curve generated by the simulated QSO using the damped random walk covariance kernel.
  The noisy observations, with the same signal-to-noise ratio as the calibrated \TargetTen light curves are shown as points.
  The mean flux of each of the light curves is shown as a horizontal line.}
  \label{fig:RM:example_simulated_qso}
\end{figure}

The resultant simulated continuum light curves are then smoothed with a top hat window of width $w$ and scaled by line scale $s$ to produce emission line light curves, one of which is shown in the middle panel of Fig~\ref{fig:RM:example_simulated_qso}.
To generate the mixture of line and continuum emission seen through a photometric filter, we scale the continuum light curve by a continuum scale $\alpha$ and add the resultant continuum to the emission line, as shown in the lower panel of Fig~\ref{fig:RM:example_simulated_qso}.
The simulated observations are then taken at the same cadence as that of \TargetTen and assuming the same signal-to-noise (shown as dots in Fig~\ref{fig:RM:example_simulated_qso}).
In order to test how dependent the lag estimate is upon the zeropoint obtained from the spectral reference calibration source, we also scale the resultant continuum+line light curve by a zeropoint offset bringing the total number of explicit parameters to 6 (where the CARMA/DRW time-scales are implicitly drawn from fits to the Kepler light-curves).
For \TargetTen, we use the distribution of these parameters fit to the Kepler sample or by inspecting the spectrum of \TargetTen, as appropriate. 
These parameter distributions are detailed in Table~\ref{tab:RM:parameter_ranges}.
\begin{table}
\centering
\begin{tabular}{lllll}
\hline
Parameter      & Distribution & Source \\ \hline
Kepler Light curve &  Choice[n=20]  & \citet{Smith2018Kepler}\\
$\log(t_{lag})$&  $\mathcal{U}(0, 300)$ & Set to cover  \\
$\log\sigma$   &  $\mathcal{N}(-2.2, 1)$ & Prior from \TargetTen\\
CARMA $\tau$   &  $\sim P(\tau_{\textrm{Smith+18}})$ & \citet{Smith2018Kepler} \\
$w$            &  $\mathcal{U}(0, 13)$  & Set to cover \\
$s$            &  $\mathcal{N}(1.70, 1.21)$  & Measured from spectrum\\
$\alpha$       &  $\mathcal{N}(1.20, 0.53)$  & Measured from spectrum\\
$\sigma_z$     &  $\mathcal{N}(0, 0.3)$  & Set from zeropoint error\\
\end{tabular}
\caption{\NumberOfSimulatedLightCurves draws were taken from these parameter distributions to create the simulated light curves for \TargetTen. 
Each draw created a different continuum lightcurve from the posterior distribution of CARMA(2, 1) fit to a randomly chosen Kepler light curve. 
The result was then propagated through a lagged smoothing window of width $w$ days, scaled by line scale $s$, and added onto the continuum at the position of the \halpha photometric filter $=\alpha * c(t)$ to create the narrow band light light curve. 
}
\label{tab:RM:parameter_ranges}
\end{table}
 
\subsection{Fitting methods}
For each of these light curves, we run the following analysis to derive the best estimate for the lag.
For \javelin, we infer the DRW parameters (amplitude, $\sigma$, and time-scale $\tau$) of the \iband continuum with \JavelinNWalkers walkers, whether generated by DRW or not. 
We use the output probability distributions as a prior for the lag estimation using both \iband and \hbeta light curves.
We run \javelin with the default settings of a logarithmic prior which begins to penalise lag values larger than a third of the observational baseline (the time between the first observation and the last), and a hard limit on lags longer than the baseline itself.
MCMC chains must have converged before any reliable parameter estimation can be performed.
The model is run until convergence is achieved, whereby MCMC is halted when the autocorrelation time for all parameters changes less than \AutocorrelationTimeTolerance and the number of iterations is larger than \AutocorrelationTimeMultiple times the largest autocorrelation time estimate, as recommended by \citet{Mackey2013Emcee}\footnote{\url{http://emcee.readthedocs.io/en/latest/user/autocorr/}}.

We find that simply using the \iband and \halpha time-series directly with the Von Neumann estimator produces biased results. 
Indeed, for light-curves with $\alpha > 0$, the Von Neumann estimator starts to underestimate the lag.
Therefore, when estimating lags with the Von Neumann estimator, we subtract the \iband continuum photometry from the \halpha narrow band photometry within the estimator.
We apply the Von Neumann algorithm detailed by \citet{Chelouche2017Methods} for 5000 samples, where each iteration samples a different realisation of the \TargetTen light-curve from its flux uncertainties (the FR/RSS scheme defined by \citealt{Peterson2004Black}) and subtracts the continuum realisation from the narrow-band realisation.

This results in a large hyper-volume of probability distributions which we can marginalise over to give us the accuracy of lag estimates as a function of known input lags, for each fitting method.

\begin{figure*}
  \centering
  \subfigure[DRW light-curves using \javelin]{
  \includegraphics[width=0.45\linewidth]{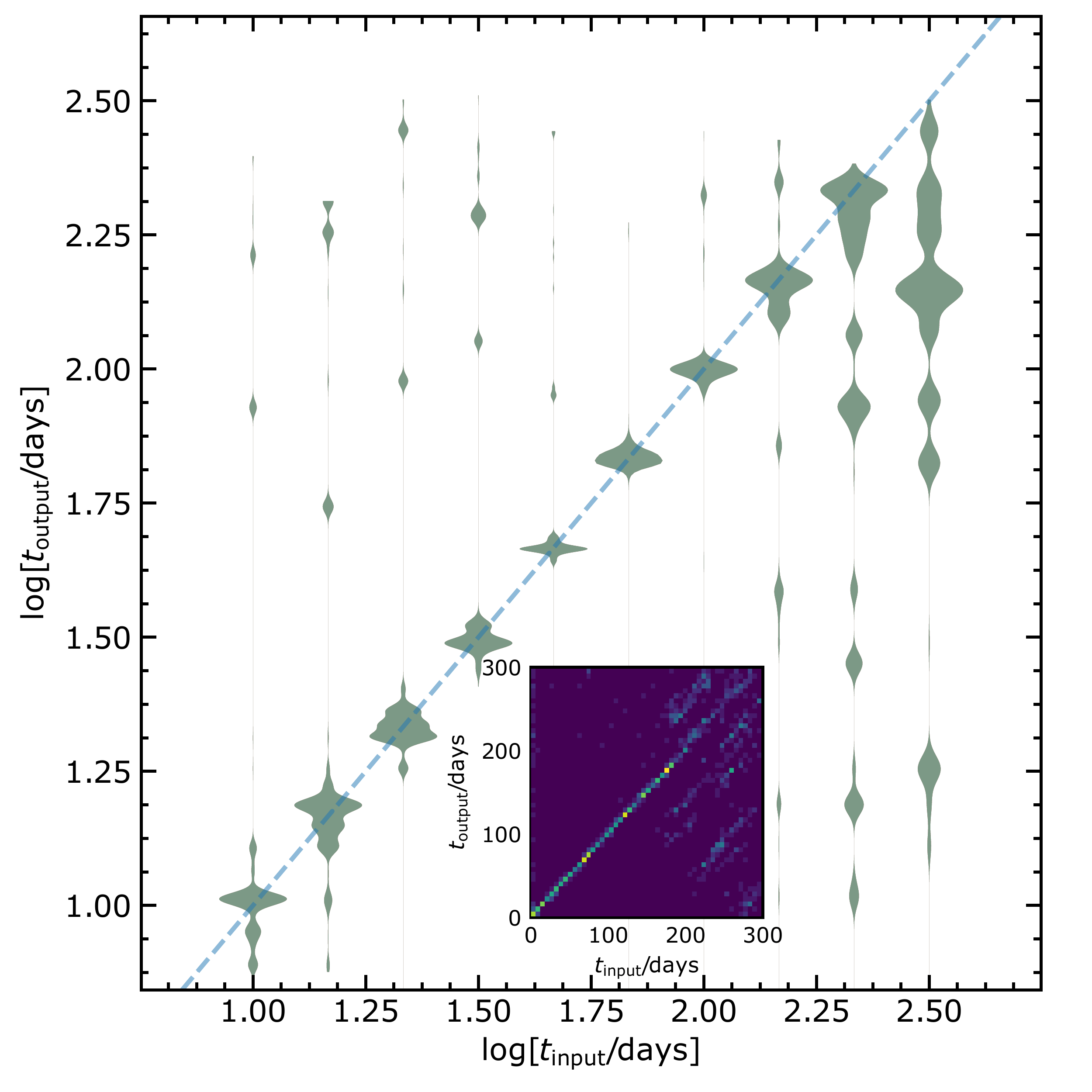}
  }
  ~
  \subfigure[CARMA(2,1) light-curves using \javelin]{
  \includegraphics[width=0.45\linewidth]{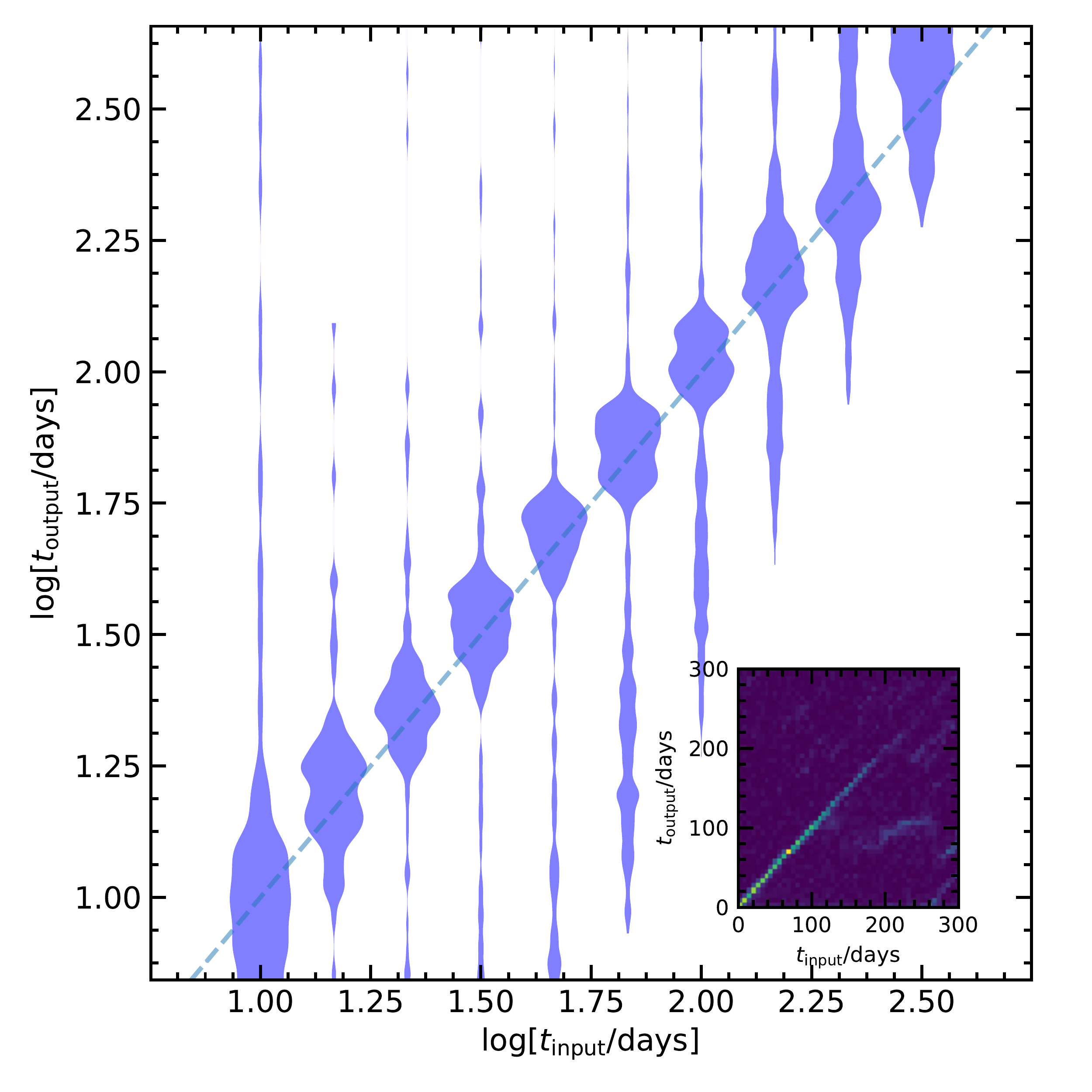}
  }
  ~
  \subfigure[CARMA(2,1) light-curves using the Von Neumann estimator]{
  \includegraphics[width=0.45\linewidth]{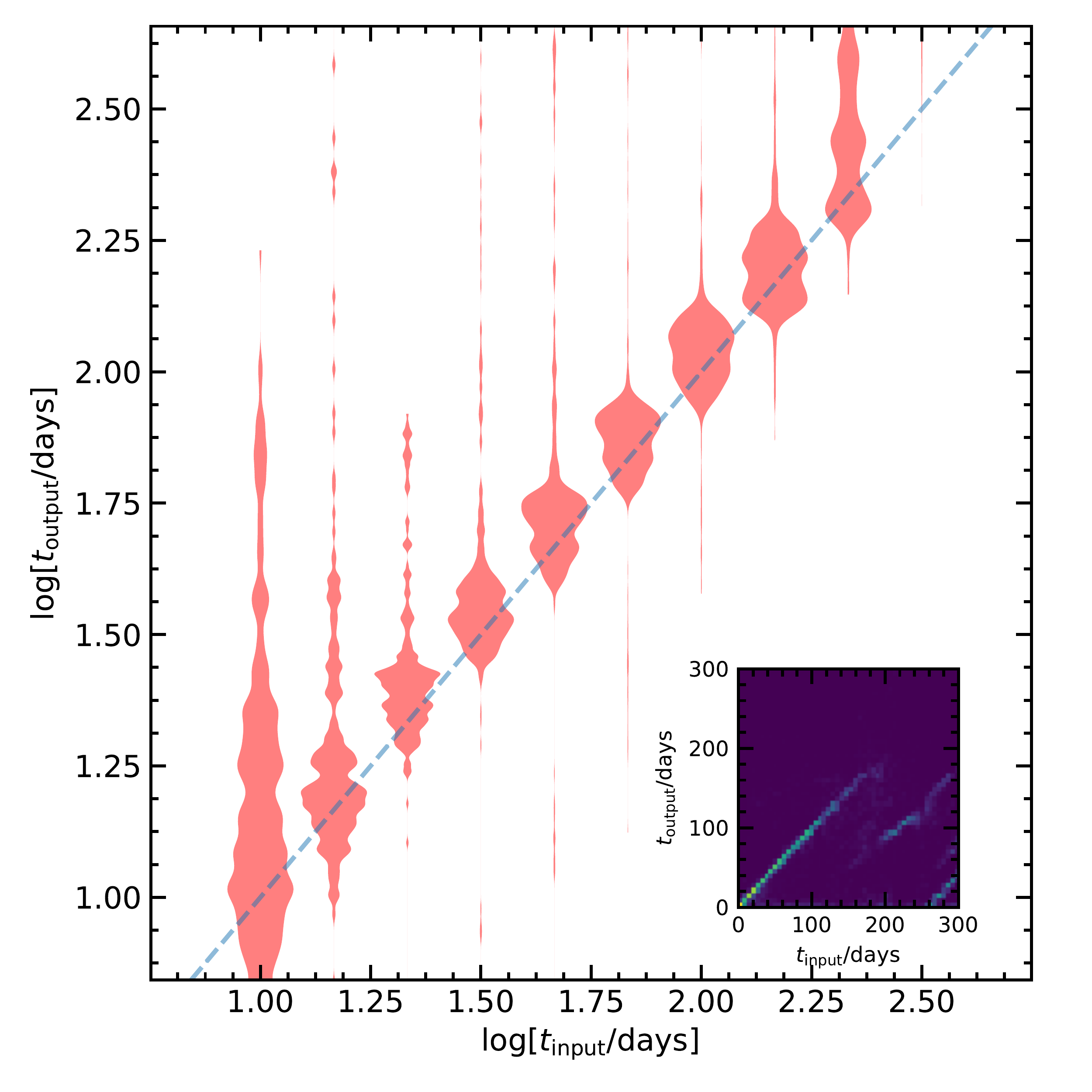}
  }
  \caption{The comparison of input and lags estimated by \javelin and the Von Neumann estimator for the simulated light curves.
  The violins at each input lag depict the distribution of best estimate lags, with their width indicating number density. 
  These best estimates are determined by the KDE procedure described in Section~\ref{sec:RM:methods:lag estimation}.
  The 1:1 relation indicating perfect recovery of input lags is shown as the dashed line.
  Each inset axes shows the 2D histogram of the same distributions to better illustrate the positions of outlying estimates. 
  The colour scale shows the relative density of results.
  \label{fig:RM:violin_plots}
  }
\end{figure*}

Due to the presence of more than one strong peak in the lag probability distributions, taking the median of an MCMC chain array may result in the parameter estimate being located in an area of low probability, between peaks, and not near a region of high probability.
Therefore, any quoted estimate and its uncertainty could be misleading.
We choose not to identify the primary peak by eye, but use a mode-finding method to identify the most probable solution within the highest-posterior-density (HPD) credible interval. 
The HPD interval is the narrowest interval that is guaranteed to contain the mode of the distribution.
We fit a kernel-density-estimate (KDE) using the \fastkde \citep{Brien2014Reducing,Brien2016Fast} algorithm which calculates the kernel's parameters objectively (\ie the hyper-parameters are informed entirely by the data and therefore it does not require user specification of bin width or kernel bandwidth), and choose the maximum value of that resultant KDE to be our best estimate for the \javelin parameters.

\section{Results}\label{sec:RM:results}
\subsection{Reliability Simulations}
Fig~\ref{fig:RM:violin_plots} shows the distributions of the KDE best estimate of the \hbeta lag based on the output lag probability distributions from \javelin (top left shows DRW as input, top right shows CARMA(2,1) as input) and the Von Neumann estimator (bottom left shows CARMA(2,1) as input).
The first observation we can make is that \javelin does indeed perform worse when the input light-curve is not a DRW process, as \javelin assumes (an average of 5 per cent error versus 1 per cent between 10 and 250 days).
We also see that the model-independent Von Neumann estimator recovers lags with an accuracy very similar to that of \javelin (4 per cent), when not assuming DRW.
In addition, all methods start to fail with lag recovery errors greater than 50 per cent above 170 days
Given that \javelin starts to penalise lag values larger than a third of the observation baseline it is perhaps not surprising that lags starting to approach the total length of the baseline itself are not as reliably recovered as those below a third of that length. 
The Von Neumann estimator does not apply such a prior and still experiences a drastic loss in accuracy beyond 170 days, suggesting that this loss is likely due to the finite baseline of the light-curve.

We also observe that there are a number of hyperparameter combinations whose recovered lags are incorrect by $>100$ days.
This occurs for combinations at all input lags and fitting methods and so we should not be surprised by spurious peaks in the probability distribution for \TargetTen at higher lags.
At all input lags and methods, we find artificial (\ie incorrect) peaks at negative lags and so we can be justified in disregarding the peaks below -100 days.
In particular, the Von Neumann estimator routinely places a large probability mass into a peak at -200 days.
We find that there is always a large peak for all fitting methods at around 0-14 days, which coincides with the average cadence of observations (\TargetTenCadence).

The KDE method allows us to assess the most likely peak without referring to the unstable maximum likelihood point, but it also implies a large uncertainty on the lag given that there are other regions of high probability which cannot be ruled out a priori.
We can address the issue in four ways: 
\begin{enumerate}
  \item Use the output lag distribution for our reliability simulations to mitigate the effect of non-linear artefacts that arise from the fitting process.
  \item Apply a prior to the lag distribution based on previous lag and luminosity measurements, and established relations \ie \citep{Bentz2013LowLuminosity}.
  \item Limit analysis to the range of lags bounded by the minima surrounding the tallest peak.
  \item Combine estimations from each fitting method, thereby mitigating the biases which are not shared by both methods.
\end{enumerate}

We perform the only the first, third and the last steps detailed above since we want our lag measurement to inform the $t_{\mathrm{rest}}-L_{5100}$ relation, which cannot be done independently if our measurement is a result of an application of a prior based on the same relation.

\subsection{Lag estimation for \TargetTen}\label{sec:RM:methods:lag estimation}
\begin{figure}
  \centering
  \includegraphics[width=\linewidth]{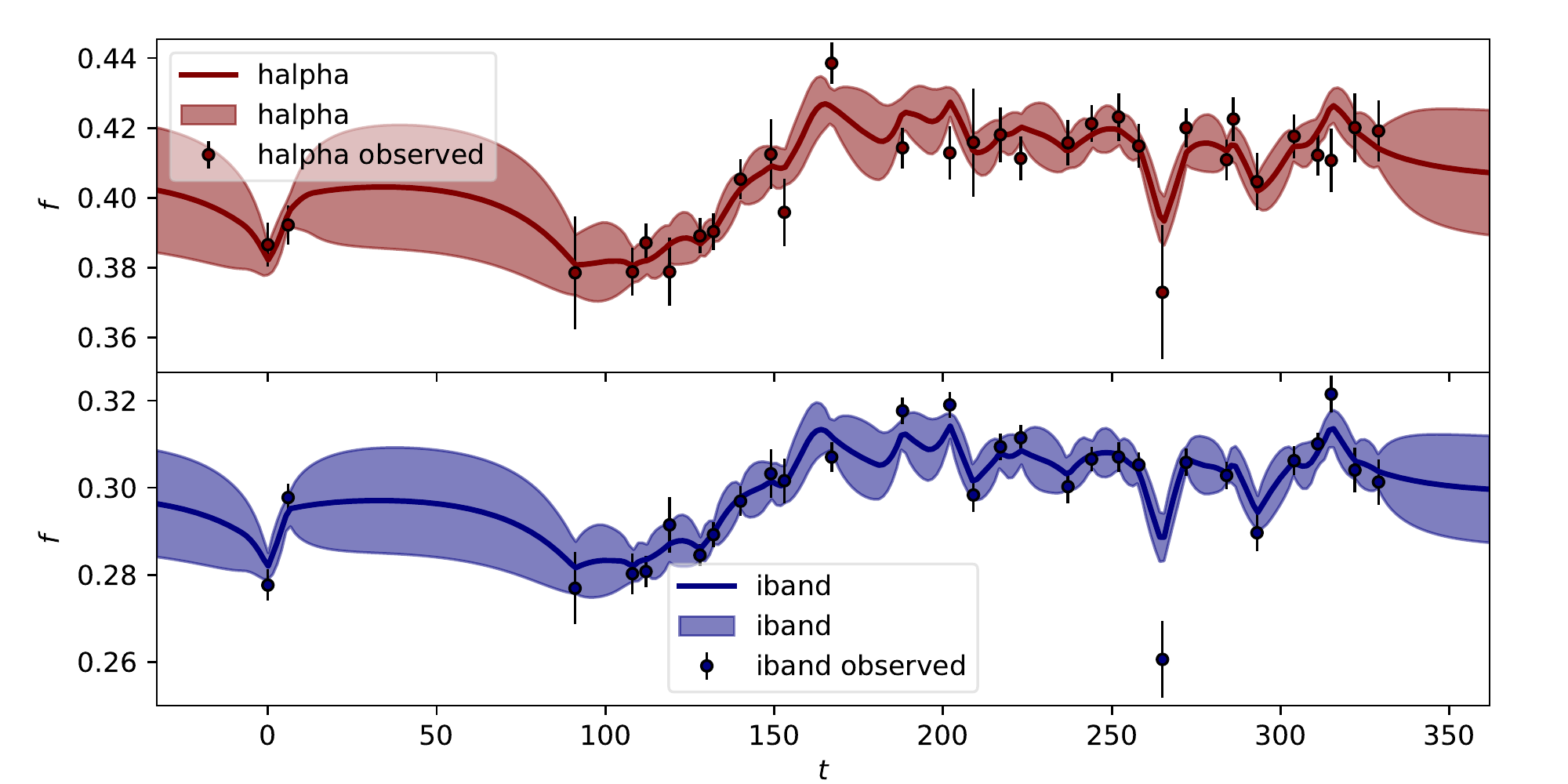}
  \caption{The posterior predictive light curves for \TargetTen in mJy.
  \textbf{Top:} The redshifted \halpha band light curve containing a mixture of \hbeta line emission and continuum emission.
  \textbf{Bottom:} The \iband continuum emission.
  The shaded regions correspond to the 68 per cent density region covered by random draws from the \javelin posterior probability distribution.
  The black error bars denote the calibrated observations for each waveband.
  \label{fig:RM:javelin_lightcurve}
  }
\end{figure}

We perform the same fitting procedure for \TargetTen as we did for our simulated light curves, using \javelin and the Von Neumann estimator. 
Fig~\ref{fig:RM:javelin_lightcurve} shows the \javelin posterior predictive distribution for the observed light curves of \TargetTen based on the burnt-in chain (\ie with the first 1000 steps for the MCMC chain removed). 
The \halpha predictive posterior light curve is the linear combination of continuum and emission line light curves where the emission line flux is only a fraction of the continuum. Manually identifying the time delay between them will be difficult.

\begin{figure}
    \centering
    \includegraphics[width=\linewidth]{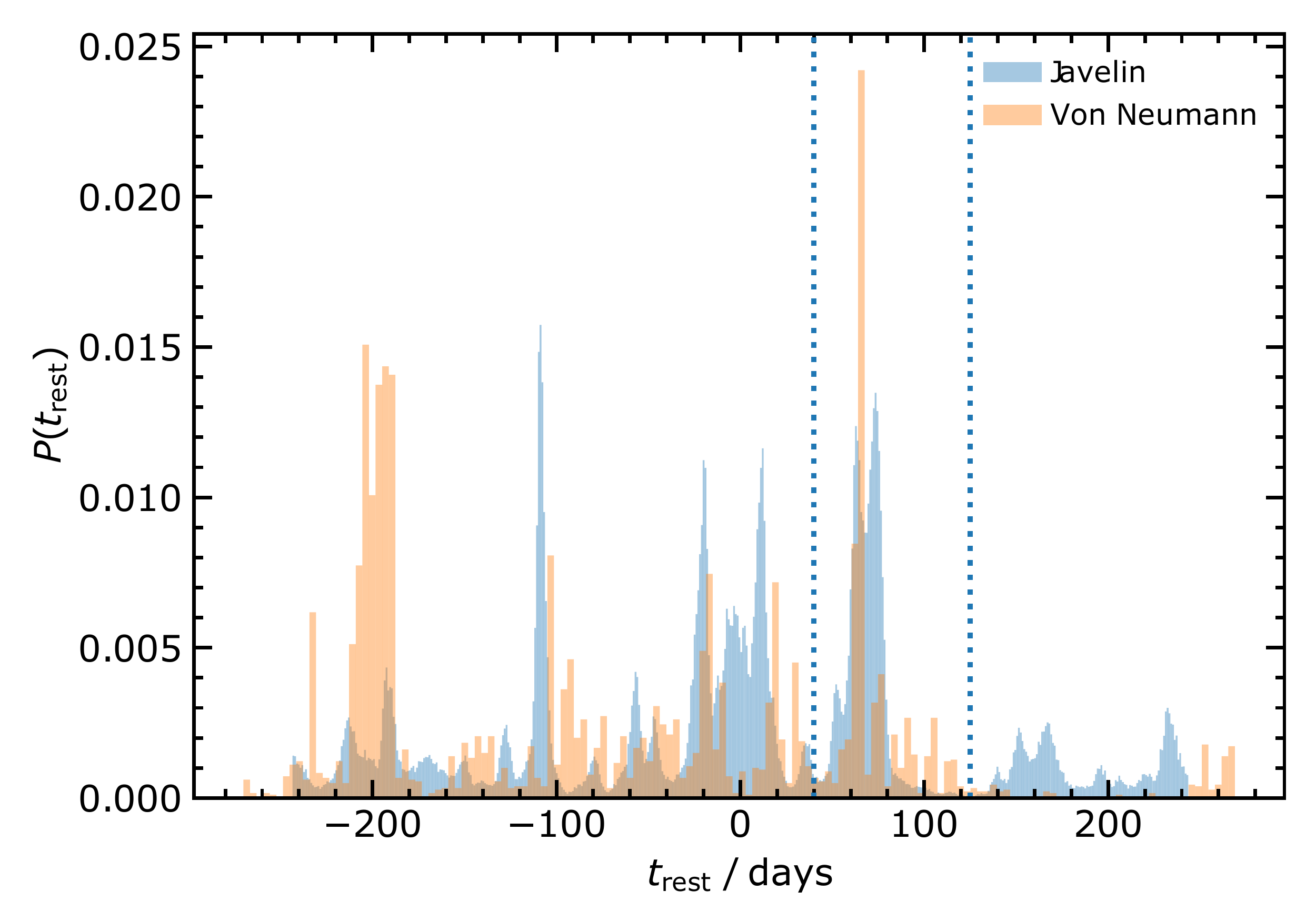}
    \caption{A comparison of the Von Neumann estimator (in orange) and \javelin (blue) probability distributions.
    The two estimates do not include artefact deconvolution.}
    \label{fig:RM:javelin_vn_comparison}
\end{figure}

Fig~\ref{fig:RM:javelin_vn_comparison} shows that the most likely positive peak from \javelin coincides with the a peak from Von Neumann estimator.
Corroboration from a model-independent method increases the likelihood of our detection being real.

\begin{figure*}
  \centering
  \subfigure[\javelin PDF with DRW artefact PDF from DRW-generated light-curves]{
    \includegraphics[width=0.45\linewidth]{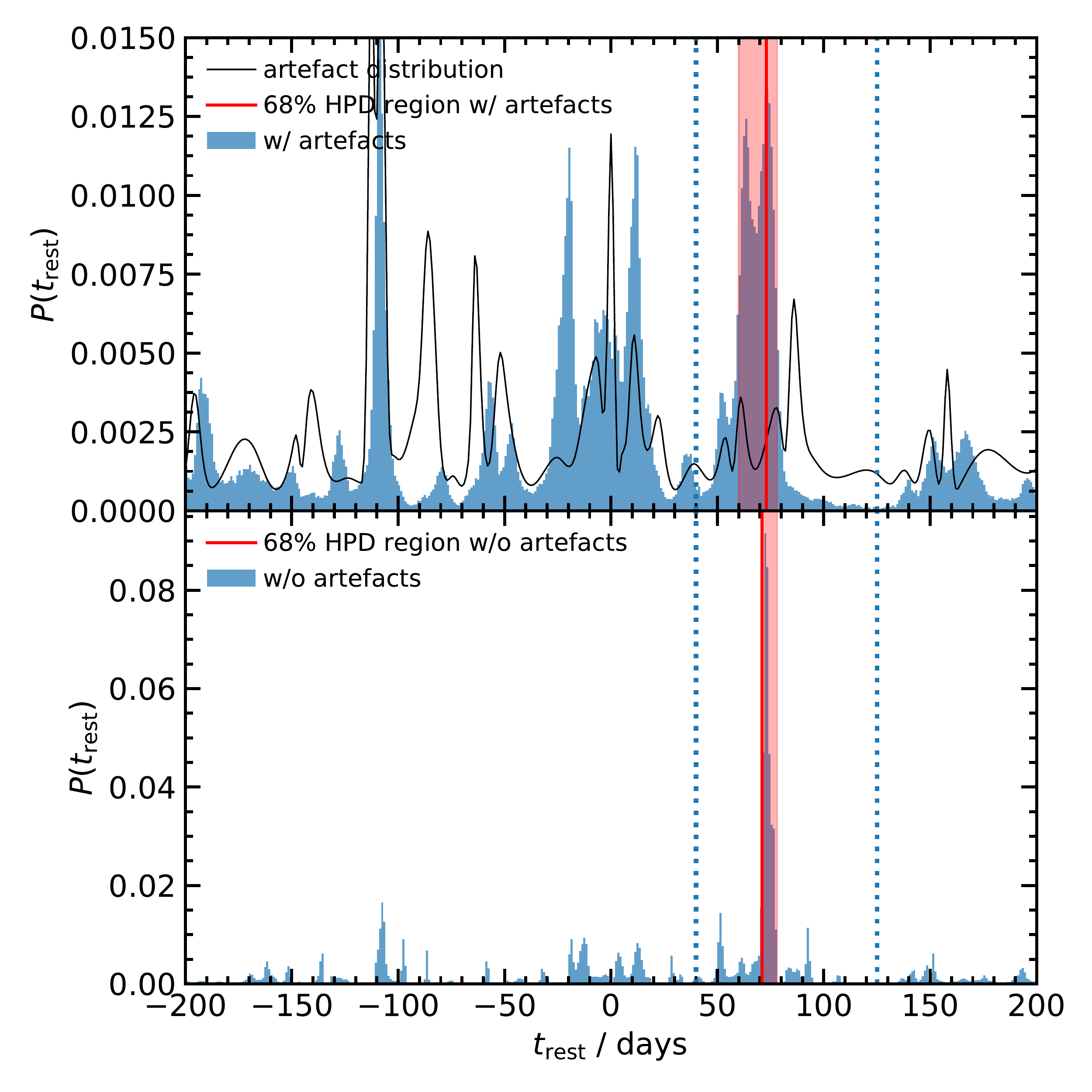}
    \label{fig:RM:lag_linear_pdf_drw}
    }
    ~
  \subfigure[\javelin PDF with CARMA(2,1) artefact PDF from CARMA(2,1)-generated light-curves]{
    \includegraphics[width=0.45\linewidth]{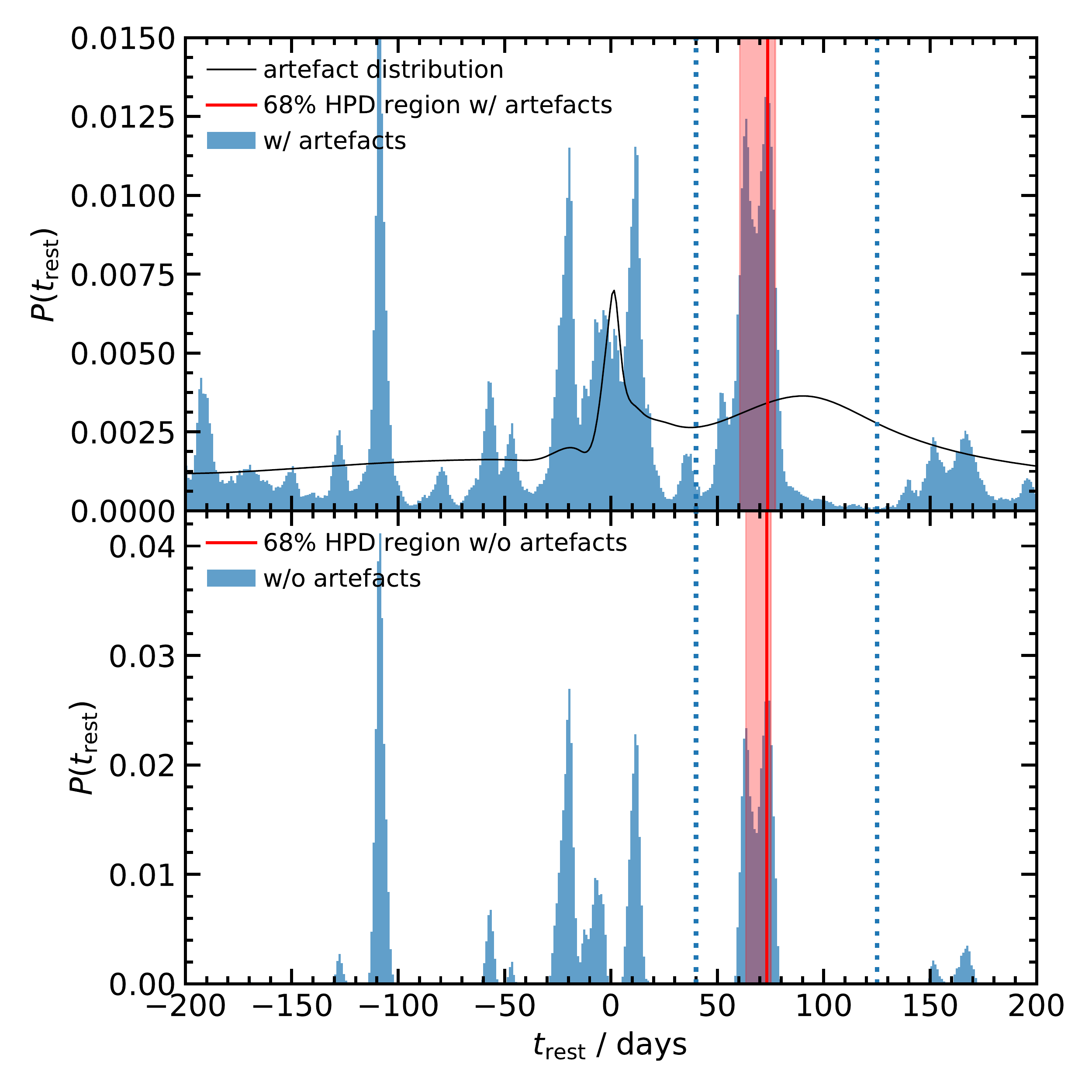}
    \label{fig:RM:lag_linear_pdf_kali}
    }
  ~
  \subfigure[Von Neumann estimated PSF with CARMA(2,1) artefact PDF from CARMA(2,1)-generated light-curves]{
    \includegraphics[width=0.45\linewidth]{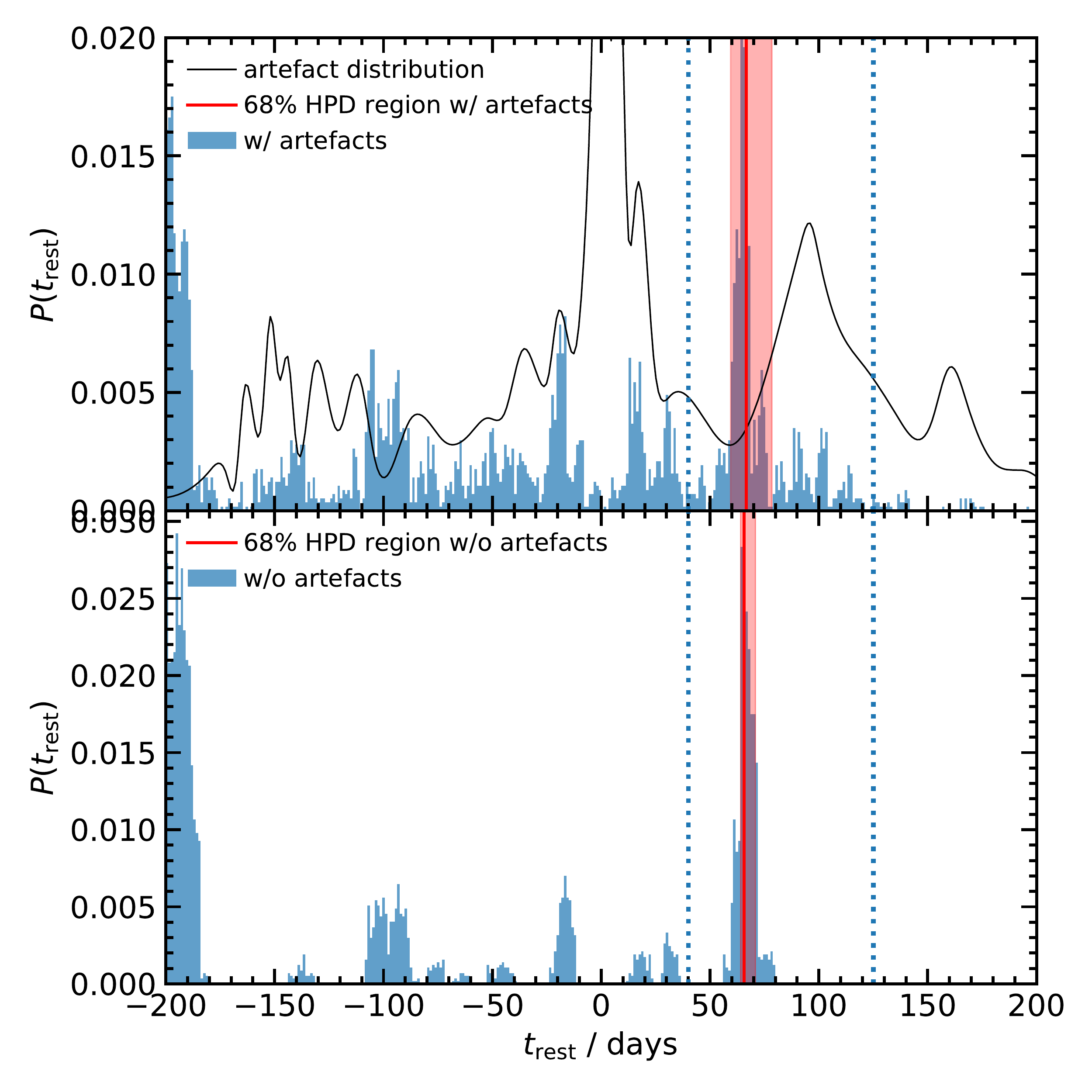}
    \label{fig:RM:lag_linear_pdf_vn}
    }
  \caption{The probability distributions for rest-frame lag of \TargetTen before and after artefact deconvolution for \javelin and the Von Neumann estimator performed on CARMA(2,1) and DRW light-curves.
  \textbf{Top panels:} The full probability distribution for rest-frame lag as the blue histogram along with the artefact distribution in black derived from simulated light-curves.  
  \textbf{Bottom panels:} The cleaned distribution of rest-frame lags for \TargetTen, where the artefact distribution is deconvolved from the output rest-frame lag distribution.
  The region marked by dashed lines indicates the region where we estimate the 68 per cent HPD interval (shaded red area), along with the mode (red line), which is determined by the position of the minima around the highest peak in the top panel (following the method performed by \citealt{Grier2017Sloan}). 
  }
  \label{fig:RM:javelin_input_output_lags_best_estimate}
\end{figure*}

However, the distribution of \hbeta lags contains more than one convincing (SNR $> 3$) peak in both methods.
Fortunately, since we have constructed a large suite of simulated light curves over a large range of DRW parameters, we can estimate the distribution of lag artefacts that results only from the fitting process and the properties of our data.
We can then use the distribution to inform us as to which peak is the ``real'' one.
For both \javelin and Von Neumann, we take the median PDF over all simulated light-curves.
This creates a distribution of lags without a peak corresponding to the true input lag, since the median at any point will suppress such a peak.
We scale the artefact distribution, an approximation of $1-P(t_{\hbeta})$, so that its median probability matches the median probability of the distribution of \TargetTen, $P(t_{\hbeta} \mid \mathcal{D})$.
Then we divide the \TargetTen lag distribution by this artefact distribution, which has the effect of suppressing spurious peaks.
The results for \javelin and the Von Neumann estimator are shown in Fig~\ref{fig:RM:javelin_input_output_lags_best_estimate}.
We can see in Fig~\ref{fig:RM:lag_linear_pdf_drw} that when the light-curves are DRW-generated, as \javelin assumes, the artefact distribution contains many peaks. 
The highest peak in the lag PDF for \TargetTen at $\sim -100$\,days is completely accounted for by DRW+\javelin effects. 
However, the much smoother distribution shown in Fig~\ref{fig:RM:lag_linear_pdf_kali} from using CARMA(2,1) light-curves, perhaps resulting from the greater inaccuracy in lag estimation, does not account for this peak. 
The artefact PDF of the Von Neumannn estimator, shown in Fig~\ref{fig:RM:lag_linear_pdf_vn}, contains many peaks, but the largest is centred around 0 days and does not account for the large probability mass found at -200 days.

\begin{figure}
    \centering
    \includegraphics[width=\linewidth]{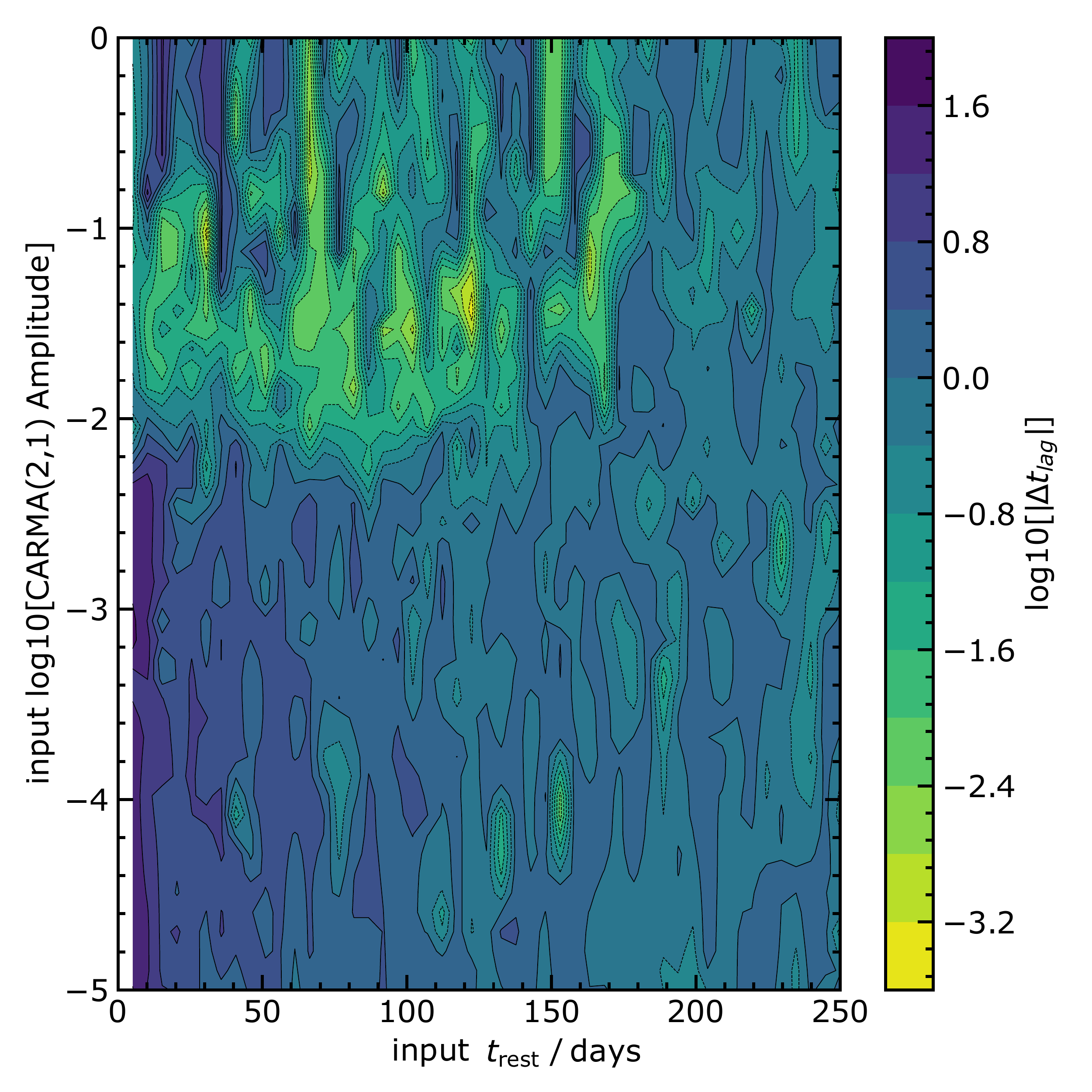}
    \caption{The distribution of logged lag-residuals over input variability amplitude and input lag. 
    The colour-bar indicates the log of the lag-residual.
    The surface was generated by linearly interpolating between the results from fitting our suite of light-curves.}
    \label{fig:RM:amplitude-lag}
\end{figure}

The accuracy of the \javelin estimations on CARMA(2,1) over input lag and input variability amplitude is shown in Fig~\ref{fig:RM:amplitude-lag}.
There is a clear region where \javelin appears to be able to recover lags: the lag must be smaller than 170 days to have the best chance of recovery and the continuum amplitude variability limit coincides with the mean fractional noise in the continuum light-curve (0.01).

\begin{figure*}
    \centering
    \includegraphics[width=\linewidth]{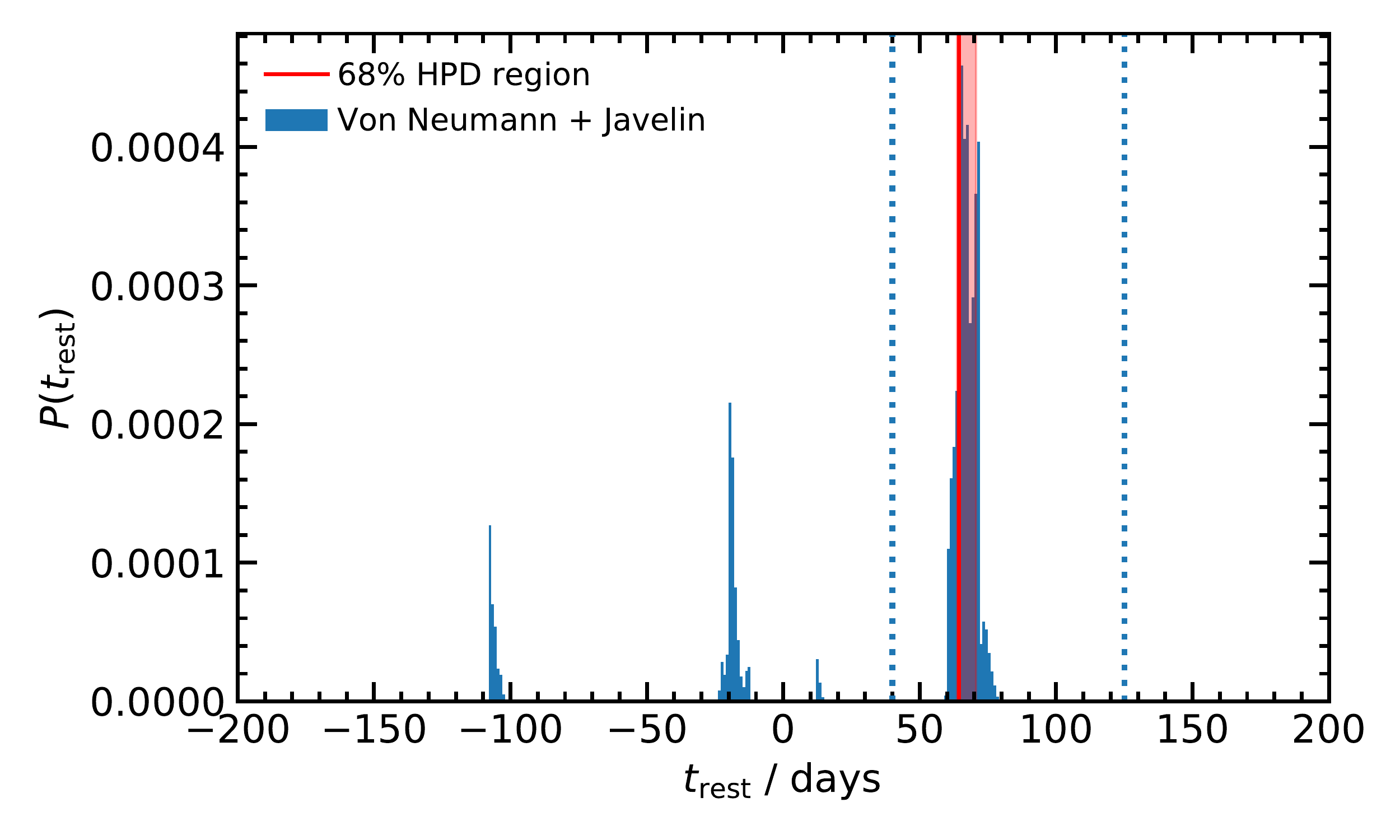}
    \caption{The probability distribution for rest-frame lag of \TargetTen after combining the artefact-deconvolved distributions of the Von Neumann estimator and \javelin.}
    \label{fig:RM:combined-estimation}
\end{figure*}

Partially following the method of \citet{Grier2017Sloan}, we select the region bounded by the minima of the tallest peak (dashed lines in Fig~\ref{fig:RM:javelin_input_output_lags_best_estimate} and \ref{fig:RM:combined-estimation}) in the distribution that still contains artefacts.
We then estimate the region of 68 per cent probability in the cases of artefact inclusion and deconvolution as shown in Fig~\ref{fig:RM:javelin_input_output_lags_best_estimate}.
In order to show that any detected lag robust to choice of model and to make use of all available data, we combine the PDFs of the deconvolved Von Neumann and \javelin lag estimations by multiplying them (shown in Fig~\ref{fig:RM:combined-estimation}). 
We do not include the PDF estimated from \javelin with the DRW-generated artefact distribution, since we have shown that this is too optimistic.
Deconvolution and combination do not entirely remove all ambiguity in the lag PDF, but it does push much of the probability mass into 3 distinct peaks at -105, -20, and +63 days.
The lack of noise and distinct peak heights makes reporting the +63 day lag more trustworthy and robust to the assumed generative time-series model (Von-Neumann doesn't assume any model and the CARMA(2,1) tests \javelin's resilience to mismatch). 

We recover an \hbeta lag for \TargetTen of \TargetTenUninformedLag without attempting to remove the influence of artefacts or combining techniques and then an \hbeta lag of \TargetTenInformedLag when we apply artefact deconvolution and method combination.
The lags estimated before and after deconvolution for each method are shown in Table~\ref{tab:RM:lags}
\begin{table}
\centering
\begin{tabular}{lll}
\toprule
{}                       & No deconvolution          & Deconvolved \\
\midrule
\javelin with DRW        &  \TargetTenUninformedLag  & $72^{+5}_{-1}$ days \\
\javelin with CARMA(2,1) &  \TargetTenUninformedLag  & $72^{+1}_{-10}$ days \\
VN with CARMA(2,1)       &  $66^{+12}_{-6}$ days     & $65^{+5}_{-1}$ days \\
Combined CARMA(2,1)      &    $-$                    & \TargetTenInformedLag \\

\bottomrule
\end{tabular}
\caption{The estimated lags for each method and model assumption, assuming that the real peak lies within the region demarked by the dashed vertical lines as described above.}
\label{tab:RM:lags}
\end{table}
 
The best KDE estimate of the lag of \TargetTen is consistent between both distributions but the uncertainty shrinks by 2.5 when we use the artefact deconvolution method to simplify the posterior and combine estimates from different techniques.

\subsection{Fits to the $t_{\mathrm{H\beta}}-L_{5100}$ Relation}\label{sec:RM:fits}
Using our derived time lag, we fit a power-law, with scatter, to the lag versus luminosity in linear space:

\begin{gather}
  t_{\mathrm{rest}}' / 1\,\mathrm{day} = 10^K[\lambda L_{\lambda} / 10^{44} \mathrm{erg s^{-1}}]^{\alpha}\\
  t_{\mathrm{rest}} \sim \mathcal{N}(\mu=t_{\mathrm{rest}}', \sigma=t_{\mathrm{rest}}'\epsilon)
\label{eq:tLrelation}
\end{gather}
where $t_{\mathrm{rest}}$ is the lag that would be observed without the effects of intrinsic scatter in the relation and $t_{\mathrm{rest}}'$ is the observed lag including that intrinsic scatter.
The normal distribution is indicated as $\mathcal{N}$.
Our fitting priors for the slope $\hat{\alpha}$, intercept $\hat{K}$, and scatter scale $\hat{\epsilon}$ are:
\begin{gather}
  \hat{\alpha} \sim \mathcal{N}(\mu=0.5, \sigma=0.75),\\
  \hat{K} \sim \mathrm{Trunc.}\mathcal{N}(\mu=1.5, \sigma=1.0, a=0, b=\infty),\\
  \log[\hat{\epsilon}] \sim \mathcal{N}(\mu=-2, \sigma=1)
\end{gather}
We correct the luminosity of our target for a host contribution of 24 per cent, as in \citet{Bentz2013LowLuminosity}. 
The details of the correction can be found in Appendix~\ref{app:host-deconvolution}.
We do not fit a straight line in log space since the uncertainties in lag and luminosity along with the scatter are not strictly Gaussian in linear space and definitely not in log space.
This subtlety may have a significant impact on the slope of the fit relation and therefore on its interpretation.
We use this opportunity to test whether the correct treatment of non-Gaussian uncertainties makes a difference to resultant fit.
We resample the uncertainty distributions of the lag estimations 1000 times per data point in order to fit the power law.
In this way, we incorporate the probability distribution from \javelin naturally whilst also treating values from the literature correctly.
We do not fit the power-law to the \citet{Grier2017Sloan} dataset since they reason that large selection effects due to limited monitoring cadence and duration may bias their lag measurements to lower values more so than the \citet{Bentz2013LowLuminosity} sample. 
Instead, we use the Clean2+ExtCorr dataset from \citet{Bentz2013LowLuminosity}, which excludes two AGN due to potentially biased time lags and corrects the influence of internal extinction of one other.
We recover the parameters listed in Table~\ref{tab:RM:fit_parameters}.

\begin{figure*}
  \centering
  \includegraphics[width=\linewidth]{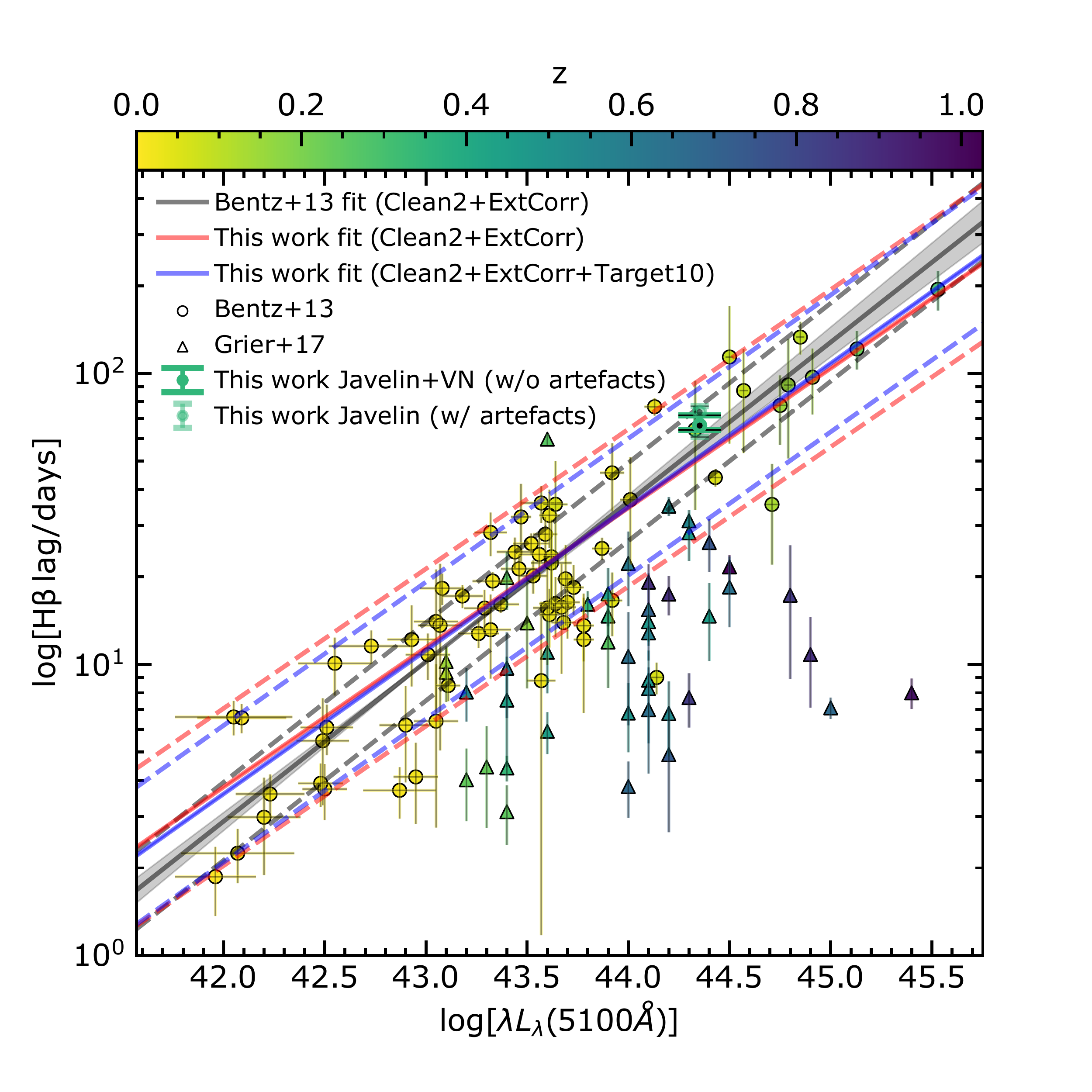}
  \caption{The rest-frame lag-luminosity relation shown for data from \citet{Bentz2013LowLuminosity} (circles), \citet{Grier2017Sloan} (triangles), and \TargetTen.
  All points are coloured by redshift. 
  The best estimate for the lag of \TargetTen is shown as a bold green circle with and without the artefact deconvolution.
  The best fit line in log space to the Clean2+ExtCorr dataset by \citep{Bentz2013LowLuminosity} is shown in grey, the best fit line in linear space to the same data is shown in red. 
  The best fit in linear space to the Clean2+ExtCorr dataset as well as \TargetTen is shown in blue.
  The scatter estimated by MCMC in all best fit lines is indicated by dashed lines.
  }
  \label{fig:RM:lag_luminosity_relation}
\end{figure*}

\begin{table}
\centering
\begin{tabular}{cccc}
\toprule
{} &                        $\hat{K}$ &                   $\hat{\alpha}$ &                  $\log[\hat{\epsilon}]$ \\
\midrule
Clean2+ExtCorr+\\Target10 &  $1.541_{-0.002}^{+0.001}$ &  $0.494_{-0.001}^{+0.001}$ &  $-0.542_{-0.005}^{+0.005}$ \\
\midrule
Clean2+ExtCorr          &  $1.539_{-0.002}^{+0.001}$ &  $0.480_{-0.001}^{+0.001}$ &  $-0.623_{-0.005}^{+0.004}$ \\
\midrule
Clean2+ExtCorr\\(Bentz+13) & $1.559\pm0.024$ &  $0.549_{-0.027}^{+0.028}$ & $\sim-1.016_{-0.187}^{+0.169}$\\
\bottomrule
\end{tabular}
\caption{Lag fit parameters for datasets with and without \TargetTen. 
The fit results from \citet{Bentz2013LowLuminosity} are included but the scatter has been approximately converted to the power law model using $\epsilon \approx 10^{\sigma} -1$  for comparison.}
\label{tab:RM:fit_parameters}
\end{table}
 Fig~\ref{fig:RM:lag_luminosity_relation} shows the fit lag-luminosity relation to the \citet{Bentz2013LowLuminosity} Clean2+ExtCorr sample.
There is no significant difference between the fits with and without \TargetTen included.
However, fitting in linear space produces a shallower relation (by $\sim0.013$) than that of \citet{Bentz2013LowLuminosity} and so, at extremes of luminosities, we find that our fit is significantly (3$\sigma$ at 41 dex) different to the log-log straight line. 
Additionally, the uncertainty in our fit parameters is much reduced when compared to \citet{Bentz2013LowLuminosity} and the scatter is larger (by about $0.5$ dex).
We also note that the impact of selection effects upon this and any fit of a t-L relation will be dependent on the cadence and duration of observations. 
This may go some way to explaining the seemingly excessive number of QSOs populating the space below the \citet{Bentz2013LowLuminosity} data points.
Furthermore, there may be an accretion rate dependency whereby the more fundamental relation is the plane of rest-frame lag, luminosity and accretion rate, as outlined by \citet{Du2016Supermassive}.
However, the explanatory power of this model is small for sources with the low accretion rates seen in the \citet{Grier2017Sloan} sample.

\begin{figure}
  \centering
  \includegraphics[width=\linewidth]{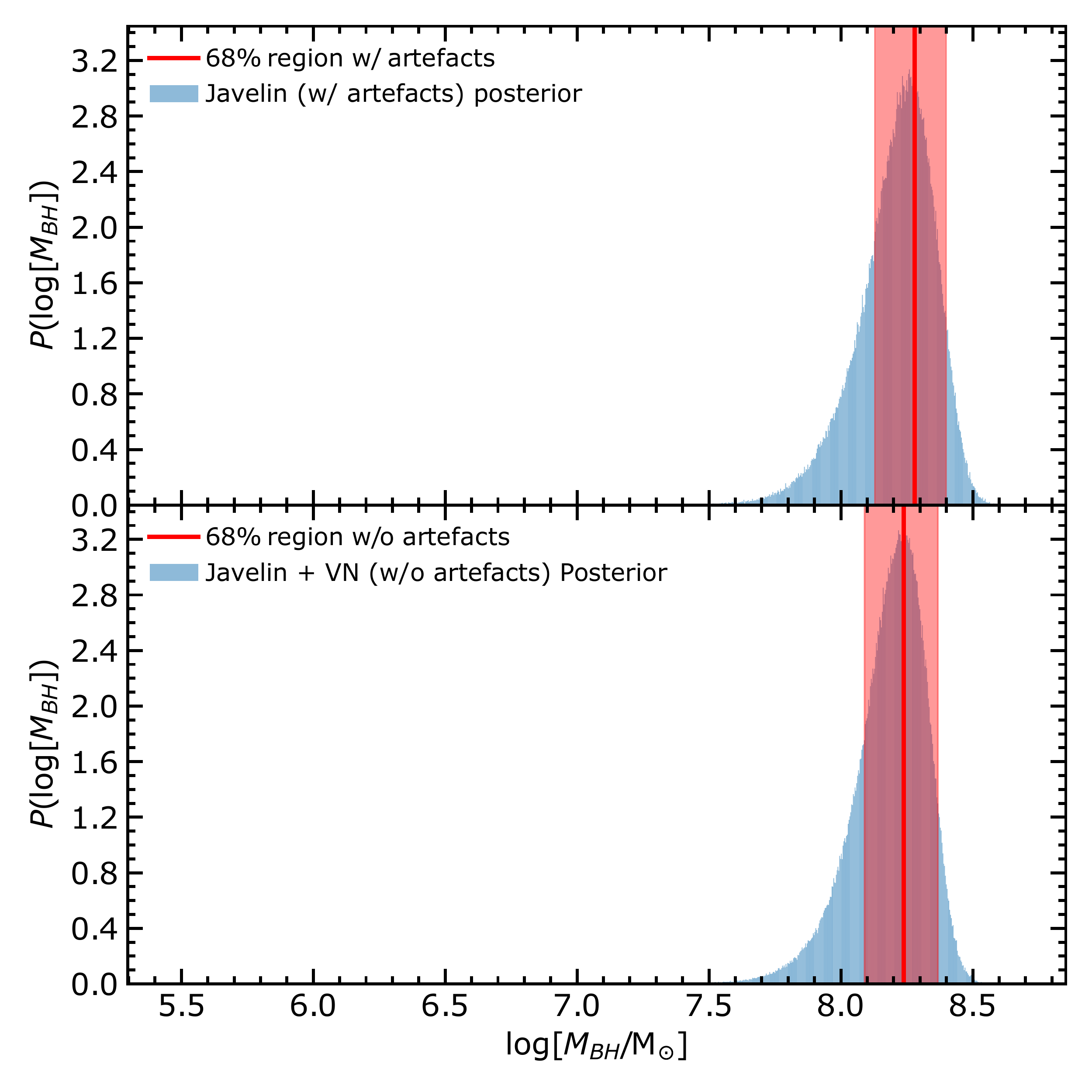}
  \caption{The probability distribution for black-hole mass before and after deconvolution of the \javelin artefact distribution.
  \textbf{Top:} The probability distribution for the black-hole mass of \TargetTen given the raw output from \javelin.
  \textbf{Bottom:} The probability distribution for the black-hole mass of \TargetTen given the deconvolved lag distribution. Both distributions incorporate uncertainties on velocity dispersion and the virial factor.
  The 68 per cent HPD region is shown in red in both cases with the best estimate indicated by the solid line.
  }
  \label{fig:RM:mass_pdf}
\end{figure}

Propagating the posterior lag distribution for \TargetTen through Equation~\ref{eq:RM:mass}, using the virial factor from \citet{Grier2013Structure} with a Gaussian distribution of $\langle f \rangle \sim \mathcal{N}(\mu=4.3, \sigma=1.1)$, we arrive at the distribution for black-hole mass shown in Fig~\ref{fig:RM:mass_pdf}.
The best estimates, with and without deconvolution of artefacts, for black-hole mass are only separated by 0.01 dex.

\begin{figure*}
  \centering
  \includegraphics[width=\linewidth]{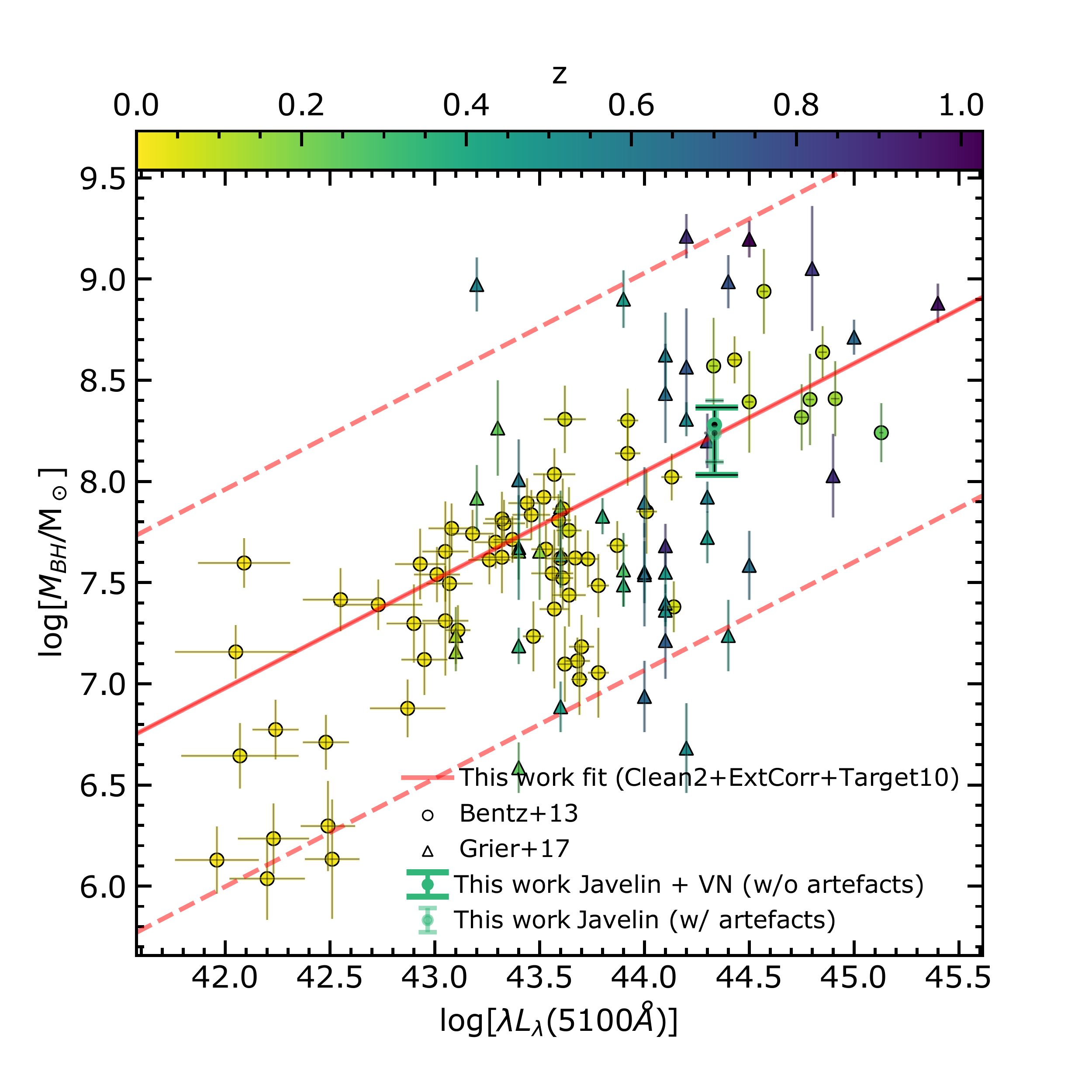}
  \caption{The black-hole mass-luminosity relation shown for the sample from \citet{Bentz2013LowLuminosity} (cirlces), \citet{Grier2017Sloan} (triangles), and \TargetTen.
  The black-hole masses for the \citet{Bentz2013LowLuminosity} sample are drawn from the \AGNMassCatalogue where possible and calculated using $f=4.3 \pm 1.1$ \citep{Grier2013Stellar}. 
  The \citet{Grier2017Sloan} masses are scaled from $f=4.47$ to $f=4.3$.
  All points are coloured by redshift. 
  The best estimate for the mass of \TargetTen is shown in green with and without the artefact deconvolution.
  The best fit in linear space to the Clean2+ExtCorr dataset as well as \TargetTen is shown in red.
  The scatter estimated by MCMC in all best fit lines is indicated by dashed lines.
  }
  \label{fig:RM:mass_luminosity_relation}
\end{figure*}

Fig~\ref{fig:RM:mass_luminosity_relation} shows the black-hole mass-luminosity relation for the \citet{Bentz2013LowLuminosity} \hbeta lags with line widths from the \AGNMassCatalogue \citep{Bentz2015Agn}. 
The parameter fits for the mass-luminosity relation are detailed in Table~\ref{tab:fit_parameters_mass}.
We find that \TargetTen is in good agreement with the \citet{Bentz2013LowLuminosity} Clean2+ExtCorr dataset.

\newcommand{\MassLuminosityScatter}{\ensuremath{-0.017_{-0.004}^{+0.006}}}

\begin{table}
\centering
\begin{tabular}{cccc}
\toprule
{} &                        $\hat{K}$ &                   $\hat{\alpha}$ &                  $\log[\hat{\epsilon}]$ \\
\midrule
Clean2+ExtCorr+\\Target10 &  $8.048_{-0.002}^{+0.002}$ &  $0.535_{-0.002}^{+0.001}$ &  \MassLuminosityScatter \\
\bottomrule
\end{tabular}
\caption{Mass fit parameters for datasets with \TargetTen, using the same parametrisation for a power-law as in Table~\ref{tab:RM:fit_parameters}.}
\label{tab:fit_parameters_mass}
\end{table}
 
\begin{figure}
  \centering
  \includegraphics[width=\linewidth]{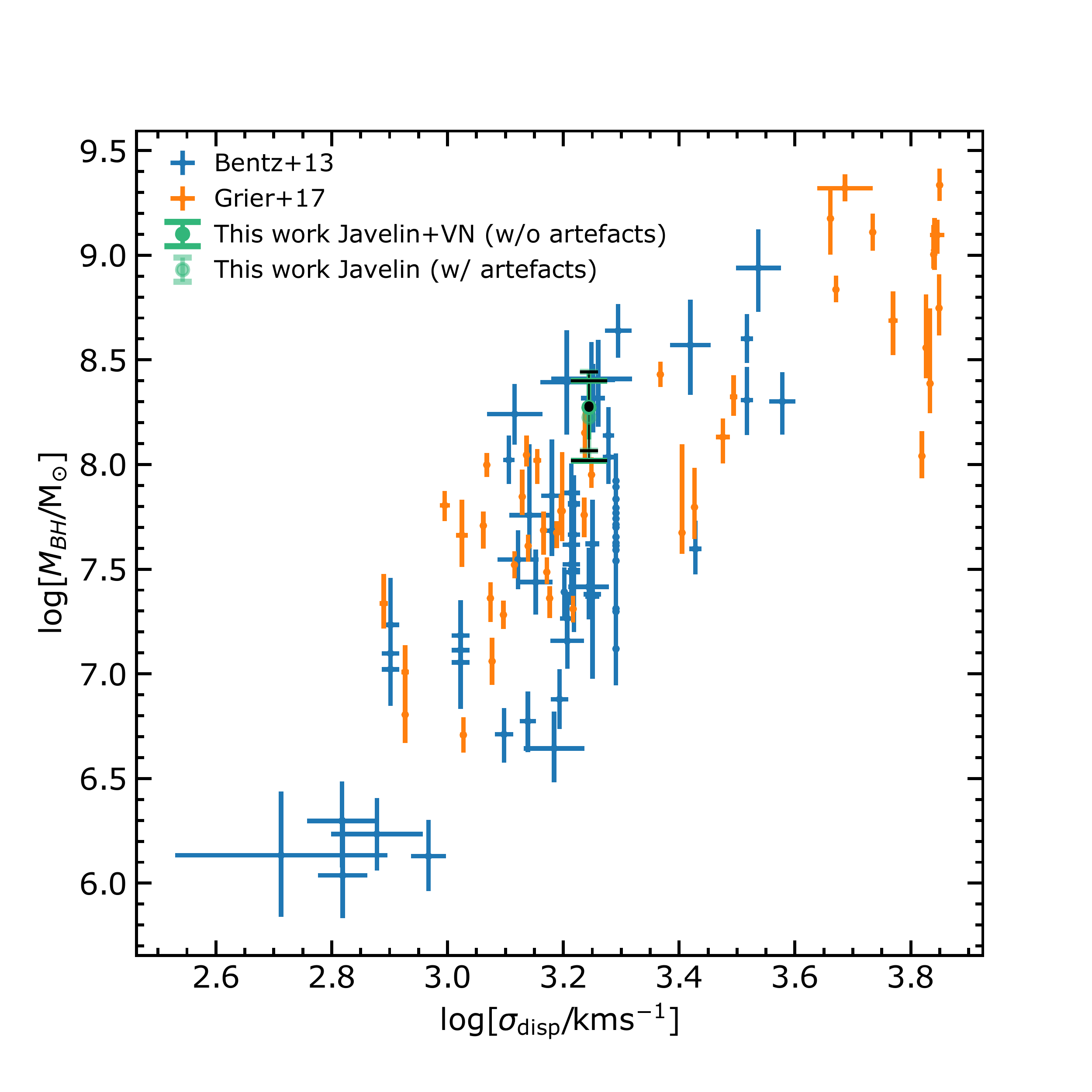}
  \caption{Estimated black-hole mass versus the \hbeta velocity dispersion. 
  The \citet{Bentz2013LowLuminosity} sample is shown in blue, \citet{Grier2017Sloan} sample is shown in orange, and \TargetTen is shown in green with and without the \citet{Bentz2013LowLuminosity} prior applied. 
  The line widths for the \citet{Bentz2013LowLuminosity} sample are retrieved from the \AGNMassCatalogue \citep{Bentz2015Agn}.
  }
  \label{fig:RM:mass_velocity_dispersion}
\end{figure}

We find that the scatter of the mass-luminosity relation (0.5 dex) is much larger than that of the lag-luminosity relation in log space.
This is unsurprising since the former combines uncertainty from the virial factor $\langle f\rangle$ as well as the scatter in line widths shown in Fig~\ref{fig:RM:mass_velocity_dispersion}, which shows the black-hole mass against broad line velocity dispersion.

However, it is still useful to note that a black-hole mass predicted from the $t-L_{5100}$ relation can be wrong by more than 0.3 dex 50 per cent of the time\footnote{Calculated from the fit line in log-space (Figure~\ref{fig:RM:mass_luminosity_relation}) with a $1\sigma$ width of $\sim0.5$ dex. $1 - P(-0.3 < t \leq +0.3) = 0.5$}.

\section{Discussion}\label{sec:RM:discussion}
\subsection{Efficiency}\label{sec:RM:discussion:efficiency}
This observing campaign totalled 17.4 hours (15.2 for \halpha and 2.2 for \iband) in total, with 5.9 hours dedicated to \TargetTen.
This is far shorter than the large majority of spectroscopic observing campaigns such as \citet{Shen2015Sloan} where the typical epoch consists of at least eight 15 minute sub-exposures rather than our one 10 minute exposure with the Liverpool Telescope per epoch. 
\citet{Grier2017Sloan} achieved an average uncertainty of $3\pm2$ days and a maximum SNR of 23.1 whereas \TargetTen has an uncertainty of $+6/-1$ days (SNR$ = \TargetTenInformedSNR$), with much of the uncertainty attributed to artificial peaks having been mitigated using our simulations (see Section~\ref{sec:RM:methods:lag estimation}).

We define efficiency as the mean SNR$_{\mathrm{lag}}$ achieved for a given observing campaign divided by the total time required.
\begin{equation}\label{eq:RM:efficiency}
  \epsilon = \frac{\sum_{i=0}^{i=n}\mathrm{SNR_{lag}}}{nt_{\mathrm{total}}\pi (D/2)^2},
\end{equation}
\noindent where $n$ is the number of observed targets (detection or not), $t_{\mathrm{total}}$ is the total observing campaign observing time, and $D$ is the primary mirror diameter.
The mirror diameters are 2.5\,m for SDSS-RM and 2\,m for this work, which uses the Liverpool Telescope.
This gives us the expected signal-to-noise for a given QSO per hour of observation per collecting area.
In order to make a fair comparison, we include the SDSS spectrum integration time required to estimate velocity dispersions for each of our targets in the total time required to observe our targets as well.

\begin{table}
\centering
\begin{tabular}{lllll}
\toprule
              &                Selection                                   & $\epsilon_{\mathrm{SDSS-RM}}$ & $\epsilon_{\mathrm{This\,work}}$ & $\frac{\epsilon_{\mathrm{This\,work}}}{\epsilon_{\mathrm{SDSS-RM}}}$ \\
This work & SDSS-RM &          $\times10^{-3}$ &                          $\times10^{-3}$ &                                                                      \\
\midrule
\multirow{ 2}{*}{$i_{AB} < 18$} & all objects & 4.4 & 14.0 & 3.2 \\
              & $i_{AB} < 18$ & 3.1 & & 4.5 \\
\cline{1-5}
\multirow{ 2}{*}{Target-10} & $\min[\Delta f^{obs}_{5100}]$ & 1.3 & 356.7 & 274.4 \\
              & $\max[\mathrm{SNR}_{lag}]$ & 17.7 & & 20.2 \\
\bottomrule
\end{tabular}
\label{tab:RM:efficiencies}
\caption{The efficiencies, calculated with different selection criteria, for SDSS-RM \citep{Shen2015Sloana,Grier2017Sloan} and this work.
The efficiencies are calculated using Equation~\ref{eq:RM:efficiency}.
We compare the efficiencies on a per object basis as well as over the whole campaign.
We compare our \TargetTen to the most similar QSO in the \citet{Grier2017Sloan} catalogue (based on $f_{5100}$) and to their most precise lag estimation (in terms of SNR$_{lag}$). 
In all cases, photometric reverberation mapping is more efficient than spectroscopic reverberation mapping.}. 
\end{table}
 
We have achieved an efficiency of $\epsilon = 14.0\times10^{-3}\mathrm{hr}^{-1}\mathrm{m}^{-2}$, whereas with spectroscopic reverberation mapping, SDSS-RM achieved $\epsilon = 4.4\times10^{-3}\mathrm{hr}^{-1}\mathrm{m}^{-2}$, where our fraction of sources with detected lags (0.2) is the same as that of \citet{Grier2017Sloan}.
This is a \EfficiencyPercentIncrease increase in efficiency over the multiplexed SDSS-RM campaign.
If we instead calculate the signal-to-noise per hour per square metre per object, SNR$/t_{obj}$, we find that on average we achieve 12 times more signal-to-noise per hour than \citet{Grier2017Sloan}.
Since the SNRs of the \citet{Grier2017Sloan} lags do not depend strongly on redshift, observed flux or luminosity, this is a fair comparison.

The efficiencies described above include targets that we have observed but not analysed and consider the whole observing campaign at once.
If we only consider \TargetTen compared to the most precise lag measured by \citet{Grier2017Sloan}, for \texttt{SDSS J142103.53+515819.5}, our efficiency rises to 18 times more signal-to-noise per hour per square metre than \citet{Grier2017Sloan}
Furthermore, if we consider the most similar target to our \TargetTen in terms of observed flux (\texttt{SDSS J140759.07+534759.8}), their efficiency drops to $\epsilon = 1.3\times10^{-3}\mathrm{hr}^{-1}\mathrm{m}^{-2}$.

\subsection{Future Applications}\label{sec:RM:discussion:future}
Having shown that reverberation mapping using photometric methods with minimal spectroscopy can be an effective means with which to measure black-hole masses, we can foresee a number of exciting applications for long term studies, which would require little extra effort to instigate.

The Liverpool Telescope \citep{Steele2004Liverpool} will soon be superseded by a new robotic successor, the Liverpool Telescope 2 \citep{Copperwheat2014Liverpool}, with first light after 2020. 
The Liverpool Telescope 2 will benefit from a 4 metre diameter as opposed to the current Liverpool Telescope's 2 metres. 
Given the efficiency of photometric reverberation mapping with the current Liverpool Telescope, the application of these methods to its successor would be an effective use of time when applied robotically and make higher redshift measurements possible.

Photometric reverberation mapping lends itself well to large surveys, which often require that the instrument make repeated visits to the same field for calibration to standard stars.
Selecting calibration fields to contain known QSOs would generate light curves with baselines as long as the survey's duration with a regular high-frequency cadence for little extra effort.
The upcoming photometric surveys of the Javalambre Physics of the Accelerating Universe Astrophysical Survey \citep[J-PAS,][]{Benitez2014JPas} and its companion calibration survey Javalambre-Photometric Local Universe Survey (J-PLUS) promise an opportunity for sustained long-term photometric reverberation mapping campaigns. 
Designed to accurately measure photometric redshifts for galaxies up to $z=1$, with its unprecedented 56 narrow band filters, J-PLUS could easily observe the continuum and a wide range of emission lines for a sample of QSOs observed during calibration exposures.
In addition, instruments such as the PAUCam \citep{Castander2012Pau,Padilla2016Pau}, providing 40 narrow-band filters in addition to the $u$,$g$,$r$,$i$,$z$, and $y$ photometric filters, could also detect lags with higher SNR and a larger range of redshifts than IO:O.
These observations could provide a far more detailed map of the broad-line region as inferred by \citet{Williams2018Lick}, and also provide a large enough dataset to perform continuum reverberation mapping \citep{Mudd2017Quasar} to estimate accretion disk sizes.

The Large Synoptic Survey Telescope \citep[LSST,][]{Marshall2017Lsst} will run a 10 year survey over 30~000 square degrees of sky with 6 broad-band photometric filters.
LSST will observe the same regions of sky with a high frequency and 3 day cadence, making pure photometric reverberation \citep{Zu2016Application} with large numbers of QSOs a realistic possibility \citep{Chelouche2014On}.
A QSO light curve dataset from LSST would probe the extremes of time-scales where the damped random walk model for QSO variability is thought to break down \citep{Zu2013Quasar} whilst also providing opportunities for continuum mapping \citep{Mudd2017Quasar}. 
However, it is currently not clear whether LSST will be able to estimate accurate lags since \citet{Chelouche2014On} do not account for photometric measurement errors, dilution of light curve variations by host galaxy contribution, seeing effects which affect the host/nucleus separation and luminosity determination. 
Indeed, the selection is restricted to objects with strong emission lines, which is not the case for narrow-band photometric reverberation mapping.

Given that we can measure lags with \TargetTenLagUncertaintyWidth uncertainty with current instrumentation, for baselines longer than $3t_{\mathrm{rest}}(1+z)$, these survey's long campaigns and high cadences, along with high precision photometry, will likely provide more than enough signal-to-noise for lag estimation for hundreds of QSOs/AGN covering a large range of lags and luminosities.
Indeed, strategic application of photometric continuum mapping and multiple narrow band filters probing multiple broad-line region radii will yield much information regarding the geometry and mass of SMBHs.

\section{Conclusions}\label{sec:RM:conclusions}
We demonstrate an efficient method for purely photometric QSO reverberation mapping at high redshift ($z=\TargetTenRedshift$) using \javelin \citep{Zu2016Application,Zu2013Quasar}.
\begin{enumerate}
  \item We observe 10 targets selected for their estimated signal-to-noise, observable time, and inferred \hbeta emission line lag (according to the $t_{lag}-L_{5100}$ relation fit in \citealt{Bentz2013LowLuminosity}).\\

  \item Observing conditions ruled out the observation of 5 of our selected targets and 4 observed targets did not have the required baseline, recommended by \citet{Shen2015Sloan}, to observe their expected lag given their luminosity.
  We therefore proceed to discuss only \TargetTenFullName (referred to as \TargetTen).\\

  \item We calibrate the \halpha and \iband light curves, using an ensemble photometry method, to SDSS AB magnitudes.
  In order to achieve as accurate an \halpha relative calibration zeropoint as possible, we use the only available SDSS-BOSS spectrum. 
  This spectrum is observed to be resolved into two components in both our \iband and \halpha exposures, and the SDSS \iband exposures. 
  Therefore, we fit a two-component Gaussian model to the source in order to transform to the same seeing as the BOSS observation before fitting a zeropoint.\\

  \item Javelin and other tools assume the frequently-used DRW model and the effect of this assumption on the accuracy of lag estimation when the light-curves are not DRW-generated is not known. 
  To make our lag robust to the choice of model and interpolation of the model, we generate \NumberOfSimulatedLightCurves simulated CARMA(2,1) \& DRW-generated light-curves based on CARMA(2,1) fits to the reprocessed Kepler light-curves \citep{Smith2018Kepler} using the same cadence and signal-to-noise measured in our calibrated light curves for \TargetTen.
  We find that although the accuracy of \javelin decreases when its base assumption is violated, it can still recover the correct input lag. 
  Indeed, a model-independent Von Neumann estimator corroborates the 63 day peak in the \javelin lag PDF.
  
  \item We find that the output lag probability distribution from photometric RM is frequently affected by multiple peaks, some at negative lag values. 
  We find that median estimate of the lag from the posterior probability distributions often reports inaccurate values and large uncertainties for lags. 
  We therefore use an HPD kernel method (Section~\ref{sec:RM:methods:lag estimation}) to automatically identify the most probable peak objectively.
  Using the HPD kernel method, we report the reliability of \javelin and Von Neumann estimated lag over 0 to 316 days.
  We are able to reliably recover the original input lag over all other nuisance parameter ranges for the simulated light curves with an average of \AverageLagRecoveryErrorLessThan deviation when the input lag is less than \ReliableLagsBelow.
  When simulating light curves based on the signal-to-noise and cadence of \TargetTen, we find that an error of no more than 0.4 mag in \halpha narrow-band zeropoint calibration is still able to recover the given input lag to within an average of \AverageLagRecoveryErrorLessThan.\\

  \item Using the simulated light curves generated from reprocessed Kepler light-curves \citep{Smith2018Kepler}, we compile a distribution of artefacts in the lag distribution produced by the \javelin and Von Neumann fitting procedure.
  We deconvolve the artefact distribution from the lag distributions of \TargetTen and combine the estimations from both \javelin and the Von Neumann estimator, measuring \hbeta lags and black-hole masses with smaller uncertainties than without artefact deconvolution. 
  We find that the best estimate of the \hbeta lag and black-hole mass do not change beyond the 68 per cent HPD credible interval when the artefact deconvolution is applied but their uncertainties shrink.
  We recover an \hbeta lag for \TargetTen of \TargetTenUninformedLag with \javelin and an \hbeta lag of \TargetTenInformedLag when we apply artefact deconvolution to both \javelin and the Von Neumann estimator and combine their results.
  Assuming an \ffactor, we measure a black-hole mass for \TargetTen of \TargetTenUninformedMass with \javelin and a black-hole mass of \TargetTenInformedMass when we apply artefact deconvolution and combination.
\end{enumerate}

In conclusion, we find that if a Damped Random Walk (DRW) model is assumed by the fitting procedure when the light-curves are generated by a different Continuous Auto-Regressive Moving Average (CARMA) process, we can still recover accurate lags (despite a small loss in reliability). 
We find that by analysing the resulting probability distribution with more in-depth techniques, we can approach the precision demonstrated by spectroscopic reverberation mapping using photometric techniques.
Furthermore, we can achieve this precision with a quarter of the total exposure time that the SDSS-RM programme required to achieve a higher average SNR with a smaller telescope. 
This results in a \EfficiencyPercentIncrease increase in efficiency over SDSS-RM.
These simple yet powerful photometric methods can be readily applied to large surveys which require regular calibration in order to build a large baseline of known QSO observations.

\section{Acknowledgements}
SCR thanks Garreth Martin and Martin Hardcastle for fruitful discussions. 
SCR thanks Vishal Kasliwal for informative descriptions of how \texttt{Kali} works and fruitful discussions about the application of the CARMA(2,1) process to the Kepler light curves.
SCR thanks Krista Lynne Smith for providing the reprocessed Kepler light curves.
This research made use of Astropy, a community-developed core Python package for Astronomy \citep{Collaboration2013Astropy}. The Astropy website is \url{http://www.astropy.org/}.
SCR acknowledges support from the UK Science and Technology Facilities Council [ST/N504105/1].
This research has made use of the University of Hertfordshire high-performance computing facility \url{http://stri-cluster.herts.ac.uk/} and the LOFAR-UK computing facility located at the University of Hertfordshire and supported by STFC [ST/P000096/1].
Funding for SDSS-III has been provided by the Alfred P. Sloan Foundation, the Participating Institutions, the National Science Foundation, and the U.S. Department of Energy Office of Science. The SDSS-III web site is \url{http://www.sdss3.org/}. SDSS-III is managed by the Astrophysical Research Consortium for the Participating Institutions of the SDSS-III Collaboration including the University of Arizona, the Brazilian Participation Group, Brookhaven National Laboratory, Carnegie Mellon University, University of Florida, the French Participation Group, the German Participation Group, Harvard University, the Instituto de Astrofisica de Canarias, the Michigan State/Notre Dame/JINA Participation Group, Johns Hopkins University, Lawrence Berkeley National Laboratory, Max Planck Institute for Astrophysics, Max Planck Institute for Extraterrestrial Physics, New Mexico State University, New York University, Ohio State University, Pennsylvania State University, University of Portsmouth, Princeton University, the Spanish Participation Group, University of Tokyo, University of Utah, Vanderbilt University, University of Virginia, University of Washington, and Yale University.

\bibliographystyle{mnras}\bibliography{main}

\begin{thebibliography}{}
\makeatletter
\relax
\def\mn@urlcharsother{\let\do\@makeother \do\$\do\&\do\#\do\^\do\_\do\%\do\~}
\def\mn@doi{\begingroup\mn@urlcharsother \@ifnextchar [ {\mn@doi@}
  {\mn@doi@[]}}
\def\mn@doi@[#1]#2{\def\@tempa{#1}\ifx\@tempa\@empty \href
  {http://dx.doi.org/#2} {doi:#2}\else \href {http://dx.doi.org/#2} {#1}\fi
  \endgroup}
\def\mn@eprint#1#2{\mn@eprint@#1:#2::\@nil}
\def\mn@eprint@arXiv#1{\href {http://arxiv.org/abs/#1} {{\tt arXiv:#1}}}
\def\mn@eprint@dblp#1{\href {http://dblp.uni-trier.de/rec/bibtex/#1.xml}
  {dblp:#1}}
\def\mn@eprint@#1:#2:#3:#4\@nil{\def\@tempa {#1}\def\@tempb {#2}\def\@tempc
  {#3}\ifx \@tempc \@empty \let \@tempc \@tempb \let \@tempb \@tempa \fi \ifx
  \@tempb \@empty \def\@tempb {arXiv}\fi \@ifundefined
  {mn@eprint@\@tempb}{\@tempb:\@tempc}{\expandafter \expandafter \csname
  mn@eprint@\@tempb\endcsname \expandafter{\@tempc}}}

\bibitem[\protect\citeauthoryear{Alexander}{Alexander}{2013}]{Alexander2013Improved}
Alexander T.,  2013, arXiv e-prints, p. arXiv:1302.1508

\bibitem[\protect\citeauthoryear{Antonucci}{Antonucci}{1993}]{Antonucci1993Unified}
Antonucci R.,  1993, \mn@doi [\aapr] {10.1146/annurev.aa.31.090193.002353}, 31,
  473

\bibitem[\protect\citeauthoryear{Barišić et~al.,}{Barišić
  et~al.}{2017}]{Barisic2017Stellar}
Barišić I.,  et~al., 2017, \mn@doi [\apj] {10.3847/1538-4357/aa8768}, 847, 72

\bibitem[\protect\citeauthoryear{Barth et~al.,}{Barth
  et~al.}{2015}]{Barth2015Lick}
Barth A.~J.,  et~al., 2015, \mn@doi [\apjs] {10.1088/0067-0049/217/2/26}, 217,
  26

\bibitem[\protect\citeauthoryear{Benitez et~al.,}{Benitez
  et~al.}{2014}]{Benitez2014JPas}
Benitez N.,  et~al., 2014, arXiv:1403.5237 [astro-ph]

\bibitem[\protect\citeauthoryear{Benson, Bower, Frenk, Lacey, Baugh  \&
  Cole}{Benson et~al.}{2003}]{Benson2003Shapes}
Benson A.~J.,  Bower R.~G.,  Frenk C.~S.,  Lacey C.~G.,  Baugh C.~M.,   Cole
  S.,  2003, \mn@doi [\apj] {10.1086/379160}, 599

\bibitem[\protect\citeauthoryear{Bentz \& Katz}{Bentz \&
  Katz}{2015}]{Bentz2015Agn}
Bentz M.~C.,  Katz S.,  2015, \mn@doi [\pasp] {10.1086/679601}, 127, 67

\bibitem[\protect\citeauthoryear{Bentz, Peterson, Netzer, Pogge  \&
  Vestergaard}{Bentz et~al.}{2009a}]{Bentz2009RadiusLuminosity}
Bentz M.~C.,  Peterson B.~M.,  Netzer H.,  Pogge R.~W.,   Vestergaard M.,
  2009a, \mn@doi [\apj] {10.1088/0004-637X/697/1/160}, 697, 160

\bibitem[\protect\citeauthoryear{Bentz et~al.,}{Bentz
  et~al.}{2009b}]{Bentz2009Lick}
Bentz M.~C.,  et~al., 2009b, \mn@doi [\apj] {10.1088/0004-637X/705/1/199}, 705,
  199

\bibitem[\protect\citeauthoryear{Bentz et~al.,}{Bentz
  et~al.}{2013}]{Bentz2013LowLuminosity}
Bentz M.~C.,  et~al., 2013, \mn@doi [\apj] {10.1088/0004-637X/767/2/149}, 767,
  149

\bibitem[\protect\citeauthoryear{Bentz et~al.,}{Bentz
  et~al.}{2014}]{Bentz2014Mass}
Bentz M.~C.,  et~al., 2014, \mn@doi [\apj] {10.1088/0004-637X/796/1/8}, 796, 8

\bibitem[\protect\citeauthoryear{Bertin \& Arnouts}{Bertin \&
  Arnouts}{1996}]{Bertin1996Sextractor}
Bertin E.,  Arnouts S.,  1996, \mn@doi [\aaps] {10.1051/aas:1996164}, 117, 393

\bibitem[\protect\citeauthoryear{Blandford \& McKee}{Blandford \&
  McKee}{1982}]{Blandford1982Reverberation}
Blandford R.~D.,  McKee C.~F.,  1982, \mn@doi [\apj] {10.1086/159843}, 255, 419

\bibitem[\protect\citeauthoryear{Brewer \& Elliott}{Brewer \&
  Elliott}{2014}]{Brewer2014Hierarchical}
Brewer B.~J.,  Elliott T.~M.,  2014, \mn@doi [\mnras] {10.1093/mnrasl/slt174},
  439, L31

\bibitem[\protect\citeauthoryear{Cackett, Chiang, McHardy, Edelson, Goad, Horne
   \& Korista}{Cackett et~al.}{2018}]{Cackett2018Accretion}
Cackett E.~M.,  Chiang C.-Y.,  McHardy I.,  Edelson R.,  Goad M.~R.,  Horne K.,
    Korista K.~T.,  2018, \mn@doi [\apj] {10.3847/1538-4357/aab4f7}, 857, 53

\bibitem[\protect\citeauthoryear{{Carroll} \& {Joner}}{{Carroll} \&
  {Joner}}{2015}]{Carroll2015Photometric}
{Carroll} C.~J.,  {Joner} M.~D.,  2015, in \aas~Meeting Abstracts \#225. p.
  144.08

\bibitem[\protect\citeauthoryear{Castander et~al.,}{Castander
  et~al.}{2012}]{Castander2012Pau}
Castander F.~J.,  et~al., 2012. p. 84466D, \mn@doi{10.1117/12.926234}

\bibitem[\protect\citeauthoryear{Chelouche \& Daniel}{Chelouche \&
  Daniel}{2012}]{Chelouche2012Photometric}
Chelouche D.,  Daniel E.,  2012, \mn@doi [\apj] {10.1088/0004-637X/747/1/62},
  747, 62

\bibitem[\protect\citeauthoryear{Chelouche, Shemmer, Cotlier, Barth  \&
  Rafter}{Chelouche et~al.}{2014}]{Chelouche2014On}
Chelouche D.,  Shemmer O.,  Cotlier G.~I.,  Barth A.~J.,   Rafter S.~E.,  2014,
  ] {10.1088/0004-637X/785/2/140}

\bibitem[\protect\citeauthoryear{Chelouche, Pozo-Nuñez  \& Zucker}{Chelouche
  et~al.}{2017}]{Chelouche2017Methods}
Chelouche D.,  Pozo-Nuñez F.,   Zucker S.,  2017, \mn@doi [ApJ]
  {10.3847/1538-4357/aa7b86}, 844, 146

\bibitem[\protect\citeauthoryear{Collaboration et~al.,}{Collaboration
  et~al.}{2013}]{Collaboration2013Astropy}
Collaboration A.,  et~al., 2013, \mn@doi [\aap] {10.1051/0004-6361/201322068},
  558, A33

\bibitem[\protect\citeauthoryear{Copperwheat, Steele, Bates, Smith, Bode,
  Baker, Peacocke  \& Thomson}{Copperwheat
  et~al.}{2014}]{Copperwheat2014Liverpool}
Copperwheat C.~M.,  Steele I.~A.,  Bates S.~D.,  Smith R.~J.,  Bode M.~F.,
  Baker I.,  Peacocke T.,   Thomson K.,  2014. p. 914511,
  \mn@doi{10.1117/12.2055527}

\bibitem[\protect\citeauthoryear{Croton et~al.,}{Croton
  et~al.}{2006}]{Croton2006Many}
Croton D.~J.,  et~al., 2006, \mn@doi [\mnras]
  {10.1111/j.1365-2966.2005.09675.x}, 365, 11

\bibitem[\protect\citeauthoryear{Dawson et~al.,}{Dawson
  et~al.}{2013}]{Dawson2013Baryon}
Dawson K.~S.,  et~al., 2013, \mn@doi [\aj] {10.1088/0004-6256/145/1/10}, 145,
  10

\bibitem[\protect\citeauthoryear{Denney et~al.,}{Denney
  et~al.}{2010}]{Denney2010Reverberation}
Denney K.~D.,  et~al., 2010, \mn@doi [\apj] {10.1088/0004-637X/721/1/715}, 721,
  715

\bibitem[\protect\citeauthoryear{Du et~al.,}{Du
  et~al.}{2015}]{Du2015Supermassive}
Du P.,  et~al., 2015, \mn@doi [\apj] {10.1088/0004-637X/806/1/22}, 806, 22

\bibitem[\protect\citeauthoryear{Du et~al.,}{Du
  et~al.}{2016a}]{Du2016Supermassive}
Du P.,  et~al., 2016a, \mn@doi [\apj] {10.3847/0004-637X/820/1/27}, 820, 27

\bibitem[\protect\citeauthoryear{Du et~al.,}{Du
  et~al.}{2016b}]{Du2016Supermassivea}
Du P.,  et~al., 2016b, \mn@doi [\apj] {10.3847/0004-637X/825/2/126}, 825, 126

\bibitem[\protect\citeauthoryear{Edri, Rafter, Chelouche, Kaspi  \& Behar}{Edri
  et~al.}{2012}]{Edri2012Broadband}
Edri H.,  Rafter S.~E.,  Chelouche D.,  Kaspi S.,   Behar E.,  2012, \mn@doi
  [\apj] {10.1088/0004-637X/756/1/73}, 756, 73

\bibitem[\protect\citeauthoryear{Eisenstein et~al.,}{Eisenstein
  et~al.}{2011}]{Eisenstein2011SdssIii}
Eisenstein D.~J.,  et~al., 2011, \mn@doi [\aj] {10.1088/0004-6256/142/3/72},
  142, 72

\bibitem[\protect\citeauthoryear{Fausnaugh et~al.,}{Fausnaugh
  et~al.}{2017}]{Fausnaugh2017Reverberation}
Fausnaugh M.~M.,  et~al., 2017, \mn@doi [\apj] {10.3847/1538-4357/aa6d52}, 840,
  97

\bibitem[\protect\citeauthoryear{Feng, Shen  \& Li}{Feng
  et~al.}{2014}]{Feng2014SingleEpoch}
Feng H.,  Shen Y.,   Li H.,  2014, \mn@doi [\apj] {10.1088/0004-637X/794/1/77},
  794, 77

\bibitem[\protect\citeauthoryear{Ferrarese \& Merritt}{Ferrarese \&
  Merritt}{2000}]{Ferrarese2000Fundamental}
Ferrarese L.,  Merritt D.,  2000, \mn@doi [\apj] {10.1086/312838}, 539, L9

\bibitem[\protect\citeauthoryear{Fine et~al.,}{Fine
  et~al.}{2013}]{Fine2013Stacked}
Fine S.,  et~al., 2013, \mn@doi [\mnras] {10.1093/mnrasl/slt069}, 434, L16

\bibitem[\protect\citeauthoryear{Foreman-Mackey, Hogg, Lang  \&
  Goodman}{Foreman-Mackey et~al.}{2013}]{Mackey2013Emcee}
Foreman-Mackey D.,  Hogg D.~W.,  Lang D.,   Goodman J.,  2013, \mn@doi [\pasp]
  {10.1086/670067}, 125, 306

\bibitem[\protect\citeauthoryear{{Gaskell} \& {Sparke}}{{Gaskell} \&
  {Sparke}}{1986}]{Gaskell1986Line}
{Gaskell} C.~M.,  {Sparke} L.~S.,  1986, \mn@doi [\apj] {10.1086/164238}, \href
  {https://ui.adsabs.harvard.edu/abs/1986ApJ...305..175G} {305, 175}

\bibitem[\protect\citeauthoryear{{Gebhardt} et~al.,}{{Gebhardt}
  et~al.}{2000}]{Gebhardt2000Relationship}
{Gebhardt} K.,  et~al., 2000, \mn@doi [\apjl] {10.1086/312840}, \href
  {http://adsabs.harvard.edu/abs/2000ApJ...539L..13G} {539, L13}

\bibitem[\protect\citeauthoryear{Graham, Driver, Petrosian, Conselice,
  Bershady, Crawford  \& Goto}{Graham et~al.}{2005}]{Graham2005Total}
Graham A.~W.,  Driver S.~P.,  Petrosian V.,  Conselice C.~J.,  Bershady M.~A.,
  Crawford S.~M.,   Goto T.,  2005, \mn@doi [\aj] {10.1086/444475}, 130, 1535

\bibitem[\protect\citeauthoryear{Graham, Onken, Athanassoula  \& Combes}{Graham
  et~al.}{2011}]{Graham2011Expanded}
Graham A.~W.,  Onken C.~A.,  Athanassoula E.,   Combes F.,  2011, \mn@doi
  [\mnras] {10.1111/j.1365-2966.2010.18045.x}, 412, 2211

\bibitem[\protect\citeauthoryear{Grier et~al.,}{Grier
  et~al.}{2012}]{Grier2012Reverberation}
Grier C.~J.,  et~al., 2012, \mn@doi [\apj] {10.1088/0004-637X/755/1/60}, 755,
  60

\bibitem[\protect\citeauthoryear{Grier et~al.,}{Grier
  et~al.}{2013a}]{Grier2013Structure}
Grier C.~J.,  et~al., 2013a, \mn@doi [\apj] {10.1088/0004-637X/764/1/47}, 764,
  47

\bibitem[\protect\citeauthoryear{Grier et~al.,}{Grier
  et~al.}{2013b}]{Grier2013Stellar}
Grier C.~J.,  et~al., 2013b, \mn@doi [\apj] {10.1088/0004-637X/773/2/90}, 773,
  90

\bibitem[\protect\citeauthoryear{Grier et~al.,}{Grier
  et~al.}{2017}]{Grier2017Sloan}
Grier C.~J.,  et~al., 2017, \mn@doi [\apj] {10.3847/1538-4357/aa98dc}, 851, 21

\bibitem[\protect\citeauthoryear{Guo et~al.,}{Guo et~al.}{2011}]{Guo2011From}
Guo Q.,  et~al., 2011, \mn@doi [\mnras] {10.1111/j.1365-2966.2010.18114.x},
  413, 101

\bibitem[\protect\citeauthoryear{Guo, Wang, Cai  \& Sun}{Guo
  et~al.}{2017}]{Guo2017Far}
Guo H.,  Wang J.,  Cai Z.,   Sun M.,  2017, \mn@doi [ApJ]
  {10.3847/1538-4357/aa8d71}, 847, 132

\bibitem[\protect\citeauthoryear{Haas, Chini, Ramolla, Nunez, Westhues,
  Watermann, Hoffmeister  \& Murphy}{Haas et~al.}{2011}]{Haas2011Photometric}
Haas M.,  Chini R.,  Ramolla M.,  Nunez F.~P.,  Westhues C.,  Watermann R.,
  Hoffmeister V.,   Murphy M.,  2011, \mn@doi [\aap]
  {10.1051/0004-6361/201117325}, 535, A73

\bibitem[\protect\citeauthoryear{Haering \& Rix}{Haering \&
  Rix}{2004}]{Haering2004On}
Haering N.,  Rix H.-W.,  2004, \mn@doi [\apj] {10.1086/383567}, 604, L89

\bibitem[\protect\citeauthoryear{Heckman \& Best}{Heckman \&
  Best}{2014}]{Heckman2014Coevolution}
Heckman T.~M.,  Best P.~N.,  2014, \mn@doi [\aapr]
  {10.1146/annurev-astro-081913-035722}, 52, 589

\bibitem[\protect\citeauthoryear{Hernitschek, Rix, Bovy  \&
  Morganson}{Hernitschek et~al.}{2015}]{Hernitschek2015Estimating}
Hernitschek N.,  Rix H.-W.,  Bovy J.,   Morganson E.,  2015, \mn@doi [\apj]
  {10.1088/0004-637X/801/1/45}, 801, 45

\bibitem[\protect\citeauthoryear{Hiner, Cales, Calderon, Treister, Canalizo,
  Urry  \& Woo}{Hiner et~al.}{2015}]{Hiner2015Probing}
Hiner K.~D.,  Cales S.,  Calderon P.,  Treister E.,  Canalizo G.,  Urry C.~M.,
   Woo J.-H.,  2015, \aas~Meeting Abstracts \#225, 225, 432.09

\bibitem[\protect\citeauthoryear{Ho}{Ho}{2008}]{Ho2008Nuclear}
Ho L.~C.,  2008, \mn@doi [\aapr] {10.1146/annurev.astro.45.051806.110546}, 46,
  475

\bibitem[\protect\citeauthoryear{Honeycutt}{Honeycutt}{1992}]{Honeycutt1992Ccd}
Honeycutt R.~K.,  1992, \mn@doi [\pasp] {10.1086/133015}, 104, 435

\bibitem[\protect\citeauthoryear{Hood, Rivera, Thackeray-Lacko, Powers,
  Stuckey, Watson  \& Hood}{Hood et~al.}{2015}]{Hood2015Photometric}
Hood C.~E.,  Rivera N.~I.,  Thackeray-Lacko B.,  Powers R.~M.,  Stuckey H.,
  Watson R.,   Hood M.~A.,  2015, in \aas~Meeting Abstracts \#225. p. 144.11

\bibitem[\protect\citeauthoryear{Kasliwal, Vogeley  \& Richards}{Kasliwal
  et~al.}{2015a}]{Kasliwal2015Are}
Kasliwal V.~P.,  Vogeley M.~S.,   Richards G.~T.,  2015a, \mn@doi [MNRAS]
  {10.1093/mnras/stv1230}, 451, 4328

\bibitem[\protect\citeauthoryear{Kasliwal, Vogeley, Richards, Williams  \&
  Carini}{Kasliwal et~al.}{2015b}]{Kasliwal2015Do}
Kasliwal V.~P.,  Vogeley M.~S.,  Richards G.~T.,  Williams J.,   Carini M.~T.,
  2015b, \mn@doi [MNRAS] {10.1093/mnras/stv1797}, 453, 2075

\bibitem[\protect\citeauthoryear{Kasliwal, Vogeley  \& Richards}{Kasliwal
  et~al.}{2017}]{Kasliwal2017Extracting}
Kasliwal V.~P.,  Vogeley M.~S.,   Richards G.~T.,  2017, \mn@doi [MNRAS]
  {10.1093/mnras/stx1420}, 470, 3027

\bibitem[\protect\citeauthoryear{Kaspi, Smith, Netzer, Maoz, Jannuzi  \&
  Giveon}{Kaspi et~al.}{2000}]{Kaspi2000Reverberation}
Kaspi S.,  Smith P.~S.,  Netzer H.,  Maoz D.,  Jannuzi B.~T.,   Giveon U.,
  2000, \mn@doi [\apj] {10.1086/308704}, 533, 631

\bibitem[\protect\citeauthoryear{Kaspi, Brandt, Maoz, Netzer, Schneider  \&
  Shemmer}{Kaspi et~al.}{2007}]{Kaspi2007Reverberation}
Kaspi S.,  Brandt W.~N.,  Maoz D.,  Netzer H.,  Schneider D.~P.,   Shemmer O.,
  2007, \mn@doi [\apj] {10.1086/512094}, 659, 997

\bibitem[\protect\citeauthoryear{Kelly, Bechtold  \& Siemiginowska}{Kelly
  et~al.}{2009}]{Kelly2009Are}
Kelly B.~C.,  Bechtold J.,   Siemiginowska A.,  2009, \mn@doi [\apj]
  {10.1088/0004-637X/698/1/895}, 698, 895

\bibitem[\protect\citeauthoryear{{Kelly}, {Becker}, {Sobolewska}  \&
  {Uttley}}{{Kelly} et~al.}{2014}]{Kelly2014Flexible}
{Kelly} B.~C.,  {Becker} A.~C.,  {Sobolewska} Malgosia~and{Siemiginowska} A.,
  {Uttley} P.,  2014, \mn@doi [\apj] {10.1088/0004-637X/788/1/33}, \href
  {https://ui.adsabs.harvard.edu/abs/2014ApJ...788...33K} {788, 33}

\bibitem[\protect\citeauthoryear{Kormendy \& Ho}{Kormendy \&
  Ho}{2013}]{Kormendy2013Coevolution}
Kormendy J.,  Ho L.~C.,  2013, \mn@doi [\aapr]
  {10.1146/annurev-astro-082708-101811}, 51, 511

\bibitem[\protect\citeauthoryear{Liu, Feng  \& Bai}{Liu
  et~al.}{2017}]{Liu2017New}
Liu H.~T.,  Feng H.~C.,   Bai J.~M.,  2017, \mn@doi [\mnras]
  {10.1093/mnras/stw3261}, 466, 3323

\bibitem[\protect\citeauthoryear{Magorrian et~al.,}{Magorrian
  et~al.}{1998}]{Magorrian1998Demography}
Magorrian J.,  et~al., 1998, \mn@doi [\aj] {10.1086/300353}, 115, 2285

\bibitem[\protect\citeauthoryear{Marshall et~al.,}{Marshall
  et~al.}{2017}]{Marshall2017Lsst}
Marshall P.,  et~al., 2017, Lsst Science Collaborations Observing Strategy
  White Paper: "Science-Driven Optimization Of The Lsst Observing Strategy,
  \url {https://zenodo.org/record/842712}

\bibitem[\protect\citeauthoryear{McLure \& Dunlop}{McLure \&
  Dunlop}{2001}]{Mclure2001Black}
McLure R.~J.,  Dunlop J.~S.,  2001, \mn@doi [\mnras]
  {10.1046/j.1365-8711.2001.04709.x}, 327, 199

\bibitem[\protect\citeauthoryear{McLure \& Dunlop}{McLure \&
  Dunlop}{2004}]{Mclure2004Cosmological}
McLure R.~J.,  Dunlop J.~S.,  2004, \mn@doi [\mnras]
  {10.1111/j.1365-2966.2004.08034.x}, 352, 1390

\bibitem[\protect\citeauthoryear{McLure \& Jarvis}{McLure \&
  Jarvis}{2002}]{Mclure2002Measuring}
McLure R.~J.,  Jarvis M.~J.,  2002, \mn@doi [\mnras]
  {10.1046/j.1365-8711.2002.05871.x}, 337, 109

\bibitem[\protect\citeauthoryear{Mejía-Restrepo, Trakhtenbrot, Lira, Netzer
  \& Capellupo}{Mejía-Restrepo et~al.}{2016}]{Restrepo2016Active}
Mejía-Restrepo J.~E.,  Trakhtenbrot B.,  Lira P.,  Netzer H.,   Capellupo
  D.~M.,  2016, \mn@doi [\mnras] {10.1093/mnras/stw568}, 460, 187

\bibitem[\protect\citeauthoryear{{Mellen}}{{Mellen}}{1952}]{Mellen1952Thermal}
{Mellen} R.~H.,  1952, \mn@doi [ASAJ] {10.1121/1.1906924}, \href
  {https://ui.adsabs.harvard.edu/abs/1952ASAJ...24..478M} {24, 478}

\bibitem[\protect\citeauthoryear{{Mudd} et~al.,}{{Mudd}
  et~al.}{2018}]{Mudd2017Quasar}
{Mudd} D.,  et~al., 2018, \mn@doi [\apj] {10.3847/1538-4357/aac9bb}, \href
  {https://ui.adsabs.harvard.edu/\#abs/2018ApJ...862..123M} {862, 123}

\bibitem[\protect\citeauthoryear{Mushotzky, Edelson, Baumgartner  \&
  Gandhi}{Mushotzky et~al.}{2011}]{Mushotzky2011Kepler}
Mushotzky R.~F.,  Edelson R.,  Baumgartner W.~H.,   Gandhi P.,  2011, \mn@doi
  [\apj] {10.1088/2041-8205/743/1/L12}, 743, L12

\bibitem[\protect\citeauthoryear{Netzer, Lira, Trakhtenbrot, Shemmer  \&
  Cury}{Netzer et~al.}{2007}]{Netzer2007Black}
Netzer H.,  Lira P.,  Trakhtenbrot B.,  Shemmer O.,   Cury I.,  2007, \mn@doi
  [\apj] {10.1086/523035}, 671, 1256

\bibitem[\protect\citeauthoryear{Onken, Ferrarese, Merritt, Peterson, Pogge,
  Vestergaard  \& Wandel}{Onken et~al.}{2004}]{Onken2004Supermassive}
Onken C.~A.,  Ferrarese L.,  Merritt D.,  Peterson B.~M.,  Pogge R.~W.,
  Vestergaard M.,   Wandel A.,  2004, \mn@doi [\apj] {10.1086/424655}, 615, 645

\bibitem[\protect\citeauthoryear{O’Brien, Collins, Rauscher  \&
  Ringler}{O’Brien et~al.}{2014}]{Brien2014Reducing}
O’Brien T.~A.,  Collins W.~D.,  Rauscher S.~A.,   Ringler T.~D.,  2014,
  \mn@doi [CS \& DA] {10.1016/j.csda.2014.06.002}, 79, 222

\bibitem[\protect\citeauthoryear{O’Brien, Kashinath, Cavanaugh, Collins  \&
  O’Brien}{O’Brien et~al.}{2016}]{Brien2016Fast}
O’Brien T.~A.,  Kashinath K.,  Cavanaugh N.~R.,  Collins W.~D.,   O’Brien
  J.~P.,  2016, \mn@doi [CS \& DA] {10.1016/j.csda.2016.02.014}, 101, 148

\bibitem[\protect\citeauthoryear{Padilla et~al.,}{Padilla
  et~al.}{2016}]{Padilla2016Pau}
Padilla C.,  et~al., 2016. \procspie.
p. 99080Z, \mn@doi{10.1117/12.2231884}

\bibitem[\protect\citeauthoryear{Pancoast, Brewer  \& Treu}{Pancoast
  et~al.}{2011}]{Pancoast2011Geometric}
Pancoast A.,  Brewer B.~J.,   Treu T.,  2011, \mn@doi [\apj]
  {10.1088/0004-637X/730/2/139}, 730, 139

\bibitem[\protect\citeauthoryear{Pancoast, Brewer, Treu, Park, Barth, Bentz  \&
  Woo}{Pancoast et~al.}{2014}]{Pancoast2014Modelling}
Pancoast A.,  Brewer B.~J.,  Treu T.,  Park D.,  Barth A.~J.,  Bentz M.~C.,
  Woo J.-H.,  2014, \mn@doi [\mnras] {10.1093/mnras/stu1419}, 445, 3073

\bibitem[\protect\citeauthoryear{Park, Kelly, Woo  \& Treu}{Park
  et~al.}{2012}]{Park2012Recalibration}
Park D.,  Kelly B.~C.,  Woo J.-H.,   Treu T.,  2012, \mn@doi [\apjs]
  {10.1088/0067-0049/203/1/6}, 203, 6

\bibitem[\protect\citeauthoryear{Peterson}{Peterson}{2004}]{Peterson2004Black}
Peterson B.~M.,  2004, \mn@doi [\prociau] {10.1017/S1743921304001358}, 2004, 15

\bibitem[\protect\citeauthoryear{Petrosian}{Petrosian}{1976}]{Petrosian1976Surface}
Petrosian V.,  1976, \mn@doi [\apj] {10.1086/182253}, 209, L1

\bibitem[\protect\citeauthoryear{{Pozo Nu{\~n}ez} et~al.,}{{Pozo Nu{\~n}ez}
  et~al.}{2019}]{Nunez2019Optical}
{Pozo Nu{\~n}ez} F.,  et~al., 2019, \mn@doi [\mnras] {10.1093/mnras/stz2830},
  \href {https://ui.adsabs.harvard.edu/abs/2019MNRAS.490.3936P} {490, 3936}

\bibitem[\protect\citeauthoryear{Pozo~Nuñez, Ramolla, Westhues, Bruckmann,
  Haas, Chini, Steenbrugge  \& Murphy}{Pozo~Nuñez
  et~al.}{2012}]{Nunez2012Photometric}
Pozo~Nuñez F.,  Ramolla M.,  Westhues C.,  Bruckmann C.,  Haas M.,  Chini R.,
  Steenbrugge K.,   Murphy M.,  2012, \mn@doi [\aap]
  {10.1051/0004-6361/201219107}, 545, A84

\bibitem[\protect\citeauthoryear{Pozo~Nuñez et~al.,}{Pozo~Nuñez
  et~al.}{2014}]{Nunez2014Modelling}
Pozo~Nuñez F.,  et~al., 2014, \mn@doi [\aap] {10.1051/0004-6361/201322736},
  568, A36

\bibitem[\protect\citeauthoryear{Pozo~Nuñez et~al.,}{Pozo~Nuñez
  et~al.}{2015}]{Nunez2015BroadLine}
Pozo~Nuñez F.,  et~al., 2015, \mn@doi [\aap] {10.1051/0004-6361/201525910},
  576, A73

\bibitem[\protect\citeauthoryear{Ramolla, Pozo, Westhues, Haas, Chini,
  Steenbrugge, Lemke  \& Murphy}{Ramolla et~al.}{2014}]{Ramolla2014Photometric}
Ramolla M.,  Pozo F.,  Westhues C.,  Haas M.,  Chini R.,  Steenbrugge K.,
  Lemke R.,   Murphy M.,  2014, \rmxaa, 45, 79

\bibitem[\protect\citeauthoryear{Runnoe, Brotherton, Shang  \& DiPompeo}{Runnoe
  et~al.}{2013}]{Runnoe2013Rehabilitating}
Runnoe J.~C.,  Brotherton M.~S.,  Shang Z.,   DiPompeo M.~A.,  2013, \mn@doi
  [\mnras] {10.1093/mnras/stt1077}, 434, 848

\bibitem[\protect\citeauthoryear{{Rybicki} \& {Kleyna}}{{Rybicki} \&
  {Kleyna}}{1994}]{Rybicki1994Study}
{Rybicki} G.~B.,  {Kleyna} J.~T.,  1994, in Reverberation Mapping of the
  Broad-Line Region in Active Galactic Nuclei. p.~85

\bibitem[\protect\citeauthoryear{Shen et~al.,}{Shen
  et~al.}{2011}]{Shen2011Catalog}
Shen Y.,  et~al., 2011, \mn@doi [\apjs] {10.1088/0067-0049/194/2/45}, 194, 45

\bibitem[\protect\citeauthoryear{Shen et~al.,}{Shen
  et~al.}{2015a}]{Shen2015Sloan}
Shen Y.,  et~al., 2015a, \mn@doi [\apjs] {10.1088/0067-0049/216/1/4}, 216, 4

\bibitem[\protect\citeauthoryear{Shen et~al.,}{Shen
  et~al.}{2015b}]{Shen2015Sloana}
Shen Y.,  et~al., 2015b, \mn@doi [\apj] {10.1088/0004-637X/805/2/96}, 805, 96

\bibitem[\protect\citeauthoryear{{Silk} \& {Rees}}{{Silk} \&
  {Rees}}{1998}]{Silk1998Quasars}
{Silk} J.,  {Rees} M.~J.,  1998, \aap, \href
  {http://adsabs.harvard.edu/abs/1998A%26A...331L...1S} {331, L1}

\bibitem[\protect\citeauthoryear{Smith, Mushotzky, Boyd, Malkan, Howell  \&
  Gelino}{Smith et~al.}{2018}]{Smith2018Kepler}
Smith K.~L.,  Mushotzky R.~F.,  Boyd P.~T.,  Malkan M.,  Howell S.~B.,   Gelino
  D.~M.,  2018, \mn@doi [\apj] {10.3847/1538-4357/aab88d}, 857, 141

\bibitem[\protect\citeauthoryear{{Speagle}}{{Speagle}}{2019}]{Speagle2019Dynesty}
{Speagle} J.~S.,  2019, arXiv e-prints, \href
  {https://ui.adsabs.harvard.edu/abs/2019arXiv190402180S} {p. arXiv:1904.02180}

\bibitem[\protect\citeauthoryear{Steele et~al.,}{Steele
  et~al.}{2004}]{Steele2004Liverpool}
Steele I.~A.,  et~al., 2004. pp 679--692, \mn@doi{10.1117/12.551456}

\bibitem[\protect\citeauthoryear{Urry \& Padovani}{Urry \&
  Padovani}{1995}]{Urry1995Unified}
Urry C.~M.,  Padovani P.,  1995, \mn@doi [\pasp] {10.1086/133630}, 107, 803

\bibitem[\protect\citeauthoryear{Vestergaard}{Vestergaard}{2004}]{Vestergaard2004Early}
Vestergaard M.,  2004, \mn@doi [\apj] {10.1086/379758}, 601, 676

\bibitem[\protect\citeauthoryear{Vestergaard \& Peterson}{Vestergaard \&
  Peterson}{2006}]{Vestergaard2006Determining}
Vestergaard M.,  Peterson B.~M.,  2006, \mn@doi [\apj] {10.1086/500572}, 641,
  689

\bibitem[\protect\citeauthoryear{Wandel, Peterson  \& Malkan}{Wandel
  et~al.}{1999}]{Wandel1999Central}
Wandel A.,  Peterson B.~M.,   Malkan M.~A.,  1999, \mn@doi [\apj]
  {10.1086/308017}, 526, 579

\bibitem[\protect\citeauthoryear{White \& Peterson}{White \&
  Peterson}{1994}]{White1994Comments}
White R.~J.,  Peterson B.~M.,  1994, \mn@doi [\pasp] {10.1086/133456}, 106

\bibitem[\protect\citeauthoryear{Williams et~al.,}{Williams
  et~al.}{2018}]{Williams2018Lick}
Williams P.~R.,  et~al., 2018, \mn@doi [\apj] {10.3847/1538-4357/aae086}, 866,
  75

\bibitem[\protect\citeauthoryear{York}{York}{2000}]{York2000Sloan}
York D.~G.,  2000, \mn@doi [\aj] {10.1086/301513}, 120, 1579

\bibitem[\protect\citeauthoryear{Zhang, Yang  \& Wu}{Zhang
  et~al.}{2018}]{Zhang2017Broadband}
Zhang H.,  Yang Q.,   Wu X.-B.,  2018, \mn@doi [\apj]
  {10.3847/1538-4357/aaa3e5}, 853, 116

\bibitem[\protect\citeauthoryear{Zu, Kochanek  \& Peterson}{Zu
  et~al.}{2011}]{Zu2011Alternative}
Zu Y.,  Kochanek C.~S.,   Peterson B.~M.,  2011, \mn@doi [\apj]
  {10.1088/0004-637X/735/2/80}, 735, 80

\bibitem[\protect\citeauthoryear{Zu, Kochanek, Kozłowski  \& Udalski}{Zu
  et~al.}{2013}]{Zu2013Quasar}
Zu Y.,  Kochanek C.~S.,  Kozłowski S.,   Udalski A.,  2013, \mn@doi [\apj]
  {10.1088/0004-637X/765/2/106}, 765, 106

\bibitem[\protect\citeauthoryear{Zu, Kochanek, Kozłowski  \& Peterson}{Zu
  et~al.}{2016}]{Zu2016Application}
Zu Y.,  Kochanek C.~S.,  Kozłowski S.,   Peterson B.~M.,  2016, \mn@doi [\apj]
  {10.3847/0004-637X/819/2/122}, 819, 122

\makeatother
\end{thebibliography}

\appendix
\section{Photometric Calibration}\label{app:calibration}
\begin{figure}
  \centering
  \includegraphics[width=\linewidth]{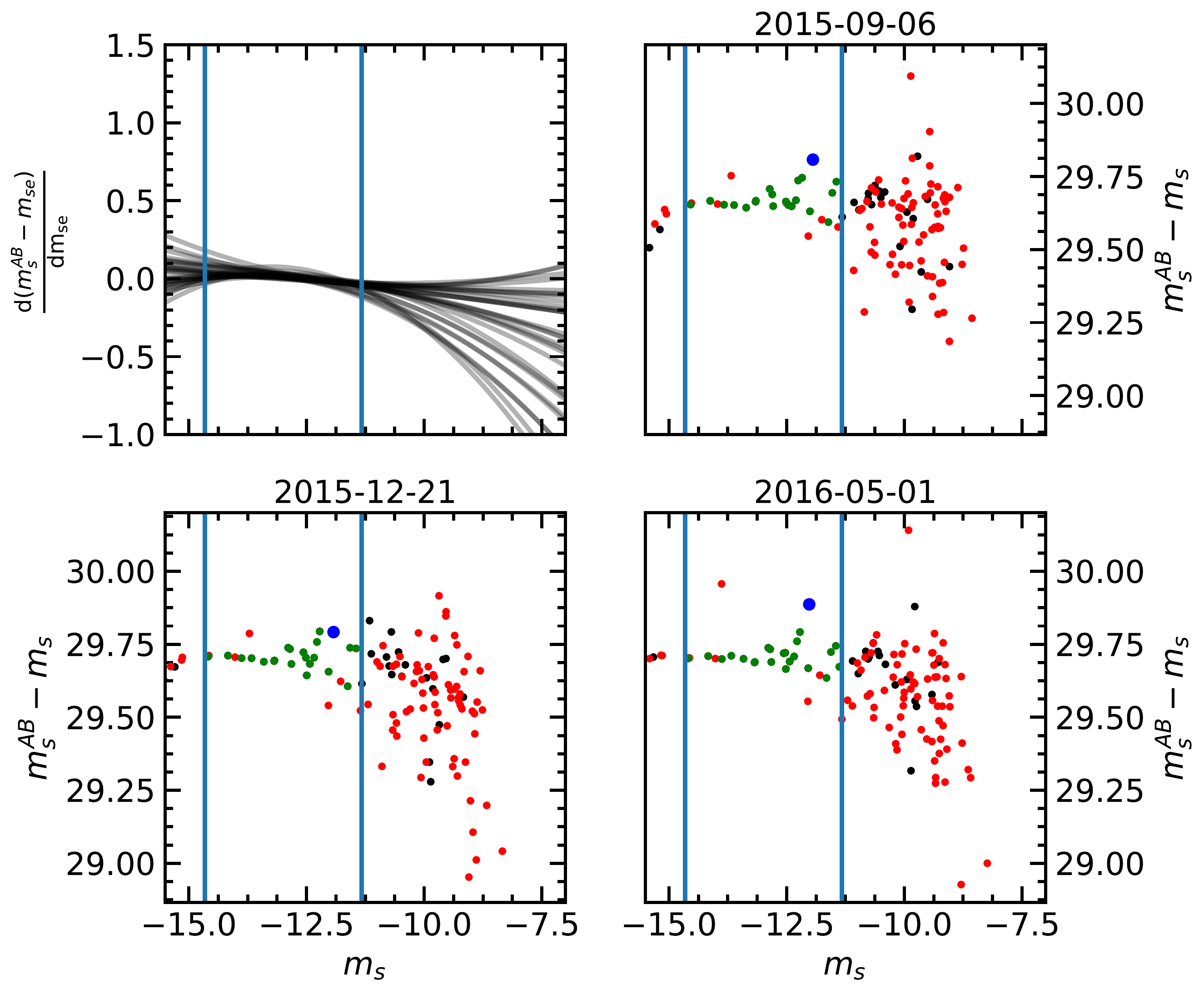}
  \caption{The selection of acceptable reference sources for \TargetTen in the \iband. 
  \textbf{Top left:} The derivative of the splines that were fit to the difference between SDSS AB Petrosian magnitude, $m_{AB}$, and \sextractor instrumental Petrosian magnitude, $m_s$, for each exposure.
  \textbf{Other quadrants:} a sample of 3 exposures are shown in the other 3 quadrants of this figure.
  The region where the gradient of all splines is less than \SplineGradientTolerance, where acceptable sources can be found, is bounded by two vertical lines.
  All sources plotted here have SDSS \textsc{clean}$=$\textsc{True} \& \sextractor \textsc{FLAGS}$=0$.
  Those sources whose extracted Petrosian magnitude extracted from the SDSS calibrated images is the same (not the same) as that extracted from the same image using \sextractor, to within \SDSSMagnitudeAggreement, are shown in black (red).
  Those sources which are accepted for use as reference sources by spline fitting (see section~\ref{sec:RM:methods:calibration}) are shown in green. 
  \TargetTen is shown in blue.}
  \label{fig:RM:iband_calibration_splines}
\end{figure}

\begin{figure}
  \centering
  \includegraphics[width=0.9\linewidth]{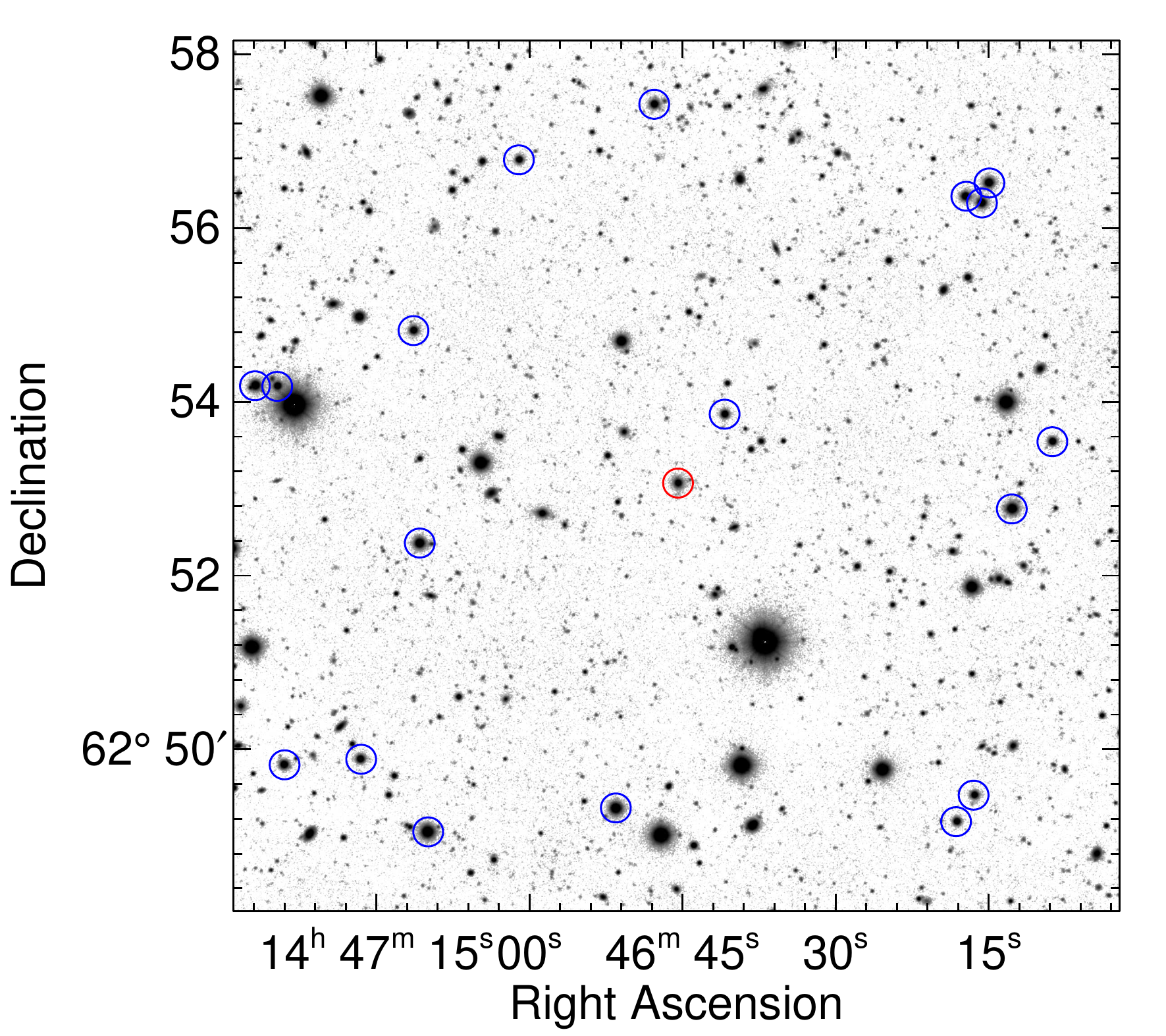}
  \caption{Stacked \iband exposures for \TargetTen. 
  The QSO is circled in red and its accepted references are circled in blue.}
  \label{fig:RM:iband_mosaic}
\end{figure}

\begin{figure}
  \centering
  \includegraphics[width=0.9\linewidth]{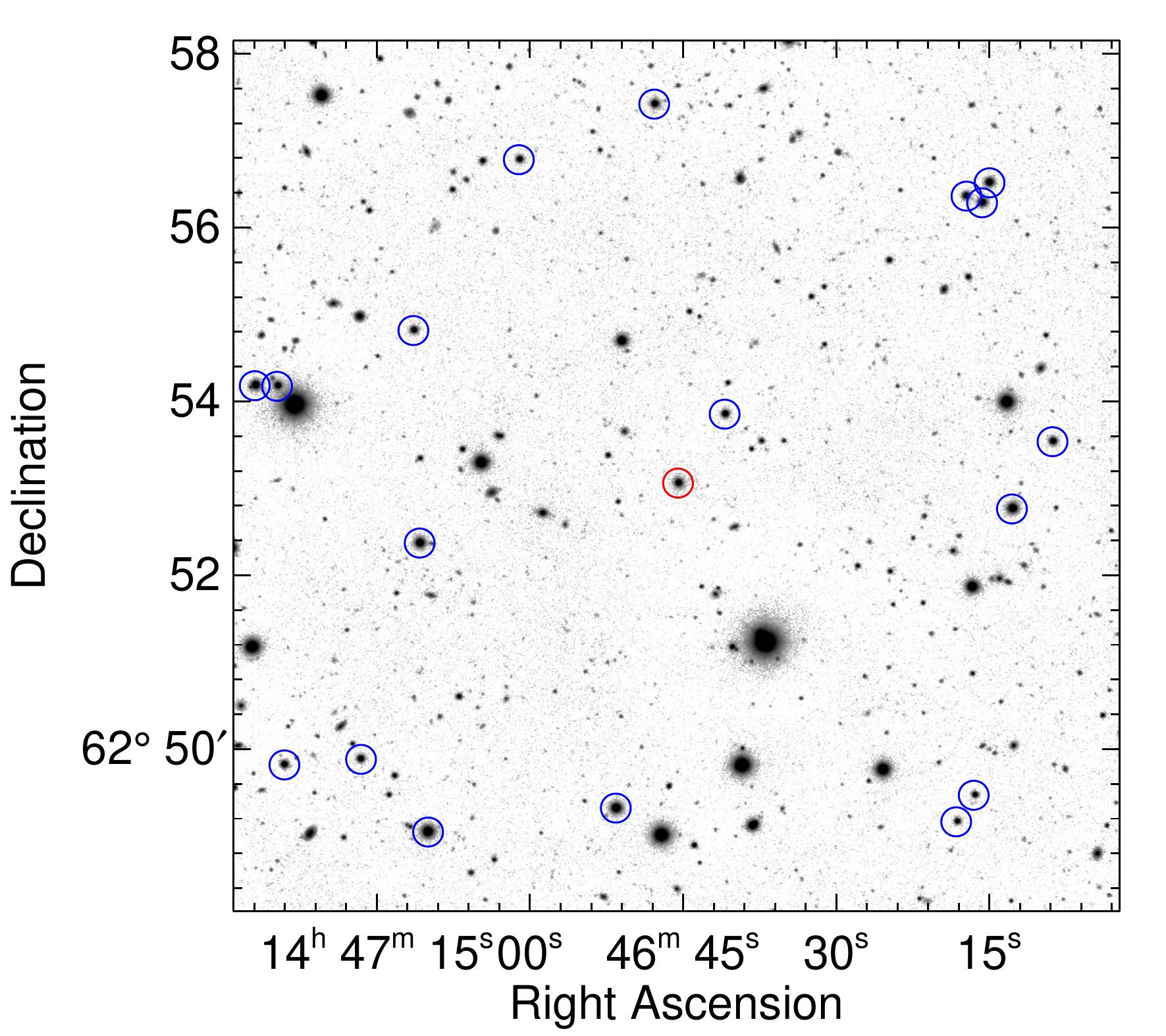}
  \caption{Stacked \halpha exposures for \TargetTen. 
  The QSO is circled in red and its accepted references are circled in blue.}
  \label{fig:RM:ha_mosaic}
\end{figure}

To further improve our set of reference sources, we perform a number of checks.
First, we perform the same aperture photometry extraction using \sextractor that we used on our own \iband exposures on the SDSS \iband exposures that contain the candidate reference sources.
If the Petrosian magnitude extracted from SDSS exposures by \sextractor does not agree  with the Petrosian magnitude quoted in the SDSS DR12 catalogue to within \SDSSMagnitudeAggreement, then we discard the source.
This leaves the sources depicted in green in Fig~\ref{fig:RM:iband_calibration_splines}.
Ideally, we would fit a single value of $m^{AB}_{s} - m_{s}$ across all instrumental magnitudes $m_s$ to measure the \iband zeropoint. 
However, as shown for the three example exposures in Fig~\ref{fig:RM:iband_calibration_splines}, the IO:O CCD can become saturated for many bright sources and faint sources are noisy.
This results in non-linearity at both high and low magnitudes.
We therefore employ a spline-based technique to select a contiguous range of \sextractor magnitudes containing ``well-behaved'' sources, where we can fit a single flat \iband zeropoint.
We fit a spline to $m^{AB}_{s} - m_{s}$ against $m_{s}$ and find the range in which the gradient of the spline is $0\pm$\SplineGradientTolerance. 
This range corresponds to the region where aperture photometry is the least affected by saturation and noise, and is shown in the first quadrant of Fig~\ref{fig:RM:iband_calibration_splines}.
We then select those candidate reference sources which have instrumental magnitudes within that range.
These sources, along with \TargetTen, are highlighted in Fig~\ref{fig:RM:iband_mosaic} and Fig~\ref{fig:RM:ha_mosaic}.

In order to estimate the exposure zeropoints and their uncertainties to the greatest accuracy, we employ an ensemble photometry technique similar to \citet{Honeycutt1992Ccd}.
We start out by fitting the instrumental magnitudes to SDSS AB magnitudes whilst also fitting a mean instrumental magnitude, $\hat{m}_r$, assuming that our reference sources are non-variable. 
This results in a log-likelihood given by

\begin{equation}
  \ln{\mathcal{L}} \propto \sum^{N_e}_{e=1}\sum^{N_r}_{r=1}(m_{er} + \hat{z}_e - \hat{m}_r)^2w_{er} 
  + 
  \sum^{N_r}_{r=1}\left(\frac{\hat{m}_r - m^{AB}_{r}}{\sigma^{AB}_{r}}\right)^2
\end{equation}

\noindent where $m_{er}$ is the instrumental magnitude for reference source $r$ in exposure $e$ with weighting $w_{er}$, $\hat{m}_r$ is the magnitude of reference source $r$ assuming that it does not vary over the course of observations, $\hat{z}_e$ is the zeropoint for exposure $e$, and $m^{AB}_{r}$ is the AB magnitude of reference source $r$ as measured by SDSS with its associated uncertainty $\sigma^{AB}_{r}$. 
We begin the fitting procedure by setting the weight $w_{er}$ for each reference source at each exposure to the instrumental magnitude uncertainty given by \sextractor, $1 / \sigma^2_{er}$. 
We then fit the quantities $\hat{m}_r$ and $\hat{z}_e$ using \textsc{emcee} \citep{Mackey2013Emcee} with 20 walkers until chain convergence is observed.  

Some reference sources may indeed vary over the course of our observations.
In addition, the instrumental uncertainty from \sextractor may be underestimated by some factor.
In order to reduce the offset to the zeropoint caused by the inclusion of varying sources, we scale the initial weighting by its probability in a fit Student-T distribution:
\begin{gather}
  \nonumber w_{er} \rightarrow \frac{p_{er}}{\sigma_{er}^2},\\
  p_{er} = T(m_{er}-\hat{m}_r | \hat{\mu}=0, \hat{\lambda}, \hat{\nu})
\end{gather}

\noindent where the inverse scale parameter, $\hat{\lambda}$, and number of degrees of freedom, $\hat{\nu}$, are both fit to the distribution of $m_{er}-\hat{m}_r$ assuming a mean of $\hat{\mu} =0$. 
The Student-T distribution fit to the distribution of deviations of the instrumental magnitudes from their estimated mean (\ie the distribution of the values of the black points in Fig~\ref{fig:RM:iband_zeropoints_and_qso_light_curve}), will update the weighting of each magnitude in each exposure and therefore assign very low weighting to sources which have larger variability over the course of our observations than others.
We iteratively run this re-weighting procedure until each flux measurement in the light curve of the target QSO no longer changes within a tolerance of \IterativeReweightingTolerance. 
This typically takes 3-5 runs of MCMC inference, updating the weighting each time.
The resulting light-curves are shown in Table~\ref{tab:RM:lightcurves}.
\begin{table}
\centering
\begin{tabular}{lll}
\toprule
t / days &      \iband / mJy &     \halpha / mJy\\
\midrule
0   &  $0.278\pm0.004$ &  $0.387\pm0.006$ \\
6   &  $0.298\pm0.003$ &  $0.392\pm0.006$ \\
91  &  $0.277\pm0.008$ &  $0.379\pm0.016$ \\
108 &  $0.280\pm0.005$ &  $0.379\pm0.007$ \\
112 &  $0.281\pm0.004$ &  $0.387\pm0.005$ \\
119 &  $0.291\pm0.006$ &  $0.379\pm0.010$ \\
128 &  $0.285\pm0.003$ &  $0.389\pm0.005$ \\
132 &  $0.289\pm0.003$ &  $0.390\pm0.005$ \\
140 &  $0.297\pm0.003$ &  $0.405\pm0.006$ \\
149 &  $0.303\pm0.006$ &  $0.412\pm0.010$ \\
153 &  $0.302\pm0.005$ &  $0.396\pm0.010$ \\
167 &  $0.307\pm0.003$ &  $0.439\pm0.006$ \\
188 &  $0.318\pm0.003$ &  $0.414\pm0.006$ \\
202 &  $0.319\pm0.003$ &  $0.413\pm0.008$ \\
209 &  $0.298\pm0.004$ &  $0.416\pm0.015$ \\
217 &  $0.309\pm0.003$ &  $0.418\pm0.008$ \\
223 &  $0.311\pm0.003$ &  $0.411\pm0.006$ \\
237 &  $0.300\pm0.004$ &  $0.416\pm0.006$ \\
244 &  $0.307\pm0.003$ &  $0.421\pm0.005$ \\
252 &  $0.307\pm0.003$ &  $0.423\pm0.007$ \\
258 &  $0.305\pm0.003$ &  $0.415\pm0.006$ \\
265 &  $0.261\pm0.009$ &  $0.373\pm0.019$ \\
272 &  $0.306\pm0.003$ &  $0.420\pm0.006$ \\
284 &  $0.303\pm0.003$ &  $0.411\pm0.006$ \\
286 &                - &  $0.423\pm0.006$ \\
293 &  $0.290\pm0.004$ &  $0.405\pm0.008$ \\
304 &  $0.306\pm0.003$ &  $0.418\pm0.006$ \\
311 &  $0.310\pm0.003$ &  $0.412\pm0.006$ \\
315 &  $0.321\pm0.004$ &  $0.411\pm0.009$ \\
322 &  $0.304\pm0.005$ &  $0.420\pm0.010$ \\
329 &  $0.301\pm0.005$ &  $0.419\pm0.009$ \\
\bottomrule
\end{tabular}
\caption{\TargetTen lightcurves for SDSS \iband and the \halpha filter.}
\label{tab:RM:lightcurves}
\end{table}
 
\section{QSO-Host Decomposition}\label{app:host-deconvolution}
We correct for the contribution of the host galaxy by fitting a host disc and QSO point source, both convolved with the SDSS \iband PSF obtained from the relevant \texttt{pSField} file eigenimages, to the SDSS \iband photometry.

\begin{figure*}
  \centering
  \includegraphics[width=\linewidth]{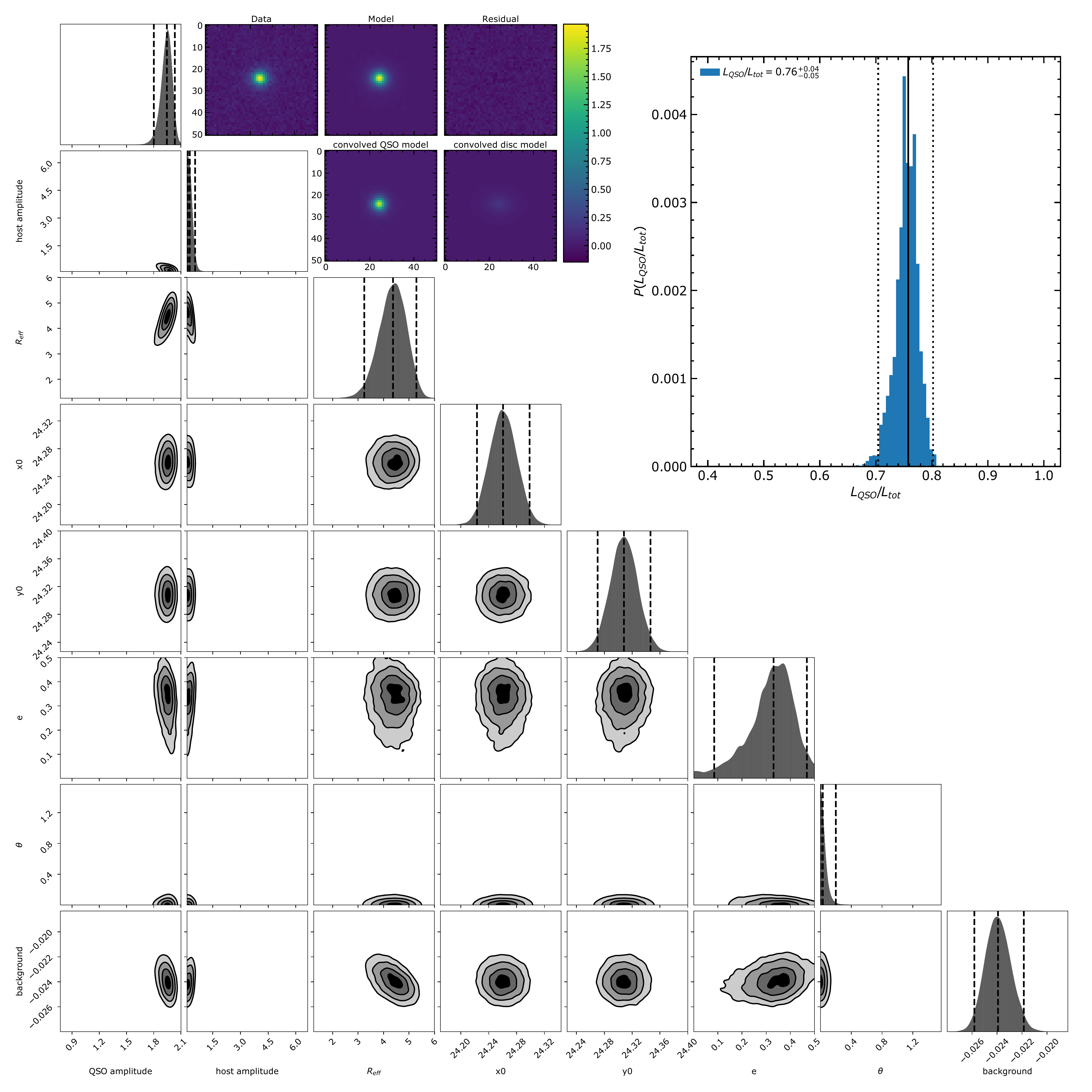}
  \caption{The posterior distribution of the QSO+Host fit to SDSS \iband data.
           The bounds of the cornerplot axes indicate the bounds of the uniform prior used in the nested sampling, except $x0$ and $y0$ for which the prior is normal with a width of 2 pixels. 
           The parameters left to right are QSO amplitude, host amplitude, effective radius of the disc, the centre point, ellipticity, orientation angle, and background.
           The inset histogram is the derived posterior distribution of the ratio of QSO luminosity to host luminosity.
           The maximum posterior image of the QSO+host model is shown in the 5 top left axes.
           The top three images show the total model and its residuals from the data.
           The bottom two show the QSO and disc components convolved with the PSF separately.}
  \label{fig:RM:qso_host_cornerplot}
\end{figure*}

\onecolumn
We use the nested sampler \texttt{Dynesty} \citep{Speagle2019Dynesty} and allow all parameters to vary including the background, orientation, ellipticity, and centre point.
We use uniform priors on each parameter as shown in Fig~\ref{fig:RM:qso_host_cornerplot} except the centre point $x0,y0$, for which we impose a normal prior distribution at the measured RA and Dec of the target with a width of 2 pixels.
As shown in Fig~\ref{fig:RM:qso_host_cornerplot}, we find strong constraints of the contribution of the host (24 per cent) and the maximum posterior model image residual shows that we have successfully modelled \TargetTen.

We also fit a QSO+disc+bulge with a Sersic index of 4, but the data does not support the additional complexity of another component, with Bayes factor of $\log[P(\textrm{data} | \textrm{disc}) / P(\textrm{data} | \textrm{disc+bulge})] = \log[B_{d,d+b}] = 2.3$ in favour of the simpler model.

\label{lastpage}
\end{document}